\documentclass[1p,number,11pt,compress]{elsarticle}
\usepackage{latexsym,amsmath,amsfonts,amssymb,graphics,color,bm,mathrsfs}
\usepackage{amsthm} % Ambiente Proofaa
\usepackage{tikz}
\usepackage[utf8]{inputenc}
\usepackage[T1]{fontenc}

\journal{Journal of Differential Equations}

\usepackage{xcolor}              % Para definir cores.
\definecolor{a1}{rgb}{0,0,0.8}   % Define a cor e o seu nome no sistema (R,G,B).
\usepackage{colortbl}	           % Para inserir cores em tabelas.
\usepackage[colorlinks=true,linktocpage=true,hyperindex=true,pageanchor=true,bookmarks=true]{hyperref} % Hiper-texto (LaTeX=>PDF).
\hypersetup{citecolor=a1,linkcolor=a1,menucolor=a1,urlcolor=a1,runcolor=a1}
\hypersetup{citebordercolor=a1,linkbordercolor=a1,urlbordercolor=a1,runbordercolor=a1}
\hypersetup{hypertexnames=false} % Nao usa a adivinhacao de links.a

\usepackage{upgreek}
\usepackage{nicefrac}
\usepackage{tikz}
\usepackage{graphicx}
\usepackage{amsmath}
\usepackage{relsize}
\DeclareMathAlphabet{\mathpzc}{OT1}{pzc}{m}{it}
\usepackage{upgreek}
\usepackage{nicefrac}
\usepackage{slashed}

\newtheorem{theorem}{\bf Theorem}[section]

\newtheorem{lem}{\bf Lemma}[section]

\newcommand{\be}{\begin{equation}}
\newcommand{\ee}{\end{equation}}

\usepackage{framed,color}
\definecolor{shadecolor}{rgb}{0.5,0.8,0.3}
\usepackage{xcolor}
\definecolor{cadmiumgreen}{rgb}{0.01, 0.42, 0.24}
\definecolor{orange}{rgb}{1.3,0.3,0.25}
\definecolor{Asparagus}{rgb}{0.53, 0.66, 0.42} 	
\definecolor{Army green}{rgb}{0.29, 0.33, 0.13}
\definecolor{Scarlet}{rgb}{1.0, 0.13, 0.0}
\definecolor{Goldmetal}{rgb}{0.47,0.39,0.14}
\definecolor{Goldyellow}{rgb}{0.88,0.5,0.15}

\DeclareMathAlphabet{\mathpzc}{OT1}{pzc}{m}{it}
\usepackage{upgreek}
\usepackage{nicefrac}

\begin{document}
\begin{frontmatter}
\title{Lorentz-equivariant flow with four delays of neutral type}
\author[fsc]{Jayme De Luca}
\ead{jayme.deluca@gmail.com}
\address[fsc]{Departamento de F\'{\i}sica,
Universidade Federal de S\~{a}o Carlos,
S\~{a}o Carlos, S\~{a}o Paulo 13565-905, Brazil}

\begin{abstract}
 We generalize electrodynamics with a second interaction in lightcone. The time-reversible equations for two-body motion define a semiflow on {\small $C^2(\mathbb{R})$} with four state-dependent delays of neutral type and nonlinear gyroscopic terms. Furthermore, if the initial segment includes velocity discontinuities, their propagation requires two energetic corner conditions defining boundary layer neighborhoods of large velocities and small denominators. Finally, we discuss a motion restricted to a straight line and a segment pair with vanishing accelerations that iterates to another constant-velocity segment pair.  
 \end{abstract}

%\pacs{05.45.-a, 02.30.Ks, 03.50.De,41.60.-m}

\begin{keyword}
  neutral differential-delay equations, state-dependent delay, ODEs, and semiflows. %\\ (*) This is arXiv:1606.07646v19.a
\end{keyword}

\end{frontmatter}

\section{Introduction}
\subsection{Significance of the problem and contents}
\label{significance}
Variational electrodynamics has neutral differential-delay equations of motion (NDDE) with state-dependent delays \cite{cinderela,JDE1,JDE2}. A hindrance to integrating an electromagnetic NDDE  forward is the non-invertibility of the linear form containing the most advanced accelerations. Here we cure the non-invertibility by adding a Lorentz-invariant unfolding term to the action functional. For the two-body problem, the resulting equation of motion is a NDDE with four state-dependent \textbf{delays}, which can be integrated, using, for example, the MATLAB function \textbf{ddensd}.  A variational principle is an economical way to derive time-reversible Lorentz-equivariant models possessing differential-delay equations of motion free of divergencies. Some important details to keep in mind are
\begin{enumerate} 
\item NDDEs are semiflows on infinite-dimensional spaces \cite{Mallet-Paret1,Mallet-Paret_bound-layer,JackHale,Nicola2}. Unlike the case of an ordinary differential equation (ODE), a NDDE needs a \textbf{trajectory segment} as the initial condition. 
\item The NDDEs studied here can start from trajectory segments possessing only two derivatives. For example, the acceleration segment can be the Takagi-van der Waerden function. The velocity will be the integral of the continuous acceleration, thus belonging to {\small $C^1(\mathbb{R})$}. Orbital segments of {\small $C^2(\mathbb{R})$} with accelerations that are  continuous and nowhere differentiable are henceforth referred to as \textit{serrated} orbits. 
\item Trajectory termination by the particles crashing at the speed of light is a difficulty when starting from {\small $C^\infty$} segments. Velocity discontinuities are instrumental to avoid head-on collisions in the case of attractive interaction. All initial segments that avoid crashing at the speed of light and produce bounded orbits without causing large far-fields have velocity discontinuities \cite{EFY2}. Likewise, the tri-dimensional motion with attractive interaction suffers from head-on collisions because of exponential instabilities \cite{Hans}. 
\item At the expense of satisfying two Weierstrass-Erdmann energetic constraints in every iterated segment, and in order to avoid head-on collisions, the initial segment must include velocity discontinuities caused by collisions-at-a-distance. The set of continuous orbital segments with a countable number of velocity discontinuities and possessing two derivatives piecewise is henceforth called {\small $\hat{C}^2(\mathbb{R}) \subset \hat{C}^\infty(\mathbb{R})$}. The head-on instability is prevented by collisions-at-a-distance inside thin boundary layers.
\item We perturb electrodynamics with a Lorentz-invariant interaction in lightcone having one control parameter {\small $\varepsilon$}.
\item A Lorentz-invariant functional with interactions in lightcone has only three interaction terms\cite{Martinez}; {\color{darkgray}(i)} the electromagnetic term, {\color{darkgray}(ii)} a term henceforth called the {\small $\varepsilon$}-strong interaction and proven here to yield a semiflow for any {\small $\varepsilon \neq 0$}, and {\color{darkgray}(iii)} the gravity-like term of \cite{Martinez}, which was left out for simplicity.
\item Our theory is called {\small $\varepsilon$}-strong variational electrodynamics, henceforth {\small $\varepsilon$}-\textbf{VE}, reducing to electrodynamics when {\small $\varepsilon=0$} \cite{cinderela, JDE2,JMP2009}. Electrodynamics is \textbf{not} a semiflow on {\small $C^2(\mathbb{R})$}, which is why the orbits studied in \cite{Driver} are {\small $C^\infty$} and 
Refs. \cite{CAM,Daniel} used a boundary-value problem. On the other hand, {\small $\varepsilon$}-\textbf{VE} yields a semiflow for any {\small $\varepsilon \neq 0$}, which is essential to integrate the NDDEs of {\small $\varepsilon$}-\textbf{VE} by the method of steps.
\item NDDEs propagate velocity discontinuities. Velocity discontinuity points are henceforth called breaking points.
\item  In {\small $\varepsilon$}-\textbf{VE}, breaking points appear inside thin boundary layers where near-luminal velocities are reached and velocity denominators are small \cite{cinderela,JDE1,JDE2}. 
\item The no-interaction theorem\cite{no-interaction} forbids an approximation of {\small $\varepsilon$}-\textbf{VE} by a Hamiltonian motion with a Coulomb-mechanical Kepler ODE, and that is particularly so near breaking points. Moreover, and again, electrodynamics has small denominators, state-dependent delays, velocity discontinuities, and boundary layers \cite{cinderela}.
\end{enumerate}

 \subsection{How this paper is divided}
In \S \ref{Basic} and in the caption of Fig. \ref{zorro} we explain our notation for indices, introduce the lightcone conditions and state the multi-purpose Lemma \ref{lemma_1} about velocity denominators. In \S \ref{Basic}-\ref{perturbfuncsec}, we introduce the Lorentz-invariant functional of {\small $\varepsilon$}-\textbf{VE}. In Fig. \ref{sew_White}, we illustrate the data for the boundary-value problem. Theorem \ref{theorem_kf44} is an inequality to estimate the parameter region of electromagnetic dominance. In \S \ref{Basic}-\ref{criticalpointsec}, we state the critical point conditions and define some quantities used throughout the paper. In \S \ref{Basic}-\ref{EulerLagrange_eqs}, we put the time-reversible Euler-Lagrange equations. In \S \ref{Basic}-\ref{weisec}, we outline the Weiestrass-Erdmann corner conditions for minimizers with velocity discontinuities. 
%*********************************
In \S \ref{ODEsec} and in the caption of Fig. \ref{figODE} we explain the method of steps.  In \S \ref{ODEsec}-\ref{reconstruction}, we discuss the reconstruction of the most advanced acceleration. In \S \ref{ODEsec}-\ref{rank_deficiency}, we discuss the rank deficiency that prevents electrodynamics to be a semiflow.  In \S \ref{ODEsec}-\ref{MATLABDDEsec}, we write the full NDDE for numerical integration by the method of steps and prove theorem \ref{semiflowtheorem} on a semigroup property when {\small $\varepsilon \neq 0$}. 
In \S \ref{ODEsec}-\ref{driversec}, we discuss the motion restricted to a straight line by the initial condition and derive the NDDE for numerical integration by the method of steps.  \S \ref{ODEsec}-\ref{driversec}  ends with theorem \ref{neutronic} exhibiting a one-parameter fixed-segment that iterates to another constant-velocity segment and a model for the neutron. \S \ref{ODEsec}-\ref{perturbed_algorithm} is designed to guide future experiments with numerical calculations and discusses domains of initial segments for the method of steps and a perturbative approach to avoid collisions.  
%*************************************************************a
In \S \ref{continuationsec}, we discuss the forward propagation of velocity discontinuities;  \S \ref{continuationsec}-\ref{tentative_continuation} discusses the \textit{a priori} propagation based on the partial-momentum sector of the Weiestrass-Erdmann corner conditions in Lemmas \ref{paraboloid_lemma} and \ref{asy_paraboloid_lemma}.  In \S \ref{continuationsec}-\ref{boundary_layer}, we discus the general case of energetic Weierstrass-Erdmann corner conditions, which are the continuity of the partial energies defining the boundary layer. In \S \ref{elastic_boundarylayer} we discuss elastic collisions starting from the case {\small $\varepsilon=0$} as a limit case to exemplify that the energetic conditions \textbf{do} have solutions. Lemma \ref{zero_symmetric_lemma} illustrates a qualitative difference between the electron-electron and the electron-proton collisions-at-a-distance. \S \ref{elastic_boundarylayer} ends with  Lemma \ref{ee_symmetric_lemma} for electron-electron collisions-at-a-distance for {\small $\varepsilon \neq 0$}. In \S \ref{Sumasection}, we put the discussions and the conclusion. The appendix has four subsections; 
 \S \ref{appendix}-\ref{referees} discusses the implications of having a semiflow for any non-zero {\small $\varepsilon$}, which is essential to integrate our NDDEs by the method of steps. Also in \S \ref{appendix}-\ref{referees} we illustrate the generic element of {\small $C^2(\mathbb{R})$} represented by orbits possessing no derivative beyond the second derivative, as indicated by the name \textit{serrated}. 
 Theorem \ref{nosemiflow_theorem} shows that a non-zero {\small $\varepsilon$} allows the serrated orbital segments to be solutions to the NDDEs of {\small $\varepsilon$}-\textbf{VE}. \S \ref{appendix}-\ref{outer-cone} has Lemma \ref{outside-cone-Lemma} on the optimality of mutual-recoil collisions-at-a-distance, \S \ref{appendix}-\ref{LG} describes the one-dimensional Lorentz group, and \S \ref{appendix}-\ref{external} has a formula to include an external electromagnetic field into the equations of motion.
% $\mathcal{M}$
 %

\section{Lorentz-invariant functional with one control parameter} 
\label{Basic}
\begin{figure}[htbp] %  figure placement: here, top, bottom, or page
   \centering
   \includegraphics[width=4.0in, height=2.0in]{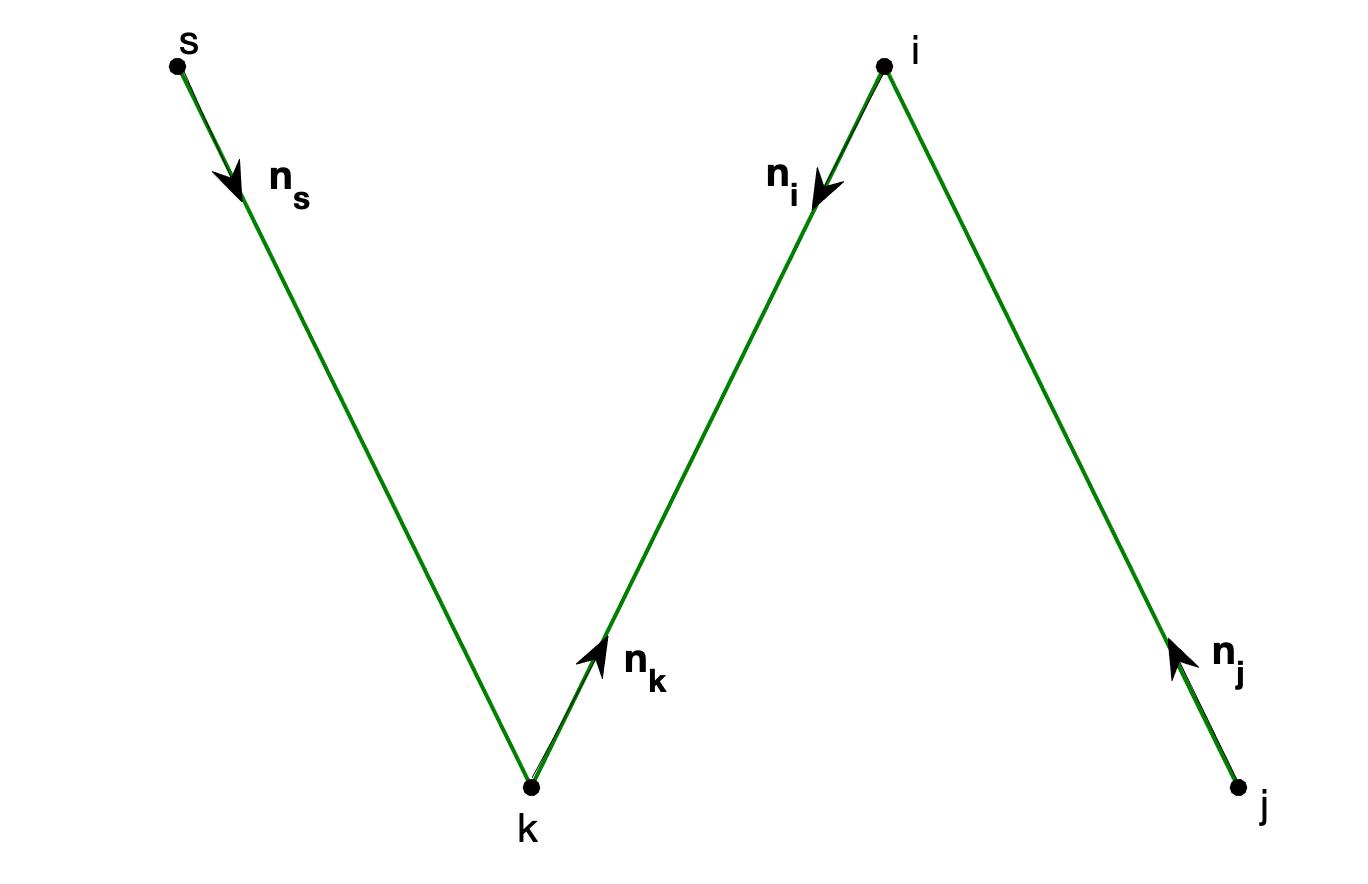} 
   \caption{Illustration of the {\small $(skij)$} convention and the unit vectors emanating from the respective positions. When used in the equation of motion for the proton at point {\small $i$}, the past electronic position is at point {\small $k$} while the future electronic position is at point {\small $j$}. On the other hand, for the equation of motion of the electron at point {\small $k$}, the past protonic position is at point {\small $s$} while the future protonic position is at point {\small $i$}.  
   \label{zorro} }
\end{figure}

Our notation for sub-indices is \textbf{character-sensitive}, as follows: the greek-letter sub-index {\small $\alpha$} designates charges in general, and, when specified, {\small $\mathpzc{\alpha}=\mathpzc{e}$} denotes the electronic quantities and {\small $\mathpzc{\alpha}=\mathpzc{p}$} denotes the protonic quantities. The speed of light is {\small $c \equiv 1$} in our unit system, and the electronic charge and electronic mass are {\small $e_\mathpzc{e} \equiv-1$} and {\small $m_{\mathpzc{e}} \equiv 1$}, respectively. The quantities {\small $(e_\mathpzc{p},m_\mathpzc{p})$} are arbitrary in our flexible setup. To describe the repulsive electromagnetic problem one can take {\small $e_\mathpzc{p}<0$} and arbitrary {\small $m_\mathpzc{p}$}. The proton is described in our unit system by {\small $m_\mathpzc{p} \simeq 1837$} and {\small $e_\mathpzc{p}=1$}. We adopt an inertial frame where every point has a time {\small $t$} defined by Einstein synchronization, and spatial coordinates {\small $\mathbf{x}\equiv (x,y,z) \in \mathbb{R}^3$} such that {\small $ (t,x,y,z)  \in \mathbb{R}^4$}. The two-body problem has coordinates {\small $(t_\mathsmaller{\mathpzc{p}}, x_\mathsmaller{\mathpzc{p}} , y_\mathsmaller{\mathpzc{p}}, z_\mathsmaller{\mathpzc{p}}, t_\mathsmaller{\mathpzc{e}}, x_\mathsmaller{\mathpzc{e}} , y_\mathsmaller{\mathpzc{e}}, z_\mathsmaller{\mathpzc{e}}) \in \mathbb{R}^4 \times \mathbb{R}^4$} and we present the equations of motion by giving the derivatives of the (three) spatial coordinates of each particle respect to its time {\small $t_{\mathsmaller{\alpha}}$}, for {\small $\alpha \in (\mathpzc{e},\mathpzc{p})$}. An \textbf{orbit} of the two-body problem is a pair of twice-differentiable functions {\small $(\mathbf{x}_\mathpzc{e}(t_\mathpzc{e}),\mathbf{x}_\mathpzc{p}(t_\mathpzc{p})) \in C^{2}(\mathbb{R})$} and satisfying the equations of motion, i.e., {\small $ \mathbf{x}_{\; \mathsmaller{\alpha}}: t_{\mathsmaller{\alpha}} \in \mathbb{R} \rightarrow (x_\mathsmaller{\alpha} , y_\mathsmaller{\alpha}, z_\mathsmaller{\alpha}) \in \mathbb{R}^3$},  {\small $\alpha \in (\mathpzc{e},\mathpzc{p})$}, while each {\small $\mathbf{x}_{\; \mathsmaller{\alpha}}(t_\mathsmaller{\alpha})$} is called the \textbf{trajectory} of charge $\alpha$. Orbits with discontinuous velocities in a countable set are studied in the class of continuous functions possessing two derivatives piecewise, henceforth {\small $\hat{C}^2(\mathbb{R})$}. Our notation becomes character-sensitive when sewing chains are involved, as illustrated in Figure \ref{zorro}. The former convention is adapted to display formulas involving one charge \textit{and} the past \textit{and} the future positions in lightcone of the other charge, in which case the sub-indices are chosen as three consecutive roman characters taken from {\small $(s,k,i,j)$} in the former order, to denote the sewing chain illustrated in Fig. \ref{zorro}. 

 Our theory is sensible for trajectories possessing a velocity with a modulus smaller than the speed of light {\small $c \equiv 1$} almost everywhere, i.e.,
{\small
\begin{eqnarray}
||\mathbf{v}_\mathsmaller{\alpha}|| < 1, \; \; \alpha \in (\mathpzc{e},\mathpzc{p}), \label{subluminal}
\end{eqnarray}
}
henceforth subluminal trajectories. For orbits satisfying (\ref{subluminal}), the future lightcone and the past lightcone of point {\small $(t_{\! \mathsmaller{\, i}},\mathbf{x}_{\! \mathsmaller{\, i}})$} are the points {\small $(t_{\! \mathsmaller{\alpha}},\mathbf{x}_{\! \mathsmaller{\alpha}}(t_\mathsmaller{\alpha}))$} defined by
{\small 
\begin{equation}
t_{\mathsmaller{\alpha}} =  t_{\! \mathsmaller{\, i}} \pm || {{\mathbf{x}_{\! \mathsmaller{\, i}} -\mathbf{x}}_{\mathsmaller{\alpha}}(t_{\mathsmaller{\alpha}})} || \equiv t_{\! \mathsmaller{\, i}} \pm r_{\! \mathsmaller{\, \alpha i}}(t_{\! \mathsmaller{\, i}},\mathbf{x}_{\! \mathsmaller{\, i}}),\label{lightcone_map}
\end{equation}
}
where the plus sign goes with {\small $\alpha =j$}, the minus sign goes with {\small $\alpha=k$}, and double bars stand for the {\small $\mathbb{R}^3$} norm. In Ref \cite{JDE2} we show that for subluminal trajectories belonging to {\small $\hat{C}^2(\mathbb{R})$}, each sign of  the implicit condition (\ref{lightcone_map}) has a unique solution defining continuous and twice differentiable maps {\small $ t_\alpha(t_{\! \mathsmaller{\, i}},\mathbf{x}_{\! \mathsmaller{\, i}}) : \mathbb{R} \times \mathbb{R}^3 \rightarrow \mathbb{R}$} and {\small $ r_{\! \mathsmaller{\,  \alpha i}}(t_{\! \mathsmaller{\, i}},\mathbf{x}_{\! \mathsmaller{\, i}}) : \mathbb{R} \times \mathbb{R}^3 \rightarrow \mathbb{R}$} \,\, for {\small $\alpha \in (k,j)$} (illustrated in Fig. \ref{zorro}), where
{\small
\begin{eqnarray}
r_{\! \mathsmaller{\,  \alpha i}} (t_{\! \mathsmaller{\, i}},\mathbf{x}_{\! \mathsmaller{\, i}}) \equiv  || {{\mathbf{x}_{\! \mathsmaller{\, i}} -\mathbf{x}}_{\mathsmaller{\alpha}}(t_{\mathsmaller{\alpha}})} || = | t_{\! \mathsmaller{\, i}}-t_{\mathsmaller{\alpha}}  |.
\label{defi}
\end{eqnarray}}
Because the square of (\ref{defi}) is the equation for a cone in {\small $\mathbb{R}^4$}, Eq. (\ref{defi}) is also called the lightcone condition without mention to the base point {\small $(t_{\! \mathsmaller{\, i}},\mathbf{x}_{\! \mathsmaller{\, i}})$}, while (\ref{lightcone_map}) is called either the future or the past lightcone \textit{of}  {\small $(t_{\! \mathsmaller{\, i}},\mathbf{x}_{\! \mathsmaller{\, i}})$}. In Ref. \cite{JDE2} we show that the lightcone time {\small $t_\mathsmaller{\alpha}$} and the inter-particle distance {\small $r_{\alpha i}$} defined respectively by (\ref{lightcone_map}) and (\ref{defi}) satisfy
{\small
\begin{eqnarray}
\frac{\partial t_\mathsmaller{\alpha}}{\partial \mathbf{x}_\mathsmaller{i}} (t_{\! \mathsmaller{\, i}},\mathbf{x}_{\! \mathsmaller{\, i}})=  \pm \nabla_\mathsmaller{\! i}\, r_{\! \mathsmaller{\!  \alpha i}}(t_{\! \mathsmaller{\, i}},\mathbf{x}_{\! \mathsmaller{\, i}}) =  \frac{ \pm \mathbf{n}_\mathsmaller{\alpha}}{(1\pm \mathbf{n}_{\! \mathsmaller{\, \alpha}}\cdot \mathbf{v}_{\! \mathsmaller{\, \alpha}})}\;, \label{denominator}  
 \end{eqnarray}} 
 where 
 {\small
\begin{eqnarray}
 \mathbf{n}_{\! \mathsmaller{\, \alpha} } & \equiv &  \frac{(\mathbf{x}_{\! \mathsmaller{\, i}} -\mathbf{x}_{\mathsmaller{\alpha}})}{ r_{\! \mathsmaller{\alpha i}}} \; . \label{defn}  
 \end{eqnarray}}
 The upper sign in (\ref{denominator}) holds when {\small $\alpha=j$}, while the lower sign holds when {\small $\alpha=k$}. The denominators of (\ref{denominator}) are henceforth called \textit{velocity denominators}, which introduce a singularity in the equations of motion derived next. Several quantities in this paper involve the denominator {\small $(1\pm \mathbf{n}_{\! \mathsmaller{\, \alpha}}\cdot \mathbf{v}_{\! \mathsmaller{\, \beta}})$}, which is non-zero for subluminal orbits. The following Lemma defines a ubiquitous quotient that is finite for subluminal orbits, even in the limit when {\small $||\mathbf{v}_\mathsmaller{\beta}|| \rightarrow 1 $}, which is a useful regularization for the numerical calculations. 

\begin{lem} For an arbitrary unit direction {\small $\mathbf{n}_\mathsmaller{\alpha} \in \mathbb{R}^3$} and {\small $||\mathbf{v}_\mathsmaller{\beta}|| < 1$} we have   
 \label{lemma_1}
{\small
\begin{eqnarray}
\negthickspace \negthickspace \negthickspace \negthickspace \frac{(1-\mathbf{v}_{\mathsmaller{\beta}}^\mathsmaller{2} )}{(1- (\mathbf{n}_\mathsmaller{\alpha}\cdot \mathbf{v}_\mathsmaller{\beta})^\mathsmaller{2} )} \leq 1. \label{ubiquitous}
\end{eqnarray}}
\end{lem}
\begin{proof} Using the Pythagora's theorem, {\small $\mathbf{v}_{\mathsmaller{\beta}}^{\mathsmaller{2}}=  (\mathbf{n}_\mathsmaller{\alpha}\cdot \mathbf{v}_\mathsmaller{\beta})^\mathsmaller{2}  + || \mathbf{n}_\mathsmaller{\alpha} \times \mathbf{v}_\mathsmaller{\beta}  ||^2 $}, the left-hand side of (\ref{ubiquitous}) can be expressed as
{\small
\begin{eqnarray}
\frac{(1-\mathbf{v}^{\mathsmaller{2}}_{\mathsmaller{\beta}} )}{(1- (\mathbf{n}_\mathsmaller{\alpha}\cdot \mathbf{v}_\mathsmaller{\beta})^\mathsmaller{2})} = \frac{(1-\mathbf{v}^{\mathsmaller{2}}_{\mathsmaller{\beta}} )}{(1-\mathbf{v}^{\mathsmaller{2}}_{\mathsmaller{\beta}} ) +|| \mathbf{n}_\mathsmaller{\alpha} \times \mathbf{v}_\mathsmaller{\beta}  ||^2 }. \label{factored}
\end{eqnarray}}
The denominator on the right-hand side of (\ref{factored}) is equal or greater than the numerator, and it is non-zero because {\small $(1-\mathbf{v}^{\mathsmaller{2}}_{\mathsmaller{\beta}} ) >0$} for sub-luminal orbits. 
\end{proof}

\subsection{Lorentz-invariant functional}
\label{perturbfuncsec}

\begin{figure}[htbp] %  figure placement: here, top, bottom, or page
   \centering
   \includegraphics[width=4.8in, height=2.4in]{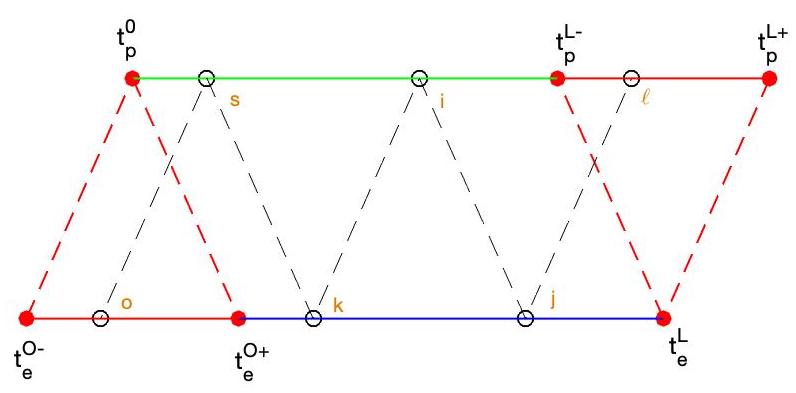} 
   \caption{ The boundary segments of {\small $H_B$} include the protonic position at the initial time {\small $t_\mathpzc{p}^O$} and the {\small $\hat{C}^{\,2}$} electronic trajectory  in the domain {\small $ [t_\mathpzc{e}^{O-}, t_\mathpzc{e}^{O+}]$} ({\color{darkgray}lower red segment}), which interval has endpoints in lightcone with the protonic position at the initial time {\small $t_\mathpzc{p}^O$}. At the other end, the boundary segments of  {\small $H_B$} include the electronic position at the end time {\small $t_\mathpzc{e}^L$}, and the {\small $\hat{C}^{\,2}$} protonic trajectory in the domain {\small $ [t_\mathpzc{p}^{L-}, t_\mathpzc{p}^{L+}]$}({\color{darkgray}upper solid red segment}), which interval is inside the lightcone of the electronic position at the end time {\small $t_\mathpzc{e}^L$}. The {\small $\hat{C}^{\, 2}$} trajectories have two continuous derivatives at every point but for a countable set of points along sewing chains of breaking points. Trajectory information on the boundary segments plus trajectory segments, henceforth referred to as the boundary chain {\small $H_{\mbox{\scriptsize{B}}}$}, is enough to formulate the variational problem. Illustrated in gold is a sewing chain of particle positions in lightcone, {\color{Goldyellow}$(o,s,k,i,j,\ell)$}, and the dashed black lines indicate the lightcones. }
\label{sew_White}
\end{figure}

% Each partial Lagrangian poses an Euler-Lagrange problem yielding a NDDE with \textit{two} state-dependent delays of neutral type for each charge\cite{cinderela, JDE1,JDE2}. The complete NDDE for the two-body problem has \textit{four} state-dependent delays of neutral type.
Here we introduce the Lorentz-invariant funcional defined by integration over the infinite-dimensional boundary data and trajectory segments of Fig. \ref{sew_White}. The most general Lorentz-invariant functional with interactions in lightcone has only three interaction terms \cite{Martinez}. Our generalization of electrodynamics uses the action written in Ref. \cite{Martinez} with renamed coefficients. After some inspection, one finds that the first term in equation 47 of Ref. \cite{Martinez} is not needed to yield a semiflow, so its coefficient was set to zero for simplicity. The coefficient of the electromagnetic sector is set to the usual product of the charges. Essential for a semiflow on {\small $C^2(\mathbb{R})$} is the third term of the functional in equation 47 of Ref. \cite{Martinez}, henceforth called the {\small $\varepsilon$}-strong interaction. We chose the third coefficient to be the parameter {\small $\varepsilon \in \mathbb{R}$}, yielding our Lorentz-invariant functional for minimization,
 {\small
\begin{eqnarray}
\negthickspace  \mathcal{A}_\varepsilon &\equiv& -\sum_{\! \mathsmaller{\, \alpha=\mathpzc{e},\mathpzc{p}}} \int_{H_B} m_{\! \mathsmaller{\, \alpha}} \sqrt{1-\mathbf{v}_{\!\mathsmaller{\alpha}}^{\! \mathsmaller{\, 2}}}\;  dt_{\! \mathsmaller{\,\alpha}} - e_{\! \mathsmaller{\, \mathpzc{e}}} e_{\! \mathsmaller{\,\mathpzc{p}}} \negthickspace \int_{H_B}\delta (s^{\! \mathsmaller{\, 2}}_{{\! \mathsmaller{\,\mathpzc{e} \mathpzc{p}}}})(1- \mathbf{v}_{\! \mathsmaller{\,\mathpzc{e}}} \cdot \mathbf{v}_{\! \mathsmaller{\, \mathpzc{p}}}) \; dt_{\! \mathsmaller{\,\mathpzc{e}}} dt_{\! \mathsmaller{\,\mathpzc{p}}} \notag \\  &&-\varepsilon  \int_{H_B}\delta (s^{\! \mathsmaller{\, 2}}_{{\! \mathsmaller{\,\mathpzc{e} \mathpzc{p}}}}) \sqrt{1- \mathbf{v}^{\! \mathsmaller{\, 2}}_{\! \mathsmaller{\,\mathpzc{e}}}} \sqrt{1- \mathbf{v}^{\! \mathsmaller{\, 2}}_{\! \mathsmaller{\,\mathpzc{p}}}} \; dt_{\! \mathsmaller{\,\mathpzc{e}}} dt_{\! \mathsmaller{\,\mathpzc{p}}},\label{Fokker}
\end{eqnarray}}
where {\small $\mathbf{v}_{\! \mathsmaller{\,\alpha}} \equiv \frac{d\mathbf{x}_{\! \mathsmaller{\, \alpha}}}{dt} |_{t_{\! \mathsmaller{\, \alpha}}}$} is the cartesian velocity of particle {\small $\alpha \in (\mathpzc{e},\mathpzc{p})$} at time {\small $t_{\! \mathsmaller{\, \alpha}}$} and {\small $s^{\! \mathsmaller{\, 2}}_{{\! \mathsmaller{\,\mathpzc{e} \mathpzc{p}}}}(t_{\! \mathsmaller{\, \mathpzc{e}}},\,t_{\! \mathsmaller{\, \mathpzc{p}}})$} is the Lorentz-invariant four-separation defined as a function of two times {\small $s^{\! \mathsmaller{\, 2}}_{{\! \mathsmaller{\,\mathpzc{e} \mathpzc{p}}}}: \mathbb{R} \times \mathbb{R} \rightarrow \mathbb{R}$} by 
{\small
\begin{eqnarray}
s^{\! \mathsmaller{\, 2}}_{{\! \mathsmaller{\,\mathpzc{e} \mathpzc{p}}}}(t_{\! \mathsmaller{\, \mathpzc{e}}},\,t_{\! \mathsmaller{\, \mathpzc{p}}})&=& ( t_{\! \mathsmaller{\, \mathpzc{e}}}-t_{\! \mathsmaller{\, \mathpzc{p}}})^\mathsmaller{2} - r_{\! \mathsmaller{\, \mathpzc{e}\mathpzc{p}}}^\mathsmaller{2} ( t_{\! \mathsmaller{\, \mathpzc{e}}},t_{\! \mathsmaller{\, \mathpzc{p}}})\,, \label{fourseparation}
\end{eqnarray}}
where 
{\small
\begin{eqnarray}
 r_{\! \mathsmaller{\, \mathpzc{e}\mathpzc{p}}}     \equiv  \Vert {{\mathbf{x}_{\! \mathsmaller{\, \mathpzc{p}}}(t_{\! \mathsmaller{\, \mathpzc{p}}}) -\mathbf{x}}_{\! \mathsmaller{\, \mathpzc{e}}}(t_{\! \mathsmaller{\, \mathpzc{e}}})} \Vert ,  \label{defrep}
\end{eqnarray}}
 is the interparticle distance as a function of two times, and the double bars in (\ref{defrep}) stand for the {\small $\mathbb{R}^3$} norm, as always in this manuscript. Still in Eq. (\ref{Fokker}), the dot represents the scalar product of {\small $\mathbb{R}^3$}, the integration variables are the particle times and the double integration is to be carried over the trajectory segments and boundary histories defined in Fig. \ref{sew_White} and indicated by {\small $H_B$}. The lightcone condition (\ref{defi}) is the condition {\small $s^{\! \mathsmaller{\, 2}}_{{\! \mathsmaller{\,\mathpzc{e} \mathpzc{p}}}}(t_{\! \mathsmaller{\, \mathpzc{e}}},\,t_{\! \mathsmaller{\, \mathpzc{p}}})=0$}, and in the following we use the standard delta-function identity of summation over the zeros of the argument (e.g. see chapter 14 of Ref. \cite{Jackson}) to integrate (\ref{Fokker}) over {\small $t_\mathsmaller{\mathpzc{e}}$} with a fixed {\small $t_\mathsmaller{\mathpzc{p}}$}, yielding  
{\small
\begin{eqnarray}
\delta(s^{\! \mathsmaller{\, 2}}_{\! \mathsmaller{\, \mathpzc{e}\mathpzc{p}}}(t_{\! \mathsmaller{\, \mathpzc{p}}},\, t_{\! \mathsmaller{\, \mathpzc{e}}} ))=\sum_{\mathfrak{z}=\pm1} \; \frac{\;\; \delta( t_{\! \mathsmaller{\, \mathpzc{e}}}   -t_\mathsmaller{\mathpzc{p}}    \mp \mathfrak{z}r_\mathsmaller{\mathpzc{ep} })}{|\frac{\partial s^{\! \mathsmaller{\, 2}}_{\mathsmaller{ \mathpzc{ep}}}}{\;\;  \partial t_{ \mathsmaller{\mathpzc{e}}} } |_{t_{ \mathsmaller{\mathpzc{e}}}=t_{ \mathsmaller{\mathpzc{p}}} \pm r_{\! \mathsmaller{\mathpzc{ep}} } }}=\frac{\;\; \delta(t_{\! \, \mathsmaller{\mathpzc{e}}}-t_{\! \mathsmaller{\, \mathpzc{p}}} - r_{\! \mathsmaller{\, \mathpzc{e}\mathpzc{p}}})}{2r_{\! \mathsmaller{\, \mathpzc{e}\mathpzc{p}}}(1 + \mathbf{n}_{ \mathsmaller{\, \mathpzc{e}}} \cdot \mathbf{v}_{\! \mathsmaller{\, \mathpzc{e}}})} +\frac{\;\; \delta(t_{ \mathsmaller{\mathpzc{e}}}-t_{\! \mathsmaller{\, \mathpzc{p}}} + r_{\! \mathsmaller{\, \mathpzc{e}\mathpzc{p}}})}{2r_{\! \mathsmaller{\, \mathpzc{e}\mathpzc{p}}}(1 - \mathbf{n}_{\! \mathsmaller{\, \mathpzc{e}}} \cdot \mathbf{v}_{\! \mathsmaller{\, \mathpzc{e}}})},  \label{jackson_delta}
\end{eqnarray} }
where {\small $\mathbf{n}_{\! \mathsmaller{\, \mathpzc{e}}}$} is defined by Eq. (\ref{defn}) with {\small $i=\mathpzc{p}$} and {\small $\alpha=\mathpzc{e}$}. In the denominators of (\ref{jackson_delta}) and henceforth, {\small $r_{\! \mathsmaller{\, \mathpzc{e}\mathpzc{p}}} $} is the distance in lightcone as a function of time {\small $t_{\mathsmaller{\, \mathpzc{p}} }$} only, and the plus sign goes when the electronic position is in the future lightcone of {\small $t_\mathsmaller{\mathpzc{p}}$}, i.e., {\small $t_{\! \, \mathsmaller{\mathpzc{e}}}=t_{\! \mathsmaller{\, \mathpzc{p}}} + r_{\! \mathsmaller{\, \mathpzc{e}\mathpzc{p}}}$}\,, while the minus sign goes when the electronic position is in the past lightcone of {\small $t_\mathsmaller{\mathpzc{p}}$}, i.e., {\small $t_{\! \, \mathsmaller{\mathpzc{e}}}=t_{\! \mathsmaller{\, \mathpzc{p}}} - r_{\! \mathsmaller{\, \mathpzc{e}\mathpzc{p}}}$}\,. We notice that {\small $\delta(s^{\! \mathsmaller{\, 2}}_{\! \mathsmaller{\, \mathpzc{e}\mathpzc{p}}}(t_{\! \mathsmaller{\, \mathpzc{p}}},\, t_{\! \mathsmaller{\, \mathpzc{e}}}))$} can be expressed by an alternative formula obtained from (\ref{jackson_delta}) by exchanging {\small $\mathpzc{e} $} and {\small $\mathpzc{p}$}, which should be used when one is integrating over $t_\mathsmaller{\mathpzc{p}}$ to derive the electronic partial Lagrangian. 
\par
Observations; (i) because of Lemma \ref{lemma_1}, our functional (\ref{Fokker}) is well defined in the domain {\small $\mathcal{D}_{\text {\tiny{Peano}}} $} of trajectory pairs where the denominators of (\ref{Fokker}) are Lebesgue integrable, i.e.,
   {\small
\begin{eqnarray}
\mathcal{D}_{\text {\tiny{Peano}}} \equiv \Big\{ \!  (\Gamma_\mathsmaller{e},\Gamma_\mathsmaller{p})  \in C^2(\mathbb{R})   \Big| \int \frac{dt_\mathsmaller{\ell}}{ r_\mathsmaller{\ell j}(1-\mathbf{v}_{\mathsmaller{\ell}}^\mathsmaller{2})} \!< \infty \; ;   \int \frac{dt_\mathsmaller{j}}{ r_\mathsmaller{j i}(1-\mathbf{v}_{\mathsmaller{i}}^\mathsmaller{2})} \!< \infty \! \Big\}, \notag \\ \label{Peano}
 \end{eqnarray}}
named after the Cauchy-Peano theorem for ODEs and (ii) the history sets to be used in the variational formulation are illustrated in red in Fig. \ref{sew_White} below.

\newpage

Next we show that the electromagnetic interaction dominates the interaction in (\ref{Fokker}) when {\small $|\varepsilon| < 1$}.

 \begin{theorem} For arbitrary $(\mathbf{v}_\mathpzc{p}, \mathbf{v}_\mathpzc{e}) \in \mathbb{R}^3$ with $|\mathbf{v}_\mathpzc{p}|\leq 1$ and $|\mathbf{v}_\mathpzc{e}| \leq 1$,  we have   
 \label{theorem_kf44}
{\small
\begin{eqnarray}
\Delta_{\mathpzc{e} \mathpzc{p} } \equiv (1-\mathbf{v}_\mathpzc{e} \cdot \mathbf{v}_\mathpzc{p})^2-(1-\mathbf{v}_\mathpzc{p}^2)(1-\mathbf{v}_\mathpzc{e}^2)& \geq & 0.  \label{ineqkf44}
\end{eqnarray}}
\end{theorem}
\begin{proof} The proof is simple and proceeds by re-arranging Eq. (\ref{ineqkf44}), yielding
{\small
\begin{eqnarray}
\Delta_{\mathpzc{e} \mathpzc{p} }& \equiv & (1-\mathbf{v}_\mathpzc{e} \cdot \mathbf{v}_\mathpzc{p})^2-(1-\mathbf{v}_\mathpzc{e}^2)(1-\mathbf{v}_\mathpzc{p}^2)\notag \\
&=& \mathbf{v}_\mathpzc{e}^2+\mathbf{v}_\mathpzc{p}^2 -2\mathbf{v}_\mathpzc{e} \cdot \mathbf{v}_\mathpzc{p}  + (\mathbf{v}_\mathpzc{e} \cdot \mathbf{v}_\mathpzc{p})^2 - \mathbf{v}_\mathpzc{e}^2\mathbf{v}_\mathpzc{p}^2 \notag \\
&=& (|\mathbf{v}_\mathpzc{e}|-|\mathbf{v}_\mathpzc{p}|)^2 +2(|\mathbf{v}_\mathpzc{e}||\mathbf{v}_\mathpzc{p}|-\mathbf{v}_\mathpzc{e} \cdot \mathbf{v}_\mathpzc{p})+ (\mathbf{v}_\mathpzc{e} \cdot \mathbf{v}_\mathpzc{p})^2 - \mathbf{v}_\mathpzc{e}^2\mathbf{v}_\mathpzc{p}^2 \notag \\
&=& (|\mathbf{v}_\mathpzc{e}|-|\mathbf{v}_\mathpzc{p}|)^2 +2|\mathbf{v}_\mathpzc{e}||\mathbf{v}_\mathpzc{p}|(1-\cos(\theta))\Big(1-|\mathbf{v}_\mathpzc{e}||\mathbf{v}_\mathpzc{p}|\cos^2(\frac{\theta}{2})\Big) \geq 0, \notag
\end{eqnarray}}
where $\theta$ is the angle between vectors $\mathbf{v}_\mathsmaller{\mathpzc{e}}$ and $\mathbf{v}_\mathsmaller{\mathpzc{p}}$.  
\end{proof}

\subsection{Critical point conditions}
\label{criticalpointsec}
Minimization of (\ref{Fokker}) poses two conditions: (i) the Euler-Lagrange equations along the piecewise {\small $C^{2}$} segments of each trajectory and (ii) the Weierstrass-Erdmann conditions at the breaking points of each trajectory \cite{cinderela,JDE1}. The critical point of (\ref{Fokker}) when {\small $i=\mathpzc{p}$} is obtained by varying the protonic trajectory with fixed endpoints and history segments illustrated in red in Fig. \ref{sew_White}, and fixing the electronic trajectory illustrated in blue in Fig. \ref{sew_White}. The partial Lagrangian of particle {\small $i=\mathpzc{p}$} is obtained by performing the integration over {\small $t_\mathpzc{e}$} in the double integral (\ref{Fokker}). During the integration over {\small $t_{\! \mathsmaller{\, e}}$}, the delta function picks the two zeros of {\small $s^2_{\! \mathsmaller{\, \mathpzc{e} \mathpzc{p}}}(t_{\! \mathsmaller{\, \mathpzc{e}}},\,t_{\! \mathsmaller{\, \mathpzc{p}}})$}, which define the two lightcones. Using (\ref{jackson_delta}) to integrate (\ref{Fokker}) over $t_{\! \mathsmaller{\, \mathpzc{e}}}$ yields a Lagrangian minimization to find {\small $\min \{ \int_{t_O}^{t_L} \mathscr{L}_{\! \mathsmaller{\, i}}(t_{\mathsmaller{i}}, \mathbf{x}_{\! \mathsmaller{\, i}}(t_\mathsmaller{i}), \mathbf{v}_{\mathsmaller{i}}(t_\mathsmaller{i}) ) dt_{\! \mathsmaller{\, i}} \}$}\, for
{\small
\begin{equation}
\mathscr{L}_{\! \mathsmaller{\, i}}(t_{\mathsmaller{i}},\mathbf{x}_{\! \mathsmaller{\, i}}\, ,\mathbf{v}_{\! \mathsmaller{\, i}}) \equiv \mathcal{K}_{\! \mathsmaller{\; i}}  - \sum_{\! \mathsmaller{\alpha= k,j}} e_{\! \mathsmaller{\, i}} \big(\,\mathcal{U}_{\! \mathsmaller{\; i}} -\mathbf{v}_{\! \mathsmaller{\, i}} \cdot \mathbf{A}_{\! \mathsmaller{\, i}}\,\big) -\varepsilon \sqrt{1-\mathbf{v}_{\! \mathsmaller{\, i}}^{\! \mathsmaller{\, 2}}}\, \mathsf{G}_{\! \mathsmaller{\, i}} \,, \label{partial_Lagrangian}
\end{equation}}
 with {\small $\mathcal{K}_{\! \mathsmaller{\; i}},\, \mathcal{U}_{\! \mathsmaller{\; i}}\,, \mathbf{A}_{\! \mathsmaller{\, i}}$} and {\small $\mathsf{G}_{\! \mathsmaller{\, i}} $} given by
 {\small
\begin{eqnarray}
\mathcal{K}_{\! \mathsmaller{\; i}} &\equiv & m_{\! \mathsmaller{\, i}}(\,1-\sqrt{1-\mathbf{v}_{\! \mathsmaller{\, i}}^{\! \mathsmaller{\, 2}}} \;\;),  \label{kinetic} \\
  \negthickspace \negthickspace  \mathbf{A}_{\! \mathsmaller{\, i}} & \equiv &  \frac{e_{\! \mathsmaller{\, k}}\mathbf{{v}}_{\! \mathsmaller{\, k}} }{2r_{\! \mathsmaller{\, ki}}(1-\mathbf{n}_{\! \mathsmaller{\, k}}\cdot \mathbf{v}_{\! \mathsmaller{k}})} +
\frac{e_{\! \mathsmaller{\, j}}\mathbf{{v}}_{\! \mathsmaller{\, j}}}{2r_{\! \mathsmaller{\, ji}}(1+\mathbf{n}_{\! \mathsmaller{\, j}}\cdot \mathbf{v}_{\! \mathsmaller{\, j}})} , \label{vectorAi} \\
\negthickspace  \negthickspace \negthickspace \negthickspace \mathcal{U}_\mathsmaller{\, i} & \equiv & \frac{e_{\! \mathsmaller{\, k}} }{2r_{\! \mathsmaller{\, ki}}(1-\mathbf{n}_{\! \mathsmaller{\, k}}\cdot \mathbf{v}_{\! \mathsmaller{\, k}})} +
\frac{e_{\! \mathsmaller{\, j}}}{2r_{\! \mathsmaller{\, ji}}(1+\mathbf{n}_{\! \mathsmaller{\, j}}\cdot \mathbf{v}_{\! \mathsmaller{\, j}})}, \label{scalarUi} \\
\negthickspace \negthickspace \mathsf{G}_{\! \mathsmaller{\, i}} & \equiv & \frac{ \sqrt{1-\mathbf{v}_{\! \mathsmaller{\, k}}^{\! \mathsmaller{\, 2}}}}{2r_{\! \mathsmaller{\, ki}}(1-\mathbf{n}_{\! \mathsmaller{\, k}}\cdot \mathbf{v}_{\! \mathsmaller{\, k}})} +
\frac{ \sqrt{1-\mathbf{v}_{\! \mathsmaller{\, j}}^{\! \mathsmaller{\, 2}}} }{2r_{\! \mathsmaller{\, ji}}(1+\mathbf{n}_{\! \mathsmaller{\, j}}\cdot \mathbf{v}_{\! \mathsmaller{\, j}})} \,, \label{scalarGi}
\end{eqnarray}}
where {\small $\mathbf{v}_{\! \mathsmaller{\, k}} \equiv d\mathbf{x}_{\! \mathsmaller{\, k}}/dt |_{t=t_{\! \mathsmaller{\, k}}}$} is the electronic velocity evaluated on the past lightcone and {\small $\mathbf{v}_{\! \mathsmaller{\, j}} \equiv d\mathbf{x}_{\! \mathsmaller{\, j}}/dt |_{t=t_{\! \mathsmaller{\, j}}}$} is the electronic velocity on the future lightcone.  Notice that {\small $\mathcal{K}_{\! \mathsmaller{\; i}},  \mathcal{U}_{\mathsmaller{\; i}},  \mathbf{A}_{\! \mathsmaller{\, i}}$} and {\small $ \mathsf{G}_{\! \mathsmaller{\, i}} $} are functions of {\small $(t_{\mathsmaller{i}}, \mathbf{x}_{\mathsmaller{i}})$} by Eqs. (\ref{lightcone_map}) and (\ref{defi}). Equation (\ref{partial_Lagrangian}) defines the protonic partial Lagrangian, while the electronic partial Lagrangian is obtained by exchanging indices with the permutation {\small $(kij) \rightarrow (ski)$} in Eq. (\ref{partial_Lagrangian}), i.e., replacing {\small $i$} with {\small $k$} and restricting the summation to the nearest neighbors {\small $\alpha \in (s,i)$} of particle {\small $k$} on the sewing chain of Fig. \ref{zorro}. Observations; (i) unlike the principle of minimal action of classical mechanics, the boundary segments used in Fig. \ref{sew_White} are \textbf{trajectory segments}\cite{CAM,Daniel} defined on a time domain that is a union of closed  intervals (a second category set) and (ii) the procedure explained above yields a different partial Lagrangian for each particle. If the partial Lagrangians were the same, like in classical mechanics, our definition (\ref{Fokker}) of variational electrodynamics would clash with the no-interaction theorem of Lorentz-equivariant dynamics \cite{no-interaction}.

\subsection{Euler-Lagrange equations of motion}
\label{EulerLagrange_eqs}

The Euler-Lagrange equations for particle {\small $i$} on the {\small $\widehat{C}^2(\mathbb{R})$} segments are
{\small
\begin{eqnarray}
(m_{\! \mathsmaller{\, i}}+\varepsilon \mathsf{G}_{\! \mathsmaller{\, i}}) \Big( \frac{ \mathbf{a}_{\! \mathsmaller{\, i}}}{\sqrt{1-\mathbf{v}_{\! \mathsmaller{\, i}}^{\! \mathsmaller{\, 2}}}}+\frac{ (\mathbf{v}_{\! \mathsmaller{\, i}} \cdot \mathbf{a}_{\! \mathsmaller{\, i}})\mathbf{a}_{\! \mathsmaller{\, i}}}{(1-\mathbf{v}_{\! \mathsmaller{\, i}}^{\! \mathsmaller{\, 2}})^{\! \mathsmaller{\,3/2}}}\Big)&=&\frac{1}{2}e_{\! \mathsmaller{\, i}}\sum_{\alpha=k,\, j} (  \mathbf{E}_{\alpha {\! \mathsmaller{\, i}}} +\mathbf{v}_{\! \mathsmaller{\, i}} \times \mathbf{B}_{\alpha i} ) \notag \\ && -\varepsilon \sqrt{1-\mathbf{v}_{\! \mathsmaller{\, i}}^{\! \mathsmaller{\, 2}}} \Big ( \nabla \mathsf{G}_{\! \mathsmaller{\, i}} +\gamma_i^{\! \mathsmaller{\, 2}}\frac{d \mathsf{G}_{\! \mathsmaller{\, i}}}{dt_\mathsmaller{i}}\mathbf{v}_\mathsmaller{i}   \Big ), \label{ELM}
\end{eqnarray}}
with  
{\small
 \begin{eqnarray}
 \mathbf{E}_{\! \mathsmaller{\, \alpha i}} &  \equiv &e_{\! \mathsmaller{\,\alpha}} \Big ( \frac{(1-\mathbf{v}^{\! \mathsmaller{\, 2}}_{\! \mathsmaller{\, \alpha}})( \mathbf{n}_{\! \mathsmaller{\, \alpha} }\pm \mathbf{v}_{\! \mathsmaller{\, \alpha}} ) } { r^{\! \mathsmaller{\,2}}_{\! \mathsmaller{ \alpha i}}(1\pm \mathbf{n}_{\! \mathsmaller{\, \alpha}}\cdot \mathbf{v}_{\! \mathsmaller{\, \alpha}})^{\! \mathsmaller{\, 3}}} +\frac{\mathbf{n}_{\! \mathsmaller{\, \alpha}}\times \big( ( \mathbf{n}_{\! \mathsmaller{\, \alpha} }\pm \mathbf{v}_{\! \mathsmaller{\, \alpha}} ) \times \mathbf{a}_{\! \mathsmaller{\, \alpha}} \big)} { r_{\! \mathsmaller{\, \alpha i }}(1\pm \mathbf{n}_{\! \mathsmaller{\, \alpha}}\cdot \mathbf{v}_{\! \mathsmaller{\, \alpha}})^{\! \mathsmaller{\, 3}} } \Big) , 
\label{electric} \\
 \mathbf{B}_{\mathsmaller{\alpha i}}& \equiv &\mp \mathbf{n}_{\! \mathsmaller{\, \alpha}}\times \mathbf{E}_{\mathsmaller{\alpha i }}\, ,
\label{magnetic}
\end{eqnarray}}
representing, respectively, the other particle's electric and the magnetic fields. When {\small $\alpha=k$} the other particle is in the past lightcone position and the lower sign applies. Otherwise, when {\small $\alpha=j$} the other particle is in the future lightcone position and the upper sign applies. In Eqs. (\ref{electric}) and (\ref{magnetic}), {\small $\mathbf{v}_{\! \mathsmaller{\, \alpha}} \equiv d \mathbf{x}_{\! \mathsmaller{\, \alpha}}/dt|_{t=t_\mathsmaller{\alpha}} $} and {\small $\mathbf{a}_{\! \mathsmaller{\, \alpha}} \equiv d \mathbf{v}_{\! \mathsmaller{\, \alpha}}/dt|_{t=t_\mathsmaller{\alpha}} $} are, respectively, the other charge's velocity and acceleration evaluated at either the retarded or at the  advanced lightcone. Notice in Eq. (\ref{ELM}) that the electromagnetic sector of the Euler-Lagrange equation involves a semi-sum of the Li\'{e}nard-Wiechert fields (\ref{electric}) and (\ref{magnetic}) combined in the Lorentz-force form \cite{Jackson}. Last, in Eqs. (\ref{electric}) and (\ref{magnetic}) the unit vector from {\small $\mathbf{x}_{\mathsmaller{\alpha}}$} to {\small $\mathbf{x}_{\! \mathsmaller{\, i}}(t_\mathsmaller{i})$} is defined by (\ref{defn}).
Multiplying (\ref{ELM}) by {\small $\mathbf{v}_i$} yields
 {\small
\begin{eqnarray}
 \frac{(m_{\! \mathsmaller{\, i}}+\varepsilon \mathsf{G}_{\! \mathsmaller{\, i}}) }{(1-\mathbf{v}_{\! \mathsmaller{\, i}}^{\! \mathsmaller{\, 2}})^{\,\mathsmaller{3/2}}}\mathbf{v}_{\! \mathsmaller{\, i}} \cdot \mathbf{a}_{\! \mathsmaller{\, i}}&=&\Big( \frac{1}{2}e_{\! \mathsmaller{\, i}}\sum_{\alpha=k,\, j}   \mathbf{E}_{\alpha {\! \mathsmaller{\, i}}} \cdot \mathbf{v}_{\! \mathsmaller{\, i}}\Big)  -\varepsilon \sqrt{1-\mathbf{v}_{\! \mathsmaller{\, i}}^{\! \mathsmaller{\, 2}}} \Big (\mathbf{v}_{\! \mathsmaller{\, i}} \cdot \nabla_\mathsmaller{\!i} \mathsf{G}_{\! \mathsmaller{\, i}} +\gamma_{\! \mathsmaller{\, i}}^{\! \mathsmaller{\, 2}} \mathbf{v}_{\! \mathsmaller{\, i}}^{\! \mathsmaller{\, 2}}\frac{d \mathsf{G}_{\! \mathsmaller{\, i}}}{dt_\mathsmaller{i}}   \Big ), \label{HB}
\end{eqnarray}}
and substituting (\ref{HB}) into the left-hand side of (\ref{ELM}) and using (\ref{scalarGi}) yields
{\small
\begin{eqnarray}
 \Big( \frac{ m_{\! \mathsmaller{\, i}}+\varepsilon \mathsf{G}_{\! \mathsmaller{\, i}}}{\sqrt{1-\mathbf{v}_{\! \mathsmaller{\, i}}^{\! \mathsmaller{\, 2}}}}\Big)\mathbf{a}_{\! \mathsmaller{\, i}}&=&\frac{1}{2}e_{\! \mathsmaller{\, i}}\sum_{\alpha=k,\, j} \Big(  \mathbf{E}_{\alpha {\! \mathsmaller{\, i}}} +\mathbf{v}_{\! \mathsmaller{\, i}} \times \mathbf{B}_{\alpha {\! \mathsmaller{\, i}}} -(\mathbf{v}_{\! \mathsmaller{\, i}} \cdot \mathbf{E}_{\alpha {\! \mathsmaller{\, i}}})\mathbf{v}_{\! \mathsmaller{\, i}} \Big) + \frac{\varepsilon}{2} \sum_{\alpha=k,\, j}\mathbf{f}_{\mathsmaller {\alpha i}} \,,  \label{BH}
\end{eqnarray}}
where 
{\small
\begin{eqnarray}
\mathbf{f}_{\mathsmaller{\alpha i}}&=&\sqrt{1-\mathbf{v}^{\mathsmaller{2}}_{\mathsmaller{i}} }\big(\nabla_\mathsmaller{\!i} G_\mathsmaller{i} +\frac{\partial G_\mathsmaller{i}}{\partial t_\mathsmaller{i}}\mathbf{v}_\mathsmaller{i}  \big)  \equiv   \frac{b_{\mathsmaller{\alpha i}}\; (\hat{\mathcal{R}}_{\mathsmaller{\alpha}}\cdot \mathbf{a}_{\mathsmaller{\alpha}}) \mathbf{u}_\mathsmaller{\alpha i}}{(1\pm \mathbf{n}_\mathsmaller{\alpha} \cdot \mathbf{v}_\mathsmaller{i})}+c_{\mathsmaller{\alpha i}}\mathbf{\Omega}^{\dagger}_{\mathsmaller{\alpha i}}\, ,\label{betaforce}\\
 \mathbf{u}_{\mathsmaller{\mathsmaller{\alpha \beta}}} &\equiv& (\mathbf{n}_{\mathsmaller{\, \mathsmaller{\alpha}}} \pm \mathbf{v}_{\! \mathsmaller{\, \beta}}) \; \; \mbox{for} \; \alpha \in (k,j) ,\;\beta \in (\alpha, i), \label{ualphai} \\
\mathbf{ \Omega}^{\dagger}_{\mathsmaller{\mathsmaller{\alpha i}}} &\equiv&\big(\frac{1-\mathbf{v}_{\! \mathsmaller{\, \alpha}}^{\! \mathsmaller{\, 2}} }{   1\pm \mathbf{n}_{\! \mathsmaller{\, \alpha}}\cdot \mathbf{v}_{\! \mathsmaller{\, \alpha}}  }\big)\mathbf{n}_\mathsmaller{\alpha} \mp (\mathbf{v}_{\mathsmaller{\, \mathsmaller{\alpha}}} + \mathbf{v}_{\! \mathsmaller{\, i}}) \; \; \mbox{for} \; \alpha \in (k,j) , \label{salphai} \\
b_{\mathsmaller{\alpha i}}&\equiv&\sqrt{\frac{1-\mathbf{v}_{\! \mathsmaller{\, i}}^{\! \mathsmaller{\, 2}} }{ 1-\mathbf{v}_{\! \mathsmaller{\, \alpha}}^{\! \mathsmaller{\, 2}} } }\;\Big(\frac{1}{r_{\mathsmaller{\alpha i}} (1\pm \mathbf{n}_{\! \mathsmaller{\, \alpha}}\cdot \mathbf{v}_{\! \mathsmaller{\, \alpha}})}  \, \Big)\Big( \frac{dt_\mathsmaller{\alpha}}{dt_\mathsmaller{i}}\Big),     \label{def_b_alphai}\\  
c_{\mathsmaller{\alpha i}}&\equiv&\sqrt{\frac{1-\mathbf{v}_{\! \mathsmaller{\, i}}^{\! \mathsmaller{\, 2}} }{ 1-\mathbf{v}_{\! \mathsmaller{\, \alpha}}^{\! \mathsmaller{\, 2}} } }\;(\frac{e_\mathsmaller{i} \mathbf{n}_\mathsmaller{\alpha}\cdot \mathbf{E}_\mathsmaller{\alpha i}}{e_\mathsmaller{i} e_\mathsmaller{\alpha}} ) \, ,     \label{def_C_alphai}\\  
\hat{\mathcal{R}}_{\mathsmaller{\alpha}} & \equiv &\Big( \frac{1-\mathbf{v}_{\! \mathsmaller{\alpha}}^{\! \mathsmaller{\, 2}}}{ 1\pm \mathbf{n}_{\! \mathsmaller{\, \alpha}}\cdot \mathbf{v}_{\! \mathsmaller{\, \alpha}} } \Big) \mathbf{n}_{\mathsmaller{\alpha}} \pm \mathbf{v}_{\mathsmaller{\alpha}} .  \label{Right_alpha}
 \end{eqnarray}}
Observations:  ({\color{darkgray}i}) the first term on the right-hand side of (\ref{betaforce}) is the linear dependence of {\small $\mathbf{f}_{\mathsmaller{\alpha i}}$} on the acceleration, while {\small $\mathbf{ \Omega}^{\dagger}_{\mathsmaller{\mathsmaller{\alpha i}}}$} is the acceleration-independent reminder of (\ref{betaforce}), ({\color{darkgray}ii}) the dagger in Eqs. (\ref{betaforce}) and (\ref{salphai}) indicates that the sign convention is reversed for the second $\pm$ of formula (\ref{salphai}), ({\color{darkgray}iii}) on the right-hand side of Eq. (\ref{def_b_alphai}) we have introduced the derivative of time $t_\mathsmaller{\alpha}$ respect to $t_\mathsmaller{i}$, as obtained by taking a derivative of the lightcone time (\ref{lightcone_map}), i.e.,
  {\small
  \begin{eqnarray}
 \negthickspace \negthickspace \negthickspace \negthickspace \negthickspace \negthickspace  \negthickspace \negthickspace \negthickspace  \negthickspace \negthickspace  \negthickspace \negthickspace \negthickspace \negthickspace \negthickspace  \negthickspace \negthickspace \negthickspace\negthickspace\negthickspace   \negthickspace \negthickspace  \negthickspace \negthickspace \negthickspace\negthickspace\frac{dt_\mathsmaller{\alpha}}{dt_\mathsmaller{i}}=\Big(\frac{dt_\mathsmaller{i}}{dt_\mathsmaller{\alpha}}\Big)^\mathsmaller{-1}  \equiv \Big(\frac{1\pm\mathbf{n}_\mathsmaller{\alpha} \cdot \mathbf{v}_\mathsmaller{i}}{1\pm\mathbf{n}_\mathsmaller{\alpha} \cdot \mathbf{v}_\mathsmaller{\alpha}} \Big), \label{time_rate}
  \end{eqnarray}}  
 ({\color{darkgray} iv}) the positivity of the right-hand side of (\ref{time_rate}) ensures that all  deviating times are monotonically increasing, ({\color{darkgray} v}) Eq. (\ref{ualphai}) is generalized to be used in several places of this manuscript. The upper sign applies when {\small $(\alpha,\beta)$} is either {\small $(j,j)$} or {\small $(j,i)$}, while the minus sign applies when {\small $(\alpha,\beta)$} is either {\small $(k,k)$} or {\small $(k,i)$}, ({\color{darkgray}vi}) in Eqs. (\ref{betaforce}), (\ref{salphai}), (\ref{def_b_alphai}), (\ref{def_C_alphai}) and (\ref{Right_alpha}), the upper sign applies when {\small $\alpha=j$}, while the lower sign applies when {\small $\alpha=k$}, ({\color{darkgray}vii}) the electric field {\small $\mathbf{E}_\mathsmaller{\alpha i}$} appearing in Eq. (\ref{def_C_alphai}) is defined by Eq. (\ref{electric}), ({\color{darkgray}viii}) Eq. (\ref{HB}) also follows from Eq. (\ref{BH}), and henceforth equation (\ref{BH}) is called the \textbf{equation of motion of particle {\small $\bf{i}$}}, as derived from the partial Lagrangian (\ref{partial_Lagrangian}), in which case the advanced index is {\small $\alpha=j$} and the retarded index is $\alpha=k$. The \textbf{equation of motion of particle {\small $\bf{k}$}} is obtained by replacing {\small $i$} with {\small $k$} in Eq. (\ref{BH}) and restricting the summation to the nearest neighbors of index {\small $k$} on the sewing chain of Fig. \ref{zorro}, {\small $\alpha \in (s,i)$}, in which case the most advanced index is {\small $\alpha=i$}, {\color{darkgray}(ix)} the future and the past lightcones exchange positions upon time-reversal, and in the next section we show that the time-reversible dynamics defines a flow on {\small $C^2(\mathbb{R}) $} when {\small $\varepsilon \neq 0$}, and {\color{darkgray}(x)} on the right-hand side of (\ref{betaforce}), the far-fields of the {\small $\varepsilon$}-sector have non-zero components along {\small $\mathbf{n}_\mathsmaller{j}$} and along the vector {\small $\mathcal{\hat{R}_\mathsmaller{\alpha}}$} defined by (\ref{Right_alpha}), i.e.,   
 {\small
 \begin{eqnarray}
\mathbf{u}_{\mathsmaller{\alpha \alpha}} \cdot \mathbf{n}_{\mathsmaller{\alpha}} &=& (1\pm\mathbf{n}_\mathsmaller{\, \alpha} \cdot \mathbf{v}_\mathsmaller{\alpha}) \geq 0, \label{defuaana} \\
 \mathbf{u}_\mathsmaller{\alpha \alpha} \cdot \hat{\mathcal{R}}_\mathsmaller{\alpha}&=&(1\pm \mathbf{n}_\mathsmaller{\, \alpha} \cdot \mathbf{v}_\mathsmaller{\alpha}) \geq 0\,, \label{defuaaRa}
  \end{eqnarray}} 
 where again {\small $\alpha  \in (j,k)$}, the upper sign applies when {\small $\alpha=j$}, and the lower sign applies when {\small $\alpha=k$}.  %%%%%%%%%%%%%%%%%%%%%%%%%%%%%%%%%:aa
 \subsection{Weierstrass-Erdmann corner conditions}
 \label{weisec}
    On breaking points, the Weierstrass-Erdmann corner conditions replace the Euler-Lagrange equation \cite{JDE1,Gelfand}. These are the continuity of partial momenta and partial energies \textbf{at} the breaking point, which involve only \textit{positions} and \textit{velocities} in lightcone.  The partial momentum derived from the partial Lagrangian (\ref{partial_Lagrangian}) is
{\small
\begin{eqnarray}
\qquad  \mathbf{P}_{\! \mathsmaller{\, i}}  \equiv  \frac{\partial \mathscr{L}_{\! \mathsmaller{\, i}} }{ \partial \mathbf{{v}}_{\! \mathsmaller{\, i}}} &=& ( m_{\! \mathsmaller{\, i}}+\varepsilon \mathsf{G}_i ) \gamma_\mathsmaller{i}  \mathbf{{v}}_{\! \mathsmaller{\, i}}   +e_{\! \mathsmaller{\, i}}{\mathbf{A}_{\! \mathsmaller{\,i}}}\,,  \label{WE1} 
\end{eqnarray} }
where {\small $\mathbf{A}_{\mathsmaller{i}}$} and {\small $\mathsf{G}_{\mathsmaller{i}}$} are defined respectively by (\ref{vectorAi}) and (\ref{scalarGi}) and
{\small
\begin{eqnarray}
\gamma_\mathsmaller{\alpha} \equiv \frac{1}{\sqrt{1-\mathbf{v}_{\mathsmaller{\alpha}}^\mathsmaller{2} }}, \label{defgamai}
\end{eqnarray}}
for $\alpha \in (i,k)$. The partial energy of the partial Lagrangian (\ref{partial_Lagrangian}) is 
{\small
\begin{equation}
\mathscr{E}_{\! \mathsmaller{\, i}}\equiv \mathbf{v}_{\! \mathsmaller{\, i}} \cdot \frac{\partial \mathscr{L}_{\! \mathsmaller{\, i}}}{ \partial \mathbf{{v}}_{\! \mathsmaller{\, i}}} -\mathscr{L}_{\! \mathsmaller{\, i}} = (m_{\! \mathsmaller{\, i}}+\varepsilon \mathsf{G}_i)\gamma_\mathsmaller{i}  + e_{\! \mathsmaller{\, i}}  \,\mathcal{U}_{\mathsmaller{\,i}}  \, ,\label{WE2}
\end{equation}}
where {\small $\mathcal{U}_{\mathsmaller{i}}$} and {\small $\mathsf{G}_{\mathsmaller{i}}$} are defined respectively by (\ref{scalarUi}) and (\ref{scalarGi}) and again, $\gamma_\mathsmaller{i}$ is defined by (\ref{defgamai}). At breaking points there is one velocity defined from the left-hand side and a different velocity defined from the right-hand side, and the acceleration does not exist. For that reason, the Euler-Lagrange equation is nonsensical \textbf{at} the breaking point. Instead, the \textbf{acceleration-independent} conditions demanding the continuity of (\ref{WE1}) and (\ref{WE2}) at the breaking point replace the Euler-Lagrange equation (\ref{BH}). One particle's velocity discontinuity must be compensated by a discontinuity of the other particle's velocity either in the past lightcone or in the future lightcone, \textbf{or both}, in order for (\ref{WE1}) and (\ref{WE2}) to be continuous at the breaking point \cite{JDE1,Gelfand}. See  Ref. \cite{Daniel} for electromagnetic velocity discontinuities and also pages 61-63 of Ref. \cite{Gelfand} for a finite-dimensional boundary-value problem for a quadratic functional having an extremum with a corner.

 \section{The method of steps}
 \label{ODEsec}

  Our NDDE must start from subluminal trajectory segments with endpoints in lightcone, i.e., consisting of \textbf{two} flights of the sewing chain, as illustrated in Fig. \ref{figODE} and henceforth called the set $\mathcal{FSH}$ of \textbf{full-swing segment pairs},
{\small
\begin{eqnarray}
\mathcal{FSH} \equiv \{ (\mathbf{x}_\mathsmaller{k}(t_\mathsmaller{k}),\mathbf{x}_\mathsmaller{i}(t_\mathsmaller{i}) ) \in \hat{C}^2( [T_\mathsmaller{0}, T_\mathsmaller{2}]) \times \hat{C}^2 ( [T_\mathsmaller{1}, T_\mathsmaller{3}])\;  |  \;  \max||\mathbf{v}_\mathsmaller{k}||<1, \; \max ||\mathbf{v}_\mathsmaller{i}||<1  \}. \label{FSH}
\end{eqnarray}
}   
We henceforth use either a subscript or a superscript {\scriptsize $\circledS$} (from segment) to indicate the dependence on the \textbf{segment} \mbox{\scriptsize $\circledS$} $\in \mathcal{FSH}$, a set defined by (\ref{FSH}). The equations of motion (\ref{BH}) can be integrated forward starting from a full-swing pair of segments {\mbox{\scriptsize $\circledS$} $ \in \mathcal{FSH}$, as explained in Fig. \ref{figODE}. The limit case when {\small $\varepsilon=0$} is henceforth called electrodynamics and discussed in theorem \ref{nosemiflow_theorem}. The method of steps is illustrated in Fig. \ref{figODE} and explained in the caption of Fig. \ref{figODE} using the most advanced acceleration of each equation of motion to construct the other particle's trajectory.  
      \begin{figure}[h!] %  figure placement: here, top, bottom, or pagek
   \centering
   \includegraphics[width=5.0in, height=2.5in]{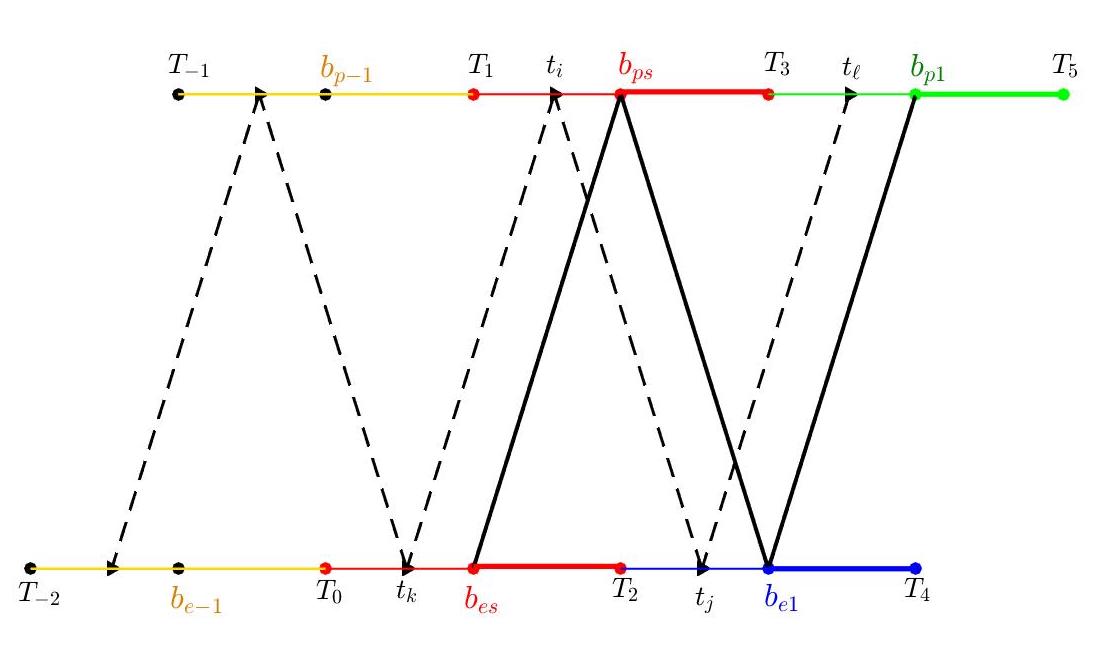} 
   \caption{The initial segments for the method of steps are the positions, the velocities and the accelerations defined on the domains {\color{darkgray}$[T_\mathsmaller{1}, T_\mathsmaller{3}]$} and {\color{darkgray}$ [T_\mathsmaller{0}, T_\mathsmaller{2}]$} with respective endpoints in lightcone. The most advanced leg of each equation of motion is used to extend the other particle's trajectory. In order to extend the electronic trajectory using the most advanced electronic acceleration at time $t_j$, the protonic equation of motion is used  with present time $t_\mathsmaller{i}$ running inside the upper red segment, {\color{darkgray}$t_\mathsmaller{i} \in [T_\mathsmaller{1}, T_\mathsmaller{3}] $}, to draw the lower blue segment. Simultaneously, the electronic equation of motion is used with present time running  from $t_\mathsmaller{j}=T_\mathsmaller{2}$ and past protonic time running from $t_\mathsmaller{i}=T_\mathsmaller{1}$. The electronic acceleration at time $t_\mathsmaller{j}$ runs on the newly created lower blue segment, while the most advanced protonic acceleration produces the upper green extension of the protonic trajectory at time $t_\mathsmaller{\ell}$. One step of integration must integrate the electronic equation of motion until $t_\mathsmaller{j}$ reaches the end of the blue segment, when the past protonic time is $t_\mathsmaller{i}=T_\mathsmaller{3}$, thus making the segments $[T_\mathsmaller{3},T_\mathsmaller{5}]$ and $[T_\mathsmaller{2},T_\mathsmaller{4}]$ simultaneously for another full swing of the sewing chain (the upper green and lower blue segments). This is our basic method of steps. The backward method of steps produces a pair of segments with domains  $[T_\mathsmaller{-1}, T_\mathsmaller{1}]$ and $[T_\mathsmaller{-2}, T_\mathsmaller{0}]$. }
   \label{figODE}
\end{figure}

 \subsection{Reconstruction of the most advanced acceleration} 
  \label{reconstruction}

 For {\small $\varepsilon=0$}, the left-hand side of (\ref{BH}) contains the acceleration in a linear form inherited from the far-field component of (\ref{electric}), i.e., {\small $\mathbf{n}_{\mathsmaller{\, \alpha}} \times \big( (\mathbf{n}_{\mathsmaller{\, \alpha}} +\mathbf{v}_{\! \mathsmaller{\, \alpha}})\times \mathbf{a}_{\! \mathsmaller{\, \alpha}}\big) $}, which linear form vanishes along the eigendirection
 {\small
 \begin{eqnarray}
\mathbf{a}_\alpha \propto  \mathbf{u}_{\mathsmaller{\mathsmaller{\alpha \alpha}}} \equiv (\mathbf{n}_{\! \mathsmaller{\, \alpha}} +\mathbf{v}_{\! \mathsmaller{\, \alpha}}), \label{nulla}
 \end{eqnarray}}
 where the last equality is definition (\ref{ualphai}) for {\small $ \mathbf{u}_{\mathsmaller{\mathsmaller{\alpha \alpha}}}$}. The acceleration can be reconstructed using a geometric identity, i.e.,
{\small
\begin{eqnarray}
\mathbf{a}_{\! \mathsmaller{\, \alpha}}=\frac{\mathbf{n}_{\! \mathsmaller{\, \alpha}}\cdot \mathbf{a}_{\! \mathsmaller{\, \alpha}} }{(1+ \mathbf{n}_{\! \mathsmaller{\, \alpha}}\cdot \mathbf{v}_{\! \mathsmaller{\, \alpha}})}(\mathbf{n}_{\! \mathsmaller{\, \alpha}} +\mathbf{v}_{\! \mathsmaller{\, \alpha}})-\frac{\mathbf{n}_{\! \mathsmaller{\, \alpha}}\times \big( (\mathbf{n}_{\mathsmaller{\, \alpha}} +\mathbf{v}_{\! \mathsmaller{\, \alpha}})\times \mathbf{a}_{\! \mathsmaller{\, \alpha}}\big) }{(1+ \mathbf{n}_{\! \mathsmaller{\, \alpha}}\cdot \mathbf{v}_{\! \mathsmaller{\, \alpha}})}. \label{geometric1}
\end{eqnarray}}
The last term on the right-hand side of (\ref{geometric1}) is proportional to the acceleration linear-form contained in the near-field of (\ref{electric}), i.e., the {\small $\frac{1}{r_\mathsmaller{ji}}$} component. Because of Eq. (\ref{magnetic}), the right-hand side of (\ref{BH}) is a function of {\small $\mathbf{E}_{\mathsmaller{ji}}$} only. The existence of the null direction {\small $(\ref{nulla})$} prevents the most advanced acceleration {\small $\mathbf{a}_{\! \mathsmaller{\, j}}$} to be reconstructed from the right-hand side of (\ref{BH}). In other words, the value of {\small $\mathbf{E}_{\mathsmaller{ji}}$} does not define the  coefficient {\small $\mathbf{n}_{\! \mathsmaller{\, j}}\cdot \mathbf{a}_{\! \mathsmaller{\, j}} $} of the first term on the right-hand side of (\ref{geometric1}). This is the reason we had to perturb electrodynamics in the first place. An identity displaying the rank-deficiency is obtained by comparing definition (\ref{Right_alpha}) with a re-arranged version of the scalar product of {\small $\mathbf{v}_{\mathsmaller{\alpha}}$} with formula (\ref{electric}), yielding

{\small
\begin{eqnarray}
\mathbf{n}_{\mathsmaller{\alpha}}\cdot \mathbf{a}_{\mathsmaller{\alpha}} &=& \hat{\mathcal{R}}_{\mathsmaller{\alpha}}\cdot \mathbf{a}_{\mathsmaller{\alpha}} +\frac{\mathbf{v}_\mathsmaller{\alpha}\cdot [\mathbf{n}_{\! \mathsmaller{\, \alpha}}\times ( \mathbf{u}_{\mathsmaller{\, \alpha \alpha}} \times \mathbf{a}_{\! \mathsmaller{\, \alpha}})] }{(1+ \mathbf{n}_{\! \mathsmaller{\, \alpha}}\cdot \mathbf{v}_{\! \mathsmaller{\, \alpha}})} \notag \\
&=&\hat{\mathcal{R}}_{\mathsmaller{\alpha}}\cdot \mathbf{a}_{\mathsmaller{\alpha}} +\frac{\mathbf{v}_\mathsmaller{\alpha} \cdot [( \mathbf{n}_\mathsmaller{\alpha}\cdot \mathbf{u}_\mathsmaller{\alpha \alpha })  \mathbf{E}_{\mathsmaller{ji}} \,-\,(\mathbf{n}_\mathsmaller{\alpha} \cdot \mathbf{E}_{\mathsmaller{ji}})\mathbf{u}_\mathsmaller{\alpha \alpha}]  }{\zeta_\mathsmaller{\alpha i} (\mathbf{n}_\mathsmaller{\alpha}\cdot \mathbf{u}_\mathsmaller{\alpha \alpha })^\mathsmaller{2} }, \label{unfolding}
\end{eqnarray}}
where
{\small
\begin{eqnarray}
\zeta_\mathsmaller{\alpha i} \equiv \frac{e_\mathsmaller{\alpha}}{r_\mathsmaller{\alpha i}(1+\mathbf{n}_\mathsmaller{\alpha}\cdot \mathbf{v}_\mathsmaller{\alpha} )^3 }. \label{defzeta}
\end{eqnarray}}
Notice on the numerator of the second equality of (\ref{unfolding}) that we have subtracted the near-field from {\small $\mathbf{E}_{\mathsmaller{ji}}$}. The reconstruction of {\small $\mathbf{a}_\mathsmaller{\alpha}$} with (\ref{geometric1}) and (\ref{unfolding}) requires the linear form {\small $\hat{\mathcal{R}}_\mathsmaller{\alpha}\cdot \mathbf{a}_\mathsmaller{\alpha} $}, as provided by the far-field perturbation (\ref{betaforce}) \textbf{plus} the information contained in the electric field (\ref{electric}).

\subsection{{\small $\varepsilon$}-strong interaction cures rank deficiency}
\label{rank_deficiency}

The equation of motion (\ref{BH}) can be re-arranged with the most advanced acceleration terms on the left-hand side, i.e.,
{\small
\begin{eqnarray}
 \digamma_{\mathsmaller{\!  j i}}^\mathsmaller{e\ell} +\varepsilon b_{\mathsmaller{j i}}\; \hat{\mathcal{R}}_{\mathsmaller{j}}\cdot \mathbf{a}_{j} \; \big( \frac{\mathbf{u}_\mathsmaller{j i}}{\mathbf{n}_\mathsmaller{j} \cdot \mathbf{u}_\mathsmaller{j i}} \big) \negthickspace \negthickspace &=& \! \! \Lambda_{\mathsmaller{ \, i}}^{\mathsmaller{\circledS}}  , \label{BH-rewrite1}
 \end{eqnarray}} 
 where 
 {\small
 \begin{eqnarray}
 \digamma_{\mathsmaller{\! \alpha i}}^\mathsmaller{ e\ell} &\equiv &  e_{\! \mathsmaller{\, i}} (1\pm \mathbf{n}_{\! \mathsmaller{\, \alpha}}\cdot \mathbf{v}_{\! \mathsmaller{\, i}})\mathbf{E}_{\! \mathsmaller{\, \alpha i}} \mp e_{\! \mathsmaller{\, i}} (\mathbf{v}_{\! \mathsmaller{\, i}} \cdot \mathbf{E}_{\! \mathsmaller{\, \alpha i}}) \mathbf{u}_\mathsmaller{\alpha i}\,, \label{defFel} \\ 
  \Lambda_{\mathsmaller{\, i}}^{\mathsmaller{\circledS}} & \equiv &  2( m_{\! \mathsmaller{\, i}}+\varepsilon \mathsf{G}_\mathsmaller{i} )\gamma_\mathsmaller{i} \mathbf{a}_{\! \mathsmaller{\, i}}
- \digamma_{\mathsmaller{\! k i}}^\mathsmaller{ e\ell} -\, \varepsilon \mathbf{\Upsilon}_\mathsmaller{i} \,, \label{defDDE}
\end{eqnarray}}
and
{\small
\begin{eqnarray}
\thickspace \thickspace \thickspace \thickspace \thickspace \thickspace \thickspace \thickspace  \mathbf{\Upsilon}_\mathsmaller{i} \equiv c_{\mathsmaller{j i}}\mathbf{\Omega}^{\dagger}_\mathsmaller{j i}+c_{\mathsmaller{k i}}\mathbf{\Omega}^{\dagger}_\mathsmaller{k i}+ b_{\mathsmaller{k i}}\; \hat{\mathcal{R}}_{\mathsmaller{k}}\cdot \mathbf{a}_{\mathsmaller{k}}  \big( \frac{\mathbf{u}_\mathsmaller{k i}}{\mathbf{n}_\mathsmaller{k} \cdot \mathbf{u}_\mathsmaller{k i}} \big).\label{def-fstar}
 \end{eqnarray}}
 Observations: {\color{darkgray}(i)} {\small $ \mathbf{E}_{\! \mathsmaller{\, \alpha i}} $} is defined by (\ref{electric}) in Eq. (\ref{defFel}), {\color{darkgray}(ii)} in Eq. (\ref{BH-rewrite1}), the term containing the most advanced acceleration {\small $\mathbf{a}_\mathsmaller{j}$} was passed to the left-hand side while the remaining terms were collected in (\ref{def-fstar}) with {\small $\mathbf{\Upsilon}_\mathsmaller{i}$} representing \textbf{two times} the remaining non-electromagnetic terms of the right-hand side of (\ref{BH}) divided by $\varepsilon$, {\color{darkgray}(iii)}  the scalar product of the left-hand side of (\ref{BH-rewrite1}) with the killing-vector  
{\small
 \begin{eqnarray}
 \Bbbk_{\mathsmaller{\alpha i}}  \equiv  \Big(\frac{1-\mathbf{v}_{\! \mathsmaller{i}}^{\! \mathsmaller{\, 2}}}{1\pm \mathbf{n}_{\! \mathsmaller{\,\alpha}}\cdot \mathbf{v}_{\! \mathsmaller{\, i}}} \Big)\mathbf{n}_{\mathsmaller{\alpha}} \pm   \mathbf{v}_{\mathsmaller{i}}\, ,  \label{left_alpha}
 \end{eqnarray}}
for $\alpha=j$ vanishes the acceleration term of the electromagnetic sector because {\small $\Bbbk_{\mathsmaller{\alpha i}} $} is a left null-vector of the acceleration form {\small $\digamma_{\mathsmaller{\! \alpha i}}^\mathsmaller{ e\ell}$}, i.e., 
 {\small
 \begin{eqnarray}
\digamma^{\mathsmaller{o}}_{\mathsmaller{\!ji}} \equiv \Bbbk_{\mathsmaller{ji}}\cdot \digamma_{\mathsmaller{\!j i}}^\mathsmaller{ e\ell}=(1-\mathbf{v}_{\mathsmaller{i}}^\mathsmaller{2})(e_\mathsmaller{i} \mathbf{n}_\mathsmaller{j} \cdot \mathbf{E}_{\mathsmaller{ji}})= \frac{e_\mathsmaller{i} e_\mathsmaller{j}(1-\mathbf{v}_{\mathsmaller{i}}^\mathsmaller{2})(1-\mathbf{v}_{\mathsmaller{j}}^\mathsmaller{2}) }{r^{\mathsmaller{2}}_{\mathsmaller{j i}}(1+ \mathbf{n}_{\! \mathsmaller{\,j}}\cdot \mathbf{v}_{\! \mathsmaller{\, j}})^\mathsmaller{2}}\,. \label{residual}
 \end{eqnarray}} 
 The acceleration-independent reminder (\ref{residual}) is due to the near-field, {\color{darkgray}(iv)} we need {\small $\hat{\mathcal{R}}_{\mathsmaller{j}}\cdot \mathbf{a}_{\mathsmaller{j}}$} to reconstruct the acceleration {\small $\mathbf{a}_\mathsmaller{j}$} using (\ref{geometric1}) and (\ref{unfolding}) with {\small $\alpha=j$}, which elusive ingredient is recovered from the scalar product of {\small $\Bbbk_{\mathsmaller{ji}}$} with (\ref{BH-rewrite1}) \textbf{only} at {\small $O(\varepsilon)$}, i.e.,
{\small
 \begin{eqnarray}
\varepsilon b_{\mathsmaller{j i}}\; \hat{\mathcal{R}}_{\mathsmaller{j}}\cdot \mathbf{a}_{j} &=& \Bbbk_{\mathsmaller{ji}}\cdot  \Lambda_{\mathsmaller{i}}^{\mathsmaller{\circledS}}-\digamma^{\mathsmaller{o}}_\mathsmaller{ji}. \label{scalarprods} 
\end{eqnarray}}  
In Eq. (\ref{scalarprods}) we have used 
 {\small
 \begin{eqnarray}
  \Bbbk_\mathsmaller{ji} \cdot \mathbf{u}_\mathsmaller{ji} =\mathbf{n}_\mathsmaller{j}\cdot \mathbf{u}_\mathsmaller{ji} =(1+ \mathbf{n}_\mathsmaller{\, j} \cdot \mathbf{v}_\mathsmaller{i}). \label{Lalphu}
 \end{eqnarray}
 }
Equation (\ref{scalarprods}) at {\small $\varepsilon=0$} is a constraint on the possible values of {\small $ \Lambda_{\mathsmaller{\, i}}^{\mathsmaller{\circledS}} $}. Otherwise, when {\small $\varepsilon \neq 0$}, the right-hand side of (\ref{scalarprods}) involves {\small $\hat{\mathcal{R}}_{\mathsmaller{j}}\cdot \mathbf{a}_{\mathsmaller{j}}$} multiplied by a non-zero coefficient, and {\small $\Lambda_{\mathsmaller{\, i}}^{\mathsmaller{\circledS}}$} is unconstrained.  In order to solve for {\small $\digamma_{\mathsmaller{\!ji}}^\mathsmaller{ e\ell}$} in terms of the past data, we subtract Eq. (\ref{scalarprods}) multiplied by {\small $\mathbf{u}_\mathsmaller{ji}/(\Bbbk_\mathsmaller{ji}\cdot \mathbf{u}_\mathsmaller{ji})$} from Eq. (\ref{BH-rewrite1}) and re-arrange, yielding 
 {\small
 \begin{eqnarray}
 \digamma_{\mathsmaller{ji}}^\mathsmaller{e\ell}&=&\frac{ \digamma^{\mathsmaller{o}}_\mathsmaller{ji} }{ (1+ \mathbf{n}_\mathsmaller{\, j} \cdot \mathbf{v}_\mathsmaller{i}) }  \mathbf{u}_\mathsmaller{ji} - \frac{\Bbbk_{\mathsmaller{ji}} \times ( \mathbf{u}_\mathsmaller{ji} \times \Lambda_{\mathsmaller{\, i}}^{\mbox{{\negmedspace \tiny{dat}}}}) }{(1+ \mathbf{n}_\mathsmaller{\, j} \cdot \mathbf{v}_\mathsmaller{i})}  \,, \label{Fel_value}
 \end{eqnarray}
 }
where again we have used (\ref{Lalphu}) and {\color{darkgray}(v)} we can calculate {\small $e_\mathsmaller{i}\mathbf{v}_{\! \mathsmaller{\, i}} \cdot \mathbf{E}_{\! \mathsmaller{\, ji}}$}\, by taking the scalar product of {\small $\mathbf{v}_\mathsmaller{i}$} with the two lines on the right-hand side of (\ref{defFel}) and dividing the result by the non-zero factor {\small $(1-\mathbf{v}_{\mathsmaller{i}}^\mathsmaller{2})$}. Substituting the resulting formula for {\small $e_\mathsmaller{i}\mathbf{v}_{\! \mathsmaller{\, i}} \cdot \mathbf{E}_{\! \mathsmaller{\, ji}}$}  back into (\ref{defFel}) we have a formula for {\small $ \mathbf{E}_\mathsmaller{ji} $ in terms of $\digamma_{\mathsmaller{ji}}^\mathsmaller{e\ell}$}, i.e.,
{\small
\begin{eqnarray}
e_\mathsmaller{i}(\mathbf{n}_\mathsmaller{j}\cdot \mathbf{u}_\mathsmaller{ji})  \mathbf{E}_\mathsmaller{ji} &=& \gamma_{\mathsmaller{i}}^{\mathsmaller{2}} \Big(  \digamma_{\mathsmaller{ji}}^\mathsmaller{e\ell} \pm (\mathbf{v}_\mathsmaller{i} \cdot\digamma_{\mathsmaller{ji}}^\mathsmaller{e\ell}) \mathbf{n}_\mathsmaller{j}  +\mathbf{v}_\mathsmaller{i} \times (\mathbf{v}_\mathsmaller{i} \times  \digamma_{\mathsmaller{ji}}^\mathsmaller{e\ell} ) \Big).
 \label{defEadvanced}
 \end{eqnarray}}
 Notice that because of (\ref{Fel_value}), Eq. (\ref{defEadvanced}) defines  {\small $ \mathbf{E}_\mathsmaller{ji} $} in terms of the past data.

\subsection{Differential-delay equation of motion with two delays of neutral type}
\label{MATLABDDEsec}
 
Numerical integration of differential-delay equations with two delays is a topic of modern interest, e.g., \cite{Daniel_Mike, Jianhong}. In order to prepare a numerical study using, for example, the function \textbf{ddensd} of MATLAB, we put together the equation of motion of particle {\small $j$} by reconstructing its acceleration with (\ref{unfolding}), (\ref{scalarprods}) and (\ref{defEadvanced}). Solving the first identity of (\ref{unfolding}) for {\small $\mathbf{a}_\mathsmaller{j}$} we obtain
 {\small
  \begin{eqnarray}
\negthickspace \negthickspace \frac{1}{r_\mathsmaller{ji}} \Big( \frac{dt_\mathsmaller{j}}{dt_\mathsmaller{i}}\Big) \mathbf{a}_\mathsmaller{j}\negthickspace \negthickspace&=& \negthickspace \negthickspace \negthickspace\frac{1}{r_\mathsmaller{ji}} \Big( \frac{dt_\mathsmaller{j}}{dt_\mathsmaller{i}}\Big)(\hat{\mathcal{R}}_\mathsmaller{j} \cdot \mathbf{a}_\mathsmaller{j}) \big( \frac{\mathbf{u}_\mathsmaller{jj}}{\mathbf{n}_\mathsmaller{j}\cdot \mathbf{u}_\mathsmaller{jj} }  \big)\, -\, \frac{1}{e_\mathsmaller{i} e_\mathsmaller{j}} \mathrm{S}_\mathsmaller{ji},  \label{equation_aj}
  \end{eqnarray}
  }
  where the gyroscopic term $\mathrm{S}_\mathsmaller{ji}$ is defined by
  {\small
  \begin{eqnarray}
 \mathrm{S}_\mathsmaller{ji} \equiv e_\mathsmaller{i} (\mathbf{n}_\mathsmaller{j}\cdot \mathbf{u}_\mathsmaller{jj}) (\mathbf{n}_\mathsmaller{j}\cdot \mathbf{u}_\mathsmaller{j i})\Big( \mathbf{E}_\mathsmaller{ji} - \hat{\mathcal{R}}_\mathsmaller{j} \cdot \mathbf{E}_\mathsmaller{ji} \big( \frac{\mathbf{u}_\mathsmaller{jj}}{\mathbf{n}_\mathsmaller{j}\cdot \mathbf{u}_\mathsmaller{jj} }  \big)\Big), \label{gyroscopic}
   \end{eqnarray}}
with {\small $\mathbf{E}_\mathsmaller{ji}$} expressed in terms of past data by (\ref{Fel_value}) and (\ref{defEadvanced}). In order to express the gyroscopic term as a function of the segment, we define the relative angular velocities by
{\small
\begin{eqnarray}
\ell_\mathsmaller{\alpha \beta} &\equiv &\mathbf{n}_\mathsmaller{\alpha} \times \mathbf{v}_\mathsmaller{\beta}, \label{defell_ab}
\end{eqnarray}} 
and define the nonlinear spin vector by
{\small
\begin{eqnarray}
\mathbb{L}_{\mathbf{E}} \equiv e_\mathsmaller{i} (\mathbf{n}_\mathsmaller{j} \cdot \mathbf{u}_\mathsmaller{ji}) \mathbf{n}_\mathsmaller{j} \times \mathbf{E}_\mathsmaller{ji} = \mathbf{n}_\mathsmaller{j} \times \digamma_{\mathsmaller{ji}}^\mathsmaller{e\ell} + \gamma^{\mathsmaller{2}}_\mathsmaller{i} (\mathbf{v}_\mathsmaller{i} \cdot \digamma_{\mathsmaller{ji}}^\mathsmaller{e\ell} )\ell_\mathsmaller{j i} \, , \label{defLE}
\end{eqnarray}
} 
where {\small $\ell_\mathsmaller{j i}$} is defined by (\ref{defell_ab}). Because of (\ref{Fel_value}), Eq. (\ref{defLE}) defines {\small $\mathbb{L}_{\mathbf{E}} $} in terms of the segment $\circledS \in \mathcal{FSH}$. Using (\ref{defLE}) to express
the gyroscopic term (\ref{gyroscopic}) we have
{\small
\begin{eqnarray}
\mathrm{S}_\mathsmaller{ji}&=&\Big((\mathbf{n}_\mathsmaller{j} \cdot \mathbf{u}_\mathsmaller{ji}) \ell_{\mathsmaller{jj}}^\mathsmaller{2} e_\mathsmaller{i}\mathbf{n}_\mathsmaller{j} \cdot \mathbf{E}_\mathsmaller{ji}\, -\, \mathbf{n}_\mathsmaller{j} \cdot \mathbf{u}_\mathsmaller{jj} \ell_\mathsmaller{j} \cdot \mathbb{L}_{\mathbf{E}} \Big)\mathbf{n}_\mathsmaller{j} -(\mathbf{n}_\mathsmaller{j} \cdot \mathbf{u}_\mathsmaller{jj}) \mathbf{n}_\mathsmaller{j} \times \mathbb{L}_{\mathbf{E}} \notag \\&&+\Big(\ell_\mathsmaller{jj} \cdot \mathbb{L}_{\mathbf{E}} +(\mathbf{n}_{\mathsmaller{j}}\cdot \mathbf{u}_\mathsmaller{ji} )(\mathbf{n}_{\mathsmaller{jj}}\cdot \mathbf{u}_\mathsmaller{jj}-\ell^{\mathsmaller{2}}_{\mathsmaller{jj}})e_\mathsmaller{i} \mathbf{n}_\mathsmaller{j}\cdot \mathbf{E}_\mathsmaller{ji} \Big)\big( \frac{\mathbf{n}_\mathsmaller{j}\times \ell_\mathsmaller{jj} }{\mathbf{n}_{\mathsmaller{jj}}\cdot \mathbf{u}_\mathsmaller{jj}} \big).
\label{gyroscopic2}
\end{eqnarray}
}
where {\small $\ell_\mathsmaller{j j}$} is defined by (\ref{defell_ab}) and {\small $\mathbb{L}_\mathbf{E}$} is defined by (\ref{defLE}). In order to display (\ref{equation_aj}) in its explicit singular form, we multiply (\ref{equation_aj}) by {\small $\varepsilon$} and use (\ref{scalarprods}) to eliminate {\small $\varepsilon \hat{\mathcal{R}}_{\mathsmaller{j}}\cdot \mathbf{a}_{j}$}, yielding
{\small
  \begin{eqnarray}
\negthickspace \negthickspace \frac{\varepsilon}{r_\mathsmaller{ji}} \Big( \frac{dt_\mathsmaller{j}}{dt_\mathsmaller{i}}\Big) \mathbf{a}_\mathsmaller{j} \negthickspace \negthickspace&=& \negthickspace \negthickspace \negthickspace \Big( \frac{dt_\mathsmaller{j}}{dt_\mathsmaller{i}}\Big)(  \frac{\Bbbk_{\mathsmaller{ji}}\cdot  \Lambda_{\mathsmaller{\, i}}^{\mbox{{\negmedspace \tiny{dat}}}}-\digamma^{\mathsmaller{o}}_{\mathsmaller{ji}} }{ b_\mathsmaller{ji}r_\mathsmaller{ji}}   ) \big( \frac{\mathbf{u}_\mathsmaller{jj}}{\mathbf{n}_\mathsmaller{j}\cdot \mathbf{u}_\mathsmaller{jj} }  \big)\, -\, \frac{\varepsilon}{e_\mathsmaller{i} e_\mathsmaller{j}} \mathrm{S}_\mathsmaller{ji}, \label{Singular_equation_aj}
  \end{eqnarray}}
where {\small $\digamma^{\mathsmaller{o}}_\mathsmaller{ji}$} is defined by (\ref{residual}).
Observations: {\color{darkgray}(i)} the equation of motion for the most advanced protonic acceleration, {\small $\mathbf{a}_\mathsmaller{\ell}$}\,, is obtained from (\ref{Singular_equation_aj}) by shifting the indices with the permutation {\small $(k i j ) \rightarrow (i j \ell)$},  {\color{darkgray}(ii)} the gyroscopic term $\mathrm{S}_\mathsmaller{ji}$ includes a longitudinal term along the direction {\small $\mathbf{n}_\mathsmaller{j}$} with a coefficient that is a nonlinear function of the transverse quantities (\ref{defell_ab}) and (\ref{defLE}), {\color{darkgray}(iii)} the quantity  {\small $\mathbf{n}_\mathsmaller{j}\cdot \mathbf{E}_\mathsmaller{ji} $} appearing in the right-hand side of (\ref{gyroscopic2}) is independent of the most advanced acceleration, and  {\color{darkgray}(iv)} both sides of the equation of motion (\ref{Singular_equation_aj}) are \textbf{linear} on the accelerations {\small $\mathbf{a}_\mathsmaller{j}(t_\mathsmaller{j})$}, {\small $\mathbf{a}_\mathsmaller{i}(t_\mathsmaller{i})$} and {\small $\mathbf{a}_\mathsmaller{k}(t_\mathsmaller{k})$}. 

\begin{theorem} Equation (\ref{Singular_equation_aj}) with {\small $\varepsilon \neq 0$} defines the most advanced acceleration in {\small $ C^2(\mathbb{R}) $} when the initial segment used to define the function {\small $\Lambda_{\mathsmaller{\, i}}^{\mathsmaller{\circledS}}$} belongs to {\small $ C^2(\mathbb{R}) $}.
\label{semiflowtheorem}
\end{theorem}
\begin{proof} The proof is the reconstruction of the most advanced acceleration {\small $\mathbf{a}_\mathsmaller{j}$} in {\small $C^2(\mathbb{R}) $} using Eq. (\ref{Singular_equation_aj}) and an initial segment belonging to {\small $C^2(\mathbb{R}) $}. An equation of motion for the most advanced protonic acceleration, {\small $\mathbf{a}_\mathsmaller{\ell}(t_\mathsmaller{\ell}) $}\,, is obtained from (\ref{Singular_equation_aj}) by shifting the indices with the permutation {\small $(k i j ) \rightarrow (i j \ell)$}.  
 \end{proof}

  NDDE (\ref{Singular_equation_aj}) can start from arbitrary full-swing segment pairs in {\small $C^2(\mathbb{R}) $}. The existence of a semiflow like (\ref{Singular_equation_aj}) is already some stability statement, because the dynamics can propagate and possibly persist in a serrated orbit of {\small $C^2(\mathbb{R})$}. Notice that the right-hand side of (\ref{Singular_equation_aj}) is independent of the most advanced accelerations {\small $\mathbf{a}_\mathsmaller{j}$} and {\small $\mathbf{a}_\mathsmaller{\ell}$}. Again, the protonic equation of motion for {\small $\mathbf{a}_\mathsmaller{\ell}$} is obtained from (\ref{Singular_equation_aj}) by shifting the indices {\small $(k i j ) \rightarrow (i j \ell)$}. This completes the method of steps explained in the caption of Fig. \ref{figODE}. Together with its index-shifted protonic equation for {\small $\mathbf{a}_\mathsmaller{\ell}$}, NDDE (\ref{Singular_equation_aj}) can be used to build the function \textbf{ddensd} of MATLAB.
  
\subsection{Unfolding Driver's degeneracy}
\label{driversec}

Here we discuss the one-dimensional electromagnetic problems with either repulsive interaction \cite{Driver,EFY1} or attractive interaction\cite{EFY2}. For motion restricted to a straight-line by the initial condition segment \cite{EFY2, Driver,EFY1}, the Euler-Lagrange equations of electrodynamics turn out to be independent of the most advanced acceleration \textit{and} the far-field interaction vanishes, a degeneracy henceforth called Driver's riddle \cite{Driver}. We start from the one-dimensional version of partial-Lagrangian (\ref{partial_Lagrangian}), i.e.,

{\small
\begin{equation}
\mathsf{L}_{\! \mathsmaller{\, i}}(t_{\mathsmaller{i}},\mathbf{x}_{\! \mathsmaller{\, i}}\, ,\mathbf{v}_{\! \mathsmaller{\, i}}) \equiv m_{\! \mathsmaller{\, i}}(\,1-\sqrt{1-\mathbf{v}_{\! \mathsmaller{\, i}}^{\! \mathsmaller{\, 2}}} \;\;) - \sum_{\! \mathsmaller{\alpha= k,j}} \frac{1}{2r_\mathsmaller{\alpha i}} \Big(e_{\! \mathsmaller{\, i}} e_\mathsmaller{\alpha}\mathsf{V\!A}_\mathsmaller{\alpha i} (\mathbf{v}_\mathsmaller{\alpha}, \mathbf{v}_\mathsmaller{i})+\varepsilon \mathsf{VG}_\mathsmaller{\alpha i}(\mathbf{v}_\mathsmaller{\alpha}, \mathbf{v}_\mathsmaller{i})\Big),\label{partial_Lagrangian1D}
\end{equation}}
 where {\small $\mathsf{V\!A}_\mathsmaller{\alpha i} (\mathbf{v}_\mathsmaller{\alpha}, \mathbf{v}_\mathsmaller{i})$} and {\small $\mathsf{VG}_\mathsmaller{\alpha i}(\mathbf{v}_\mathsmaller{\alpha}, \mathbf{v}_\mathsmaller{i})$} are given by
 {\small
\begin{eqnarray}
  \mathsf{V\!A}_\mathsmaller{\alpha i} (\mathbf{v}_\mathsmaller{\alpha}, \mathbf{v}_\mathsmaller{i}) & \equiv &  \frac{(1-\mathbf{v}_\mathsmaller{i} \cdot\mathbf{{v}}_{\! \mathsmaller{\, \alpha}}) }{(1\pm \mathbf{n}_{\! \mathsmaller{\, \alpha }}\cdot \mathbf{v}_{\! \mathsmaller{\alpha}})} , \label{VA1D} \\
\mathsf{VG}_\mathsmaller{\alpha i}(\mathbf{v}_\mathsmaller{\alpha}, \mathbf{v}_\mathsmaller{i}) & \equiv & \frac{\sqrt{1-\mathbf{v}^{\mathsmaller{2}}_{\mathsmaller{i}} }\,\sqrt{1-\mathbf{v}_{\mathsmaller{\alpha}}^{\mathsmaller{2}}} }{(1\pm \mathbf{n}_{\! \mathsmaller{\, \alpha }}\cdot \mathbf{v}_{\! \mathsmaller{\, \alpha}})}. \label{VG1D} 
\end{eqnarray}}
A significant simplification is achieved using \textit{relative} velocity angles {\small $\phi_{\mathsmaller{\alpha}} \in \mathbb{R}$} for the one-dimensional velocity of each charge, i.e.,
{\small
\begin{eqnarray}
  \mathbf{v}_\mathsmaller{\alpha} \equiv \tanh{\phi_\mathsmaller{\alpha}} \, \mathbf{n}_\mathsmaller{\alpha} \label{def_hyperphi}
\end{eqnarray}}
and using (\ref{def_hyperphi}) to express  (\ref{defgamai}) we have
{\small
\begin{eqnarray}
\gamma_\mathsmaller{\alpha}=\cosh(\phi_\mathsmaller{\alpha}). \label{gamaphi}
\end{eqnarray}
}
We henceforth assume that the position of each particle falls on its respective side along the light-cone direction {\small $\hat{\mathbf{x}} $}, with particle {\small $i$} standing on the right-hand side with a positive coordinate while particle {\small $k$} stands on the left-hand side with a negative coordinate. The former setup leads to {\small $\mathbf{n}_\mathsmaller{j}=\mathbf{n}_\mathsmaller{k} \equiv \hat{\mathbf{x}} $} and {\small $\mathbf{n}_\mathsmaller{i} = \mathbf{n}_\mathsmaller{s}=\mathbf{n}_\mathsmaller{\ell} \equiv-\hat{\mathbf{x}}$}. The one-dimensional version of {\small $\mathsf{G}_\mathsmaller{i}$} as defined by (\ref{scalarGi}) is
{\small
\begin{eqnarray}
\mathsf{G}_\mathsmaller{i}&=&\Big( \frac{e^{\phi_\mathsmaller{k}}}{2r_\mathsmaller{ki}}+\frac{e^{-\phi_\mathsmaller{j}}}{2r_\mathsmaller{ji}}\Big). \label{simpliG}
\end{eqnarray} }
 In order to describe the details of both the repulsive and the attractive case, we define the relative parameter by
{\small
\begin{eqnarray}
\varepsilon_{*} \equiv -\big( \frac{\varepsilon}{e_\mathpzc{e} e_\mathpzc{p}}\big)\label{defeps_star}.
\end{eqnarray}}
The Euler-Lagrange equation of (\ref{partial_Lagrangian1D}) written in the time-symmetric form is 
{\small
\begin{eqnarray}
\negthickspace \negthickspace (m_\mathsmaller{i}+ \frac{\varepsilon e^{\phi_\mathsmaller{k}}}{2r_\mathsmaller{ki}}+ \frac{\varepsilon e^{\mathsmaller{-}\phi_\mathsmaller{j}}}{2r_\mathsmaller{ji}})\dot\phi_\mathsmaller{i} \negthickspace &=&\negthickspace \big(  \frac{\varepsilon e^{\phi_\mathsmaller{k}}}{2r_{\mathsmaller{ki}}} \big)  \big( \frac{dt_\mathsmaller{k}}{dt_\mathsmaller{i}}\big) \dot \phi_\mathsmaller{k}+\big( \frac{\varepsilon e^{\mathsmaller{-}\phi_\mathsmaller{j}}}{2r_{\mathsmaller{ji}}}\big)  \big( \frac{dt_\mathsmaller{j}}{dt_\mathsmaller{i}}\big) \dot \phi_\mathsmaller{j} \notag \\
&&\negthickspace -\frac{ e_\mathsmaller{i}e_\mathsmaller{k}  e^{2\phi_\mathsmaller{k}} }{ 2r^{\mathsmaller{2}}_{\mathsmaller{ki}} \cosh{\phi_\mathsmaller{i}}}\Big( 1-\varepsilon_\mathsmaller{*} \cosh{(\phi_\mathsmaller{i} +\phi_\mathsmaller{k})} \Big)\notag \\
&&\negthickspace -\frac{ e_\mathsmaller{i}e_\mathsmaller{j}  e^{\mathsmaller{-}2\phi_\mathsmaller{j}} }{ 2r^{\mathsmaller{2}}_{\mathsmaller{ji}} \cosh{\phi_\mathsmaller{i}}}\Big( 1 -\varepsilon_\mathsmaller{*} \cosh{(\phi_\mathsmaller{i}+\phi_\mathsmaller{j})}  
\Big), \notag \\ \label{1D_unfold}
\end{eqnarray}}
where {\small $(dt_\mathsmaller{\alpha}/dt_\mathsmaller{i})$} is defined by (\ref{time_rate}) and, when expressed by velocity angles for {\small $\alpha \in (k,j)$} standing for a nearest neighbor of {\small $i$},  it becomes
{\small
\begin{eqnarray}
\frac{dt_\mathsmaller{\alpha}}{dt_\mathsmaller{i}} =\frac{e^{\mp \phi_\mathsmaller{\alpha}}e^{\mp \phi_\mathsmaller{i}} \cosh{\phi_\mathsmaller{\alpha}}}{ \cosh{\phi_\mathsmaller{i}}}. \label{1D-timerate}
\end{eqnarray}}
Observations: {\color{darkgray}(i)} the dot in {\small $\dot \phi_\mathsmaller{\alpha} \equiv \frac{d\phi_\mathsmaller{\alpha}}{dt_\mathsmaller{\alpha}}$} indicates derivative respect to time {\small $t_\mathsmaller{\alpha}$}, as always in this manuscript, {\color{darkgray}(ii)} the equation for the most advanced protonic acceleration {\small $\dot \phi_\mathsmaller{\ell}$} is obtained by shifting the indices {\small $(k i j ) \rightarrow (i j \ell)$}, {\color{darkgray}(iii)} Eq. (\ref{1D_unfold}) has the acceleration terms proportional to {\small $\varepsilon/r_\mathsmaller{ik}$} and {\small $\varepsilon/r_\mathsmaller{ij}$}. When {\small $\varepsilon=0$} Eq. (\ref{1D_unfold}) lacks the most advanced acceleration, giving rise to the riddle found in Refs.\cite{EFY2,Driver,EFY1}.
 \par
In order to use Eq. (\ref{1D_unfold}) in the method of steps, one should solve it for the most advanced derivative, i.e.,
{\small
\begin{eqnarray}
\varepsilon \Big( \frac{  e^{-2\phi_\mathsmaller{j}}\cosh{\phi_\mathsmaller{j}} }{r_\mathsmaller{ji}} \Big) \dot \phi_\mathsmaller{j} &=& \Big(2m_\mathsmaller{i}+ \frac{\varepsilon e^{\phi_\mathsmaller{k}}}{r_\mathsmaller{ki}} + \frac{\varepsilon e^{\mathsmaller{-}\phi_\mathsmaller{j}}}{r_\mathsmaller{ji}} \Big)  \cosh{\phi_\mathsmaller{i} }\; \dot\phi_\mathsmaller{i}   \notag \\ &&-\varepsilon  \Big( \frac{ e^{2\phi_\mathsmaller{k} }\cosh{\phi_\mathsmaller{k}} }{r_{\mathsmaller{ki}}}    \Big) \; \dot \phi_\mathsmaller{k} \notag \\
& &+\frac{ e_\mathsmaller{i}e_\mathsmaller{j} e^{-2\phi_\mathsmaller{j}}}{  r^{\mathsmaller{2}}_{\mathsmaller{ji}}}\Big( 1-\varepsilon_\mathsmaller{*}\cosh{(\phi_\mathsmaller{j}+\phi_\mathsmaller{i})}  
\Big) \notag \\ & &+\frac{ e_\mathsmaller{i}e_\mathsmaller{k}  e^{2\phi_\mathsmaller{k}}}{  r^{\mathsmaller{2}}_{\mathsmaller{ki}}}  \Big( 1-\varepsilon_\mathsmaller{*}\cosh{(\phi_\mathsmaller{i}+\phi_\mathsmaller{k})}  
\Big). \label{matlab1D}
\end{eqnarray}}
 As with (\ref{1D_unfold}), the equation for {\small $\dot \phi_\mathsmaller{\ell}$} corresponding to (\ref{matlab1D}) is obtained by shifting the indices with {\small $(k i j ) \rightarrow (i j \ell)$}.  
 For {\small $0<\varepsilon_\mathsmaller{*}\leq 1$}, the last two lines of (\ref{matlab1D}) vanish identically if the velocity of relative approximation reaches the (singular) value 
{\small
\begin{eqnarray}
\cosh(\phi_\mathsmaller{e}+\phi_\mathsmaller{p}) =\frac{1}{\varepsilon_{*}} \equiv \cosh{2\phi_\mathsmaller{\varepsilon}}. \label{defineutron}
\end{eqnarray}}
If the approximation velocities lock in condition (\ref{defineutron}), the segment iterates to another constant-velocity segment, as follows.
\begin{theorem} For {\small $0<\varepsilon_\mathsmaller{*} \leq 1$}, the method of steps for (\ref{matlab1D}) has a one-parameter family of fixed velocity-segments {\small $\phi_\mathsmaller{e}(t_\mathsmaller{e})=\phi_\mathsmaller{e}$} and {\small $\phi_\mathsmaller{p}(t_\mathsmaller{p})=\phi_\mathsmaller{p}$} such that,
\label{neutronic} 
{\small 
\begin{eqnarray}
&&\dot \phi_\mathsmaller{e}(t_\mathsmaller{e})=\dot \phi_\mathsmaller{\mathsmaller{p}}(t_\mathsmaller{p})=0, \label{fix1} \\
&&\phi_\mathsmaller{e}+\phi_\mathsmaller{p}=\pm 2\phi_\mathsmaller{\varepsilon}, \label{fix2}
\end{eqnarray}}
for {\small $t_\mathsmaller{e} \in [T_0, T_2]$} and {\small $t_\mathsmaller{p} \in [T_1, T_2]$}, as illustrated by the red segments of Fig. \ref{figODE}.
\end{theorem}
\begin{proof} The proof is by inspection that (\ref{matlab1D}) calculates {\small $\dot \phi_\mathsmaller{e}(t_\mathsmaller{j})=0$} for the whole blue segment of Fig. \ref{figODE}, and likewise for the protonic green segment of Fig. \ref{figODE}. 
\end{proof}

The motion can continue to satisfy (\ref{fix1}) indefinitely. Figure \ref{figODE} illustrates {\small $[T_\mathsmaller{3},T_\mathsmaller{5}]$} and {\small $[T_\mathsmaller{2},T_\mathsmaller{4}]$} having the same length of {\small $[T_\mathsmaller{1}, T_\mathsmaller{3}]$} and {\small $ [T_\mathsmaller{0}, T_\mathsmaller{2}]$} for simplicity, but \textbf{if} the charges keep drifting apart at constant velocities, the iterated time intervals will keep increasing. It is further possible to introduce velocity discontinuities in the initial segment; Fig. \ref{figODE} illustrates the case of one discontinuity at mid-segment, after which mid-times the velocities jump to another pair of phase-locked velocities. Sub-segments {\small $[T_\mathsmaller{1},b_{ps}]$} and {\small $[T_\mathsmaller{0},b_{es}]$} are drawn with thinner red lines while sub-segments {\small $[b_{ps},T_\mathsmaller{3}]$} and 
{\small $[b_{es}, T_\mathsmaller{2}]$} are drawn with thicker red lines in Fig. \ref{figODE}. \textbf{If} the velocities in lightcone satisfy (\ref{fix1}) and (\ref{fix2}) on both sides of the breaking points {\color{red}$b_\mathsmaller{es}$} and {\color{red}$b_\mathsmaller{ps}$} illustrated in Fig. \ref{figODE}, the orbit will continue with piecewise constant velocities for the entire blue and green segments, as seen by inspecting Fig. \ref{figODE}. Such motion of {\small $\varepsilon$}-\textbf{VE} would have vanishing far-fields because it is a motion with vanishing accelerations. The charges interact only by exchanging photonic kicks at breaking points, as discusses in  \S \ref{appendix}-\ref{referees}. It is beyond the present work to investigate magnitudes for such motions to be possible, and if such motions can be bounded, periodic, or somehow stable.

\subsection{Domains of initial histories}
 \label{perturbed_algorithm}

   This section is designed to guide future numerical experiments. The most advanced accelerations, {\small $\mathbf{a}_\mathsmaller{j}$} and {\small $\mathbf{a}_\mathsmaller{\ell}$}, given respectively by (\ref{equation_aj}) and its index-shifted formula, are piecewise continuous functions that can be integrated by the method of steps of Fig. \ref{figODE}. For straight-line orbits, the equations of motion reduce to Eqs. (\ref{matlab1D}) and its index-shifted formula. In order to have a unique extension by the method of steps, the segment needs to belong to a subset of  {\small $\hat{C}^2(\mathbb{R}) \times \hat{C}^2(\mathbb{R})  \subset \mathcal{FSH} $} where the accelerations are Lipschitz continuous. An obvious non-empty initial-segment set to start from when {\small $\varepsilon=0$} is the set of full-swing segments of circular orbits\cite{Hans,Schild}, because circular orbits are globally defined. As noticed in Ref. \cite{Martinez}, circular orbits still exist for {\small $\varepsilon \neq 0$} because time-reversal exchanges {\small $\mathbf{v}_\mathsmaller{\alpha}$} into {\small $\mathsmaller{-}\mathbf{v}_\mathsmaller{\alpha}$} in Eq.  (\ref{BH}), but the extended circular orbits should be unstable like the circular orbits of \cite{Hans}. 
   If {\small $\varepsilon \neq 0$}, the method of steps of Fig. \ref{figODE} acts as a \textbf{semiflow} on the set of full-swing segments of {\small $\varepsilon$}-circular orbits: for every {\small $\tau \in \mathbb{R}^+$}  and for every full-swing element {\small $\phi_{FS}  \in C^2(\mathbb{R}) $} with {\small $t_\mathsmaller{p} \in [T_1,T_3]$} and {\small $t_\mathsmaller{e} \in [T_0,T_2]$}, the method of steps takes {\small $\phi_{FS}$} to (another) full-swing element, i.e., {\small $\psi_\tau : (\tau, \phi) \rightarrow C^2(\mathbb{R}) $}. The function {\small $\psi_\tau$} has the semigroup property that {\small $\psi_0=I$} and {\small $\psi_{\tau_\mathsmaller{1}} \circ \,\psi_{\tau_\mathsmaller{2}} =\psi_{\tau_\mathsmaller{1}+\tau_\mathsmaller{2}} $}. Moreover, restricted to the set of full-swing circular segments we can use the method of steps for {\small $\tau \in \mathbb{R}$}, i.e., {\small $\psi_\tau$} has the \textbf{group} property and it defines a \textbf{flow}. 
 \par
   In the following we attempt to construct larger initial-segment sets where the method of steps operates either as a semiflow or as a flow. Inspecting Eq. (\ref{equation_aj}) we find that {\small $\mathbf{a}_\mathsmaller{j}$} is bounded if {\small $ r_{\mathsmaller{ji}}\,\mathbf{n}_\mathsmaller{j} \cdot \mathbf{E}_{\mathsmaller{ji}}$} is bounded. Analogously, using the index-shifted (\ref{equation_aj}) we find that {\small $\mathbf{a}_\mathsmaller{\ell}$} is bounded whenever {\small $ r_\mathsmaller{\ell j}\,\mathbf{n}_\mathsmaller{\ell} \cdot \mathbf{E}_{\mathsmaller{\ell j}}$} is bounded. Inside the Peano domain (\ref{Peano}), our Eq. (\ref{equation_aj}) is a segment-dependent ODE with a bounded right-hand side. Using Lemma \ref{lemma_1} and Eq. (\ref{equation_aj}) we also find that, whenever the orbit belongs to the Peano domain (\ref{Peano}), one can define a new independent variable {\small $\xi_\mathsmaller{j} \in \mathbb{R}$} for the {\small $\mathbf{a}_\mathsmaller{j}$} equation of motion and a new  independent variable {\small $\xi_\mathsmaller{\ell} \in \mathbb{R}$} for the {\small $\mathbf{a}_\mathsmaller{\ell}$} equation, i.e.,
{\small
\begin{eqnarray}
\xi_\mathsmaller{\ell} \equiv \int \frac{dt_\mathsmaller{\ell}}{ r_\mathsmaller{\ell j}(1-\mathbf{v}_{\mathsmaller{\ell}}^\mathsmaller{2})} \;\; \;  \mbox{and} \; \; \xi_\mathsmaller{j} \equiv \int \frac{dt_\mathsmaller{j}}{ r_\mathsmaller{j i}(1-\mathbf{v}_{\mathsmaller{i}}^\mathsmaller{2})},  \label{Lebesgue}
 \end{eqnarray}}
 \textbf{If} the minimizer orbit belongs to (\ref{Peano}), we can integrate the method of steps until a full-swing. Notice that (\ref{equation_aj}) with {\small $\varepsilon \neq 0 $} is an accomplishment in itself above the Coulomb problem, since the integration of (\ref{equation_aj}) is simpler than the gravitational problem, as follows: {\color{darkgray}(i)} the singularity in the {\small $\varepsilon \neq 0 $} case is proportional to {\small $1/r_\mathsmaller{ji}$} (versus {\small $1/r_{\mathsmaller{ji}}^\mathsmaller{2}$} for the gravitational problem), and {\color{darkgray}(ii)} a change of the integration variable like (\ref{Lebesgue}) would not work for the gravitational problem. Actually, a similar coordinate transformation is just one of the many transformations involved in the Levi-Civita regularization\cite{Aarseth}.  

 \par
 A Lipschitz set of initial segments must avoid a head-on collision\cite{EFY2}. The Lipschitz condition \cite{Cheeger} places a bound on the right-hand side of (\ref{matlab1D}), limiting the {\small $\phi_\mathsmaller{\alpha}$} by {\small $\max\{|\phi_\mathsmaller{\alpha}|\}<B_\mathsmaller{\alpha}(\min\{r_\mathsmaller{ep}\})$}. Collisions act as Lipschitz moderators that keep the delay bounded away from zero by bouncing the trajectories. A way to cure arbitrary full-swing segments is to introduce collisions-at-a-distance as perturbations designed to avoid both the physical collision\cite{Cheeger}  and a luminal velocity by enforcing
 {\small
  \begin{eqnarray}
0< \delta_\mathsmaller{1} < \min \Big( r_\mathsmaller{ji} (1-\mathbf{v}_{\mathsmaller{j}}^\mathsmaller{2}), r_\mathsmaller{ \ell j} (1-\mathbf{v}_{\mathsmaller{\ell}}^\mathsmaller{2}) \Big) . \label{bounds} 
 \end{eqnarray}}
 Because of Lemma \ref{lemma_1}, if the orbit satisfies (\ref{bounds}), the general equation of motion (\ref{equation_aj}) has finite denominators defining the terms for {\small $ r_{\mathsmaller{ji}}\, \mathbf{n}_\mathsmaller{j} \cdot \mathbf{E}_{\mathsmaller{ji}}$} and {\small $ r_\mathsmaller{\ell i}\, \mathbf{n}_\mathsmaller{\ell} \cdot \mathbf{E}_{\mathsmaller{\ell i}}$}. As the integration marches forward, whenever the orbit violates (\ref{bounds}), we can halt the integration and modify the history by introducing a collision-at-a-distance with a convenient velocity discontinuity designed to enforce (\ref{bounds}). Integration must be re-calculated with the last two iterates of the modified history, while the remaining segments should either be discarded or used in a perturbation theory analogous to the one described in Ref. \cite{Brault}. The best type of collision-at-a-distance to use in a perturbation theory is the mutual-recoil collision-at-a-distance, as described in the outer-cone Lemma \ref{outside-cone-Lemma} of \S \ref{appendix}-\ref{outer-cone}.

\section{Forward propagation of velocity discontinuities }
\label{continuationsec}
\subsection{A priori propagation based on the continuity of the partial momenta}
\label{tentative_continuation}
Because the equations of motion (\ref{Singular_equation_aj}) and the Weierstrass-Erdmann conditions (\ref{WE1}) and (\ref{WE2}) are time-reversible, the method of steps can be carried either forward or \textbf{backward}. The forward and the backward iteration of the method of steps are illustrated in Fig. \ref{figODE} and we henceforth restrict to full-swing segment pairs that can be continued in either the backward or the forward direction \textbf{at least once}. For segments belonging to {\small $\hat{C}^2(\mathbb{R})$}, the method of steps proceeds piecewise;  Fig. \ref{figODE} illustrates a generic breaking point defined by a pair of velocity discontinuities in lightcone at times {\small $(t_\mathsmaller{i}=b_{es},t_\mathsmaller{k}=b_{ps})$} respectively on the electronic and on the protonic trajectories. At the breaking point we must halt the forward integration to try and see \textbf{if} the four continuity conditions (\ref{WE1}) and (\ref{WE2}) are satisfied. We start by showing that the continuity of (\ref{WE1}) is sufficient to propagate the velocity of each particle across the breaking point.  Notice that (\ref{WE1}) involves \textbf{three} velocity vectors, {\small $\mathbf{v}_\mathsmaller{i}$}, {\small $\mathbf{v}_\mathsmaller{k}$} and {\small $\mathbf{v}_\mathsmaller{j}$}, which allows one to consider {\small $\mathbf{v}_\mathsmaller{i}$} and {\small $\mathbf{v}_\mathsmaller{k}$} as given by the initial segment and solve (\ref{WE1}) for the most advanced velocity {\small $\mathbf{v}_\mathsmaller{j}$} falling on the iterated segment. The \textit{a priori} velocity {\small $\mathbf{v}_\mathsmaller{j}$} turns out to be subluminal on both sides of the breaking point when {\small $\mathbf{v}_\mathsmaller{i}$} and {\small $\mathbf{v}_\mathsmaller{k}$} are inside proper domains, as explained in the following.
\par
The continuous partial momentum (\ref{WE1}) splits in a term containing the most advanced velocity {\small $\mathbf{v}_\mathsmaller{j}$} plus a term containing only {\small $\mathbf{v}_\mathsmaller{i}$} and {\small $\mathbf{v}_\mathsmaller{k}$}, i.e., 
{\small
\begin{eqnarray}
\mathbf{P}_\mathsmaller{i} \equiv m_\mathsmaller{i}\gamma_\mathsmaller{i}\mathbf{v}_\mathsmaller{i} - \mbox{\Large $a$}_\mathsmaller{ki}(\mathbf{v}_\mathsmaller{k},\mathbf{v}_\mathsmaller{i}) -\mbox{\Large $a$}_\mathsmaller{ji}(\mathbf{v}_\mathsmaller{j},\mathbf{v}_\mathsmaller{i}),
\label{Psplit}
\end{eqnarray}}
where
{\small
\begin{eqnarray}
\mbox{\Large $a$}_\mathsmaller{ji}(\mathbf{v}_\mathsmaller{j})& \equiv&
        \frac{-1}{2r_\mathsmaller{ji}(1+ \mathbf{n}_{\! \mathsmaller{\, j}}\cdot \mathbf{v}_{\! \mathsmaller{\, j}})} (e_\mathsmaller{i}e_\mathsmaller{j} \mathbf{v}_{\! \mathsmaller{\, j}} \,+  \varepsilon \gamma_\mathsmaller{i} \sqrt{1-\mathbf{v}_{\mathsmaller{j}}^{\mathsmaller{2}}}\; \, \mathbf{v}_\mathsmaller{i} ) \,, \label{def_fat_Ai} \\ \mbox{\Large $a$}_\mathsmaller{ki} (\mathbf{v}_\mathsmaller{k}) & \equiv&   \frac{-1}{2r_\mathsmaller{ki}(1- \mathbf{n}_{\! \mathsmaller{\, k}}\cdot \mathbf{v}_{\! \mathsmaller{\, k}})}( e_\mathsmaller{i} e_\mathsmaller{k} \mathbf{v}_{\! \mathsmaller{\, k}} \,+ \varepsilon \gamma_\mathsmaller{i} \sqrt{1-\mathbf{v}_{\mathsmaller{k}}^{\mathsmaller{2}}} \; \, \mathbf{v}_\mathsmaller{i} )\,. \label{def_aki}
\end{eqnarray}}
The value of {\small $\mathbf{P}_\mathsmaller{i}$} in (\ref{Psplit}) is a product of the integration explained in \S \ref{ODEsec} using the initial segment until the breaking points {\color{red}$b_\mathsmaller{es}$} and {\color{red}$b_\mathsmaller{ps}$} illustrated in Fig. \ref{figODE}, and it is thus a function of the segment. At the breaking point {\color{red}$b_\mathsmaller{ps}$} for {\small $\mathbf{v}_\mathsmaller{i}$}, the discontinuities of {\small $\mathbf{v}_\mathsmaller{i}$} and {\small $\mathbf{v}_{\mathsmaller{k}}$} require that the velocity {\small $\mathbf{v}_\mathsmaller{j}$} be discontinuous according to
{\small
\begin{eqnarray}
\mbox{\Large $a$}_\mathsmaller{ji} \equiv \Big( m_\mathsmaller{i}\gamma_\mathsmaller{i}\mathbf{v}_\mathsmaller{i} -\mathbf{P}_\mathsmaller{i} \Big)-\mbox{\Large $a$}_\mathsmaller{ki} \label{breakAij}.
\end{eqnarray} }
 Equation (\ref{breakAij}) defines {\small $\mbox{\Large $a$}_\mathsmaller{ji}$} as a function of  {\small $\mbox{\Large $a$}_\mathsmaller{ki}$} on both sides of {\color{red}$b_\mathsmaller{ps}$}.  In the following we invert definitions (\ref{def_fat_Ai}) and (\ref{def_aki}) in order to express {\small $\mathbf{v}_\mathsmaller{j}$} as a function of  {\small $\mbox{\Large $a$}_\mathsmaller{\alpha i}$}, {\small $\varepsilon$}, the present velocity, and either the future or the past velocity, e.g., {\small $\mathbf{v}_\mathsmaller{j}=\vec{\rho}\,(\mbox{\Large $a$}_\mathsmaller{ji}, \varepsilon, \mathbf{n}_\mathsmaller{j},\mathbf{v}_\mathsmaller{i}, \mathbf{v}_\mathsmaller{k}) $}. %For economy of notation we henceforth drop the dependence on $ \varepsilon, \mathbf{n}_\mathsmaller{j},\mathbf{v}_\mathsmaller{i},\mathbf{v}_\mathsmaller{k},\mathbf{P}_\mathsmaller{i}$.a
The generalized expression for {\small $\mathbf{v}_{\mathsmaller{\alpha}}$} is used in many places of the manuscript and it is also used to express {\small $\mathbf{v}_\mathsmaller{s}$} as a function of {\small $\mathbf{v}_\mathsmaller{i}$} and {\small $\mathbf{v}_{\mathsmaller{k}}$} using backward integration, i.e.,
  {\small
\begin{eqnarray}
\mathbf{v}_\mathsmaller{\alpha} \equiv \vec{\rho} \,( \mbox{\Large $a$}_\mathsmaller{\alpha i} ) + \varepsilon_{*} \gamma_\mathsmaller{i} \sqrt{1-\mathbf{v}_{\mathsmaller{\alpha}}^\mathsmaller{2} }\; \vec{q}\,(\mbox{\Large $a$}_\mathsmaller{\alpha i},\mathbf{n}_\mathsmaller{j},\mathbf{v}_\mathsmaller{i}), \label{def_vr_epsilonstar}
\end{eqnarray}}
where {\small $\varepsilon_\mathsmaller{*}$} is defined by (\ref{defeps_star}), the functions {\small $\vec{\rho} \,(\mbox{\Large $a$}_\mathsmaller{\alpha i},\mathbf{n}_\mathsmaller{\alpha}, r_\mathsmaller{\alpha i}) \colon \mathbb{R}^3 \times  \mathbb{R}^3 \times \mathbb{R} \rightarrow  \mathbb{R}^3$} and {\small $\vec{q}\,(\mbox{\Large $a$}_\mathsmaller{\alpha i},\mathbf{n}_\mathsmaller{\alpha },\mathbf{v}_\mathsmaller{i}) \colon \mathbb{R}^3 \times \mathbb{R}^3 \times \mathbb{R}^3 \rightarrow \mathbb{R}^3$} are defined by 
{\small
\begin{eqnarray}
 \vec{\rho} \,(\mbox{\Large $a$}_\mathsmaller{\alpha i},\mathbf{n}_\mathsmaller{\alpha}, r_\mathsmaller{\alpha i})  &\equiv&
        \frac{2r_\mathsmaller{\alpha i} \mbox{\Large $a$}_\mathsmaller{\alpha i}}{(-e_\mathsmaller{i}e_\mathsmaller{\alpha}\mp 2r_\mathsmaller{\alpha i} \mathbf{n}_{\! \mathsmaller{\, j}}\cdot \mbox{\Large $a$}_\mathsmaller{\alpha i}    )} =  \frac{ \mbox{\Large $a$}_\mathsmaller{\alpha i}}{(\frac{-e_\mathsmaller{i} e_\mathsmaller{\alpha}}{2r_\mathsmaller{\alpha i} } \mp  \mathbf{n}_{\! \mathsmaller{\, \alpha}} \cdot \mbox{\Large $a$}_\mathsmaller{\alpha i}    )}, \label{def_vr_zeroeps}\\ \vec{q} \,(\mbox{\Large $a$}_\mathsmaller{\alpha i}, \mathbf{n}_\mathsmaller{\alpha},\mathbf{v}_\mathsmaller{i}) & \equiv & \mathbf{v}_\mathsmaller{i} \pm(\mathbf{n}_{\! \mathsmaller{\, \alpha}}\cdot \mathbf{v}_{\! \mathsmaller{\, i}} )\, \vec{\rho}\,(\mbox{\Large $a$}_\mathsmaller{\alpha i}) , \label{defq}\end{eqnarray}}
and $\gamma_\mathsmaller{i}$ is defined by (\ref{defgamai}). Observations: {\color{darkgray}(i)} for economy of notation, we henceforth abbreviate the list of arguments of (\ref{def_vr_zeroeps}) and (\ref{defq}), keeping only the first argument, i.e, {\small $\vec{\rho} \, (\mbox{\Large $a$}_\mathsmaller{\alpha i} )$} and {\small $\vec{q} \, (\mbox{\Large $a$}_\mathsmaller{\alpha i} )$}, {\color{darkgray}(ii)} the plus sign in every definition below (\ref{def_vr_zeroeps}) holds when {\small $\alpha=j$}. On the contrary, when (\ref{def_vr_epsilonstar}), (\ref{def_vr_zeroeps}) and (\ref{defq}) are used to express {\small $\mathbf{v}_\mathsmaller{s}$} with the backward shift of indices, {\small $(kij) \rightarrow (ski)$}, the minus sign holds because index {\small $s$} is in the past of {\small $\alpha=k$} and {\color{darkgray}(iii)} Eq.  (\ref{def_vr_epsilonstar}) is invariant under a time-reversal operation {\small $(kij) \rightarrow (iks)$} if every {\small $\pm$} sign is exchanged. 

%Solving Eq. (\ref{def_fat_Ai}) for $\mathbf{v}_\mathsmaller{j}$ when $\varepsilon=0$ and $e_\mathsmaller{i}e_\mathsmaller{j}=-1$ yields 

\begin{lem} Equation (\ref{def_vr_epsilonstar}) with {\small $\varepsilon_{*}=0$} defines a \textbf{unique} electronic velocity {\small $\mathbf{v}_\mathsmaller{\alpha}$} for each {\small $\mbox{\Large $a$}_\mathsmaller{ji} \in \mathbb{R}^3$} inside a paraboloid of revolution domain in {\small $\mathbb{R}^3$}. 
\label{paraboloid_lemma}
\end{lem}

\begin{proof} Squaring (\ref{def_vr_epsilonstar}) for a subluminal orbit  when {\small $\varepsilon_{*}=0$} yields
{\small
\begin{eqnarray}
1-\mathbf{v}_{\mathsmaller{\alpha}}^\mathsmaller{2}=1-||\vec{\rho} \, (\mbox{\Large $a$}_{\mathsmaller{\alpha i}})||^{\mathsmaller{2}} \geq 0. \label{preparaboloid}
\end{eqnarray} }
We choose an orthogonal Cartesian system with the {\small $\hat{\bf{y}}$} axis along the {\small $\mathbf{n}_{\! \mathsmaller{\, \alpha}}$} direction to evaluate the inequality on the right-hand side (\ref{preparaboloid}), finding that {\small $ \mbox{\Large $a$}_\mathsmaller{y} \equiv \pm \mathbf{n}_\mathsmaller{\alpha}\cdot \mbox{\Large $a$}_\mathsmaller{\alpha i}$} is defined from the subluminal condition {\small $||\rho_ {\mathsmaller{\mbox{\tiny $a$}}}(\mbox{\Large $a$}_{\mathsmaller{\alpha i}})||^{\mathsmaller{2}}<1$} as
  {\small
  \begin{eqnarray}
  \mbox{\Large $a$}_\mathsmaller{y} \leq \frac{1}{4r_\mathsmaller{ij}} -r_\mathsmaller{\alpha i}( \mbox{\Large $a$}^{\mathsmaller{2}}_\mathsmaller{x}+ \mbox{\Large $a$}^{\mathsmaller{2}}_\mathsmaller{z}) \,. \label{parabola_revolut}
    \end{eqnarray}
  } 
Condition (\ref{parabola_revolut}) defines the interior of a paraboloid of revolution for $\mbox{\Large $a$}_\mathsmaller{\alpha i} \in \mathbb{R}^3$.  
\end{proof}

Otherwise, when {\small $\varepsilon_{*} \in \mathbb{R}$}, the squared modulus of (\ref{def_vr_epsilonstar}) yields a quadratic equation for {\small $\sqrt{1-\mathbf{v}_{\mathsmaller{\alpha}}^\mathsmaller{2} }$}\,, namely
{\small
\begin{eqnarray}
(1-\mathbf{v}_{\mathsmaller{\alpha}}^\mathsmaller{2})+2\mathsf{B}_\mathsmaller{ i} \sqrt{1-\mathbf{v}_{\mathsmaller{\alpha}}^\mathsmaller{2} }-\mathsf{C}_\mathsmaller{ i}=0\,, \label{quadratic}
\end{eqnarray}}
 where 
 {\small
 \begin{eqnarray}
 \mathsf{B}_\mathsmaller{i}(\mbox{\Large $a$}_\mathsmaller{ji})& \equiv& \frac{\varepsilon_{*} \gamma_\mathsmaller{i}\, \vec{q}\,(\mbox{\Large $a$}_\mathsmaller{\alpha i})  \cdot \vec{\rho} \, (\mbox{\Large $a$}_\mathsmaller{\alpha i})}{(1+\varepsilon_{*}^\mathsmaller{2} \gamma_{\mathsmaller{i}}^\mathsmaller{2} ||\,\vec{q}\, (\mbox{\Large $a$}_\mathsmaller{\alpha i}) \,||^\mathsmaller{2})}\,, \label{defB} \\
 \mathsf{C}_\mathsmaller{i} (\mbox{\Large $a$}_\mathsmaller{ji})&\equiv& \frac{1-||\vec{\rho}\, (\mbox{\Large $a$}_\mathsmaller{\alpha i})||^\mathsmaller{2}}{(1+\varepsilon_{*}^\mathsmaller{2} \gamma_{\mathsmaller{i}}^\mathsmaller{2}||\, \vec{q}\,(\mbox{\Large $a$}_\mathsmaller{\alpha i}) \,||^\mathsmaller{2})} \,. \label{defC} 
 \end{eqnarray}}
 Observations: {\color{darkgray} (i)} it can be found by inspection that the root with the minus sign, {\small $(1-\mathbf{v}_{\mathsmaller{j}}^\mathsmaller{2})\equiv -\mathsf{B}_\mathsmaller{i} - \sqrt{ \mathsf{B}_{\mathsmaller{i}}^{\mathsmaller{2}}  +\mathsf{C}_\mathsmaller{i} }$},  is never inside {\small $[0,1)$} and {\color{darkgray}(ii)} whenever {\small $\mathsf{B}_\mathsmaller{i}>0 $} and {\small $0< \mathsf{C}_\mathsmaller{i}<1$}, the plus root, {\small $(1-\mathbf{v}_{\mathsmaller{j}}^\mathsmaller{2}) \equiv -\mathsf{B}_\mathsmaller{i} + \sqrt{\mathsf{B}_{\mathsmaller{i}}^{\mathsmaller{2}}+\mathsf{C}_\mathsmaller{i} }$}, is inside {\small $ [0,1)$}. For {\small $\varepsilon_{*}=0$} the former reduces to the interior of the paraboloid of Lemma \ref{paraboloid_lemma}. The domain where the plus root {\small $(1-\mathbf{v}_{\mathsmaller{j}}^\mathsmaller{2})=-\mathsf{B}_\mathsmaller{i} + \sqrt{\mathsf{B}^{\mathsmaller{2}}_{\mathsmaller{i}}+\mathsf{C}_\mathsmaller{i} }$}\, belongs to {\small $[0,1)$} extends beyond the paraboloid of Lemma \ref{paraboloid_lemma} when {\small $\varepsilon_{*} \neq 0$} because we can have a positive discriminant {\small $\Delta_\mathsmaller{i} \equiv \mathsf{B}_{\mathsmaller{i}}^{2} + \mathsf{C}_\mathsmaller{i}> 0$} while {\small $B_\mathsmaller{i}<0$}. 
 
 \begin{lem}
 Equation (\ref{def_vr_epsilonstar}) with {\small $\varepsilon_{*} \neq 0 $} defines a \textbf{unique} electronic velocity {\small $\mathbf{v}_\mathsmaller{\alpha}$} for each {\small $\mbox{\Large $a$}_\mathsmaller{\alpha i} \in \mathbb{R}^3$} belonging to the interior of an asymmetric paraboloid domain.
\label{asy_paraboloid_lemma}
\end{lem} 
\begin{proof} As mentioned below Eq. (\ref{defC}), the unique root is Bhaskara's formula with the \textbf{plus} sign. The condition to be in $[0,1)$ is $0< -\mathsf{B}_\mathsmaller{i} + \sqrt{\mathsf{B}_{\mathsmaller{i}}^{\mathsmaller{2}}+\mathsf{C}_\mathsmaller{i} }<1 $. The upper bound condition, $-\mathsf{B}_\mathsmaller{i} + \sqrt{\mathsf{B}_{\mathsmaller{i}}^{\mathsmaller{2}}+\mathsf{C}_\mathsmaller{i} }<1 $, yields the trivial identity $0<|| \vec{\rho}+\varepsilon_{*} \gamma_\mathsmaller{i} \vec{q} \, ||^\mathsmaller{2}$ by use of (\ref{defB}) and (\ref{defC}), while the discriminant condition $\Delta_\mathsmaller{i} \equiv B_{\mathsmaller{i}}^{2}+C_\mathsmaller{i} \geq 0$ yields
{\small
\begin{eqnarray}
\rho^\mathsmaller{2}\leq 1+\varepsilon_{*}^\mathsmaller{2} \gamma_{\mathsmaller{i}}^\mathsmaller{2} q^\mathsmaller{2} +\varepsilon_{*}^\mathsmaller{2} \gamma_{\mathsmaller{i}}^\mathsmaller{2} \Big( ( \vec{q} \cdot \vec{\rho} \, )^\mathsmaller{2} - \rho^\mathsmaller{2} q^\mathsmaller{2} \Big) , \label{extended_bound}
\end{eqnarray}}
thus extending condition (\ref{preparaboloid}) of Lemma \ref{paraboloid_lemma}. Using (\ref{defq}) to re-arrange (\ref{extended_bound}) yields
{\small
\begin{eqnarray}
\rho^{\mathsmaller{2}} \leq 1+\frac{ \varepsilon_{*}^\mathsmaller{2} \gamma_{\mathsmaller{i}}^\mathsmaller{2} ( \mathbf{n}_\mathsmaller{\alpha}\cdot \mathbf{v}_\mathsmaller{i} \pm \vec{\rho} \cdot \mathbf{v}_\mathsmaller{i} )^\mathsmaller{2}}{1+\varepsilon_{*}^\mathsmaller{2}  \gamma_{\mathsmaller{i}}^\mathsmaller{2} ||\ell_\mathsmaller{\alpha i} ||^\mathsmaller{2}}, \label{extended2}
\end{eqnarray}}
where {\small $\ell_\mathsmaller{\alpha i}$} is defined by Eq. (\ref{defell_ab}). Without loss of generality we take the velocity vector {\small $\mathbf{v}_\mathsmaller{i}$} in a plane {\small $xy$} with the {\small $\hat{\bf{y}}$} axis defined along the {\small $\mathbf{n}_{\! \mathsmaller{\, j}}$} direction, i.e., 
{\small
\begin{eqnarray}
\mathbf{v}_\mathsmaller{i} \equiv ||\mathbf{v}_\mathsmaller{i}|| (\cos{(\vartheta_\mathsmaller{i})} \,\hat{\mathbf{x}} +\sin{(\vartheta_\mathsmaller{i})} \,\hat{\mathbf{y}} ) .\label{v_angle}
\end{eqnarray}}
Substituting (\ref{def_vr_zeroeps}) and (\ref{v_angle}) into (\ref{extended2}) yields
 {\small
  \begin{eqnarray}
  \mbox{\Large $a$}_{\mathsmaller{y}} \leq \Big( \frac{ 1+\varepsilon_{*}^{\!\mathsmaller{2}}\gamma_{\mathsmaller{i}}^{\mathsmaller{2}} \mathbf{v}_{\mathsmaller{i}}^{\mathsmaller{2}}   }{4r_\mathsmaller{\alpha i} } \Big) -r_\mathsmaller{\alpha i}\mbox{\Large $a$}^{\mathsmaller{2}}_\mathsmaller{z} -r_\mathsmaller{\alpha i} \frac{\Big(\mbox{\Large $a$}_\mathsmaller{x} -\frac{\varepsilon_{*}^{\mathsmaller{2}}\gamma_{\mathsmaller{i}}^{\mathsmaller{2}} \mathbf{v}_{\mathsmaller{i}}^{\mathsmaller{2}}\sin{\vartheta_\mathsmaller{i}}}{4r_\mathsmaller{\alpha i}}  \Big)^\mathsmaller{2}}{(1+\varepsilon_{*}^{\mathsmaller{2}}\gamma_{\mathsmaller{i}}^{\mathsmaller{2}} \mathbf{v}_{\mathsmaller{i}}^{\mathsmaller{2}}\cos^{\mathsmaller{2}}{\vartheta_\mathsmaller{i}} )} \,, \label{parabola_asymm}
    \end{eqnarray}
  } 
  which is the interior of an asymmetric paraboloid domain for {\small $\mbox{\Large $a$}_\mathsmaller{\alpha i} \in \mathbb{R}^3$}. Notice that (\ref{parabola_asymm}) reduces to (\ref{parabola_revolut}) when {\small $\varepsilon_{*}=0$}, which is Lemma \ref{paraboloid_lemma}.
 \end{proof}
 The protonic velocity at point {\small $\ell$} of the sewing-chain {\small $(o,s,k,i,j,\ell)$} of Fig. \ref{sew_White} is obtained  from (\ref{parabola_asymm}) by shifting the indices by one position forward. Condition (\ref{parabola_asymm}) and the analogous condition for the protonic velocity yield the a priori (tentative) forward propagation of the sewing chain, before ever testing \textbf{if} the Weierstrass-Erdmann partial energy (\ref{WE2}) is continuous for either particle. Equation (\ref{def_vr_epsilonstar}) of Lemma \ref{asy_paraboloid_lemma} defines the most advanced velocity of either particle for {\small $\alpha \in (j,\ell)$}. 
Assuming that the discontinuities of the initial segment are consistent with Lemma \ref{asy_paraboloid_lemma}, Eq. (\ref{def_vr_epsilonstar}) determines {\small $\mathbf{v}_\mathsmaller{j}$ and $\mathbf{v}_\mathsmaller{\ell}$} uniquely on the right-hand sides of times {\small $t_\mathsmaller{\ell}$ and $t_\mathsmaller{j}$}. Otherwise the initial segment predicts \textbf{inconsistent} superluminal velocities and it can not be continued forward. Having found consistent a priori {\small $\mathbf{v}_\mathsmaller{j}$ and $\mathbf{v}_\mathsmaller{\ell}$} on both sides of the breaking point, we still have to satisfy the energetic conditions (\ref{WE2}) in order to have a minimizer with a corner for the action functional (\ref{Fokker}).

 \subsection{General energetic corner conditions and boundary layer}
 \label{boundary_layer}
 
 The boundary-value formulation of \S \ref{Basic}-\ref{perturbfuncsec} is explained in the caption of (\ref{sew_White}). There are some remaining traces of the boundary-value problem that must be calculated before use in the continuity conditions; the distances in lightcone {\small $r_\mathsmaller{ij}$} and $r_\mathsmaller{jl}$, and the normals {\small $\mathbf{n}_\mathsmaller{j}$} and {\small $\mathbf{n}_\mathsmaller{\ell}$} from times {\small $t_\mathsmaller{j}$} and {\small $t_\mathsmaller{s}$} illustrated in Fig. \ref{figODE}. As explained in \S \ref{continuationsec}-\ref{tentative_continuation}, Lemma \ref{asy_paraboloid_lemma} uses (\ref{WE1}) to propagate the sewing chain. The provisional sewing chain still has to satisfy the Weierstrass-Erdmann energetic conditions (\ref{WE2}) at the breaking points {\color{blue}$b_\mathsmaller{e1}$} and {\color{Asparagus}$b_\mathsmaller{p1}$} illustrated in Fig. \ref{figODE}. For the electromagnetic case, the energetic conditions (\ref{WE2}) calculated with the a priori propagated velocity become particularly simple conditions involving the velocities at points {\color{red}$b_\mathsmaller{es}$} and {\color{red}$b_\mathsmaller{ps}$}, as obtained next. We start by separating (\ref{scalarUi}) into a continuous part plus a jumping part, i.e.,
 {\small
 \begin{eqnarray}
  e_\mathsmaller{i} \mathcal{U}_\mathsmaller{\, i} & = & \Big( \frac{e_\mathsmaller{i} e_{\! \mathsmaller{\, k}} }{2r_{\! \mathsmaller{\, ki}}} +\frac{e_\mathsmaller{i} e_{\! \mathsmaller{\, j}}}{2r_{\! \mathsmaller{\, ji}}} \Big) + \Big( \frac{e_\mathsmaller{i} e_{\! \mathsmaller{\, k}} \,\mathbf{n}_{\! \mathsmaller{\, k}}\cdot \mathbf{v}_{\! \mathsmaller{\, k}}}{2r_{\! \mathsmaller{\, ki}}(1-\mathbf{n}_{\! \mathsmaller{\, k}}\cdot \mathbf{v}_{\! \mathsmaller{\, k}})} -\frac{e_\mathsmaller{i} e_{\! \mathsmaller{\, j}} \,\mathbf{n}_{\! \mathsmaller{\, j}}\cdot \mathbf{v}_{\! \mathsmaller{\, j}}}{2r_{\! \mathsmaller{\, ji}}(1+\mathbf{n}_{\! \mathsmaller{\, j}}\cdot \mathbf{v}_{\! \mathsmaller{\, j}})} \Big), \label{identityU}
 \end{eqnarray} 
}
where the first term on the right-hand side of (\ref{identityU}) is continuous at the breaking point because the lightcone distances add to the continuous function 
{\small
\begin{eqnarray}
\bar{u}_\mathsmaller{i} \equiv  \Big( \frac{e_\mathsmaller{i} e_{\! \mathsmaller{\, k}} }{2r_{\! \mathsmaller{\, ki}}} +\frac{e_\mathsmaller{i} e_{\! \mathsmaller{\, j}}}{2r_{\! \mathsmaller{\, ji}}} \Big). \label{continuousAVG}
 \end{eqnarray}
}
The scalar product $\mathbf{n}_\mathsmaller{j} \cdot \mathbf{P}_{\! \mathsmaller{\, i}}$ calculated by (\ref{WE1}) yields 
{\small
\begin{eqnarray}
\mathbf{n}_\mathsmaller{j} \cdot \mathbf{P}_{\! \mathsmaller{\, i}}  = \Big( \frac{ m_{\! \mathsmaller{\, i}}+\varepsilon \mathsf{G}_\mathsmaller{i}  }{ \sqrt{1 - \mathbf{v}_{\! \mathsmaller{\, i}}^{\! \mathsmaller{\, 2}}}  } \Big) \mathbf{n}_\mathsmaller{j} \cdot \mathbf{{v}}_{\! \mathsmaller{\, i}}   +\Big( \frac{e_\mathsmaller{i} e_{\! \mathsmaller{\, k}} \,\mathbf{n}_{\! \mathsmaller{\, j}}\cdot \mathbf{v}_{\! \mathsmaller{\, k}}}{2r_{\! \mathsmaller{\, ki}}(1-\mathbf{n}_{\! \mathsmaller{\, k}}\cdot \mathbf{v}_{\! \mathsmaller{\, j}})} +\frac{e_\mathsmaller{i} e_{\! \mathsmaller{\, j}} \,\mathbf{n}_{\! \mathsmaller{\, j}}\cdot \mathbf{v}_{\! \mathsmaller{\, j}}}{2r_{\! \mathsmaller{\, ji}}(1+\mathbf{n}_{\! \mathsmaller{\, j}}\cdot \mathbf{v}_{\! \mathsmaller{\, j}})} \Big).  \label{identityP}
\end{eqnarray} }
Using (\ref{identityU}) to calculate the partial energy (\ref{WE2}), adding the result to (\ref{identityP}), and subtracting the continuous function (\ref{continuousAVG}) yields a combination that is independent of the most advanced velocity {\small $\mathbf{v}_\mathsmaller{j}$} \textit{and} continuous at the breaking point, i.e.,
{\small
\begin{eqnarray}
\mathbb{C}_\mathsmaller{i}\equiv \mathscr{E}_{\! \mathsmaller{\, i}}+\mathbf{n}_\mathsmaller{j} \cdot \mathbf{P}_{\! \mathsmaller{\, i}} -\bar{u}_\mathsmaller{i}&=&\Big( \frac{1+\mathbf{n}_\mathsmaller{j} \cdot \mathbf{v}_\mathsmaller{i} }{  \sqrt{1-\mathbf{v}^{\mathsmaller{2}}_\mathsmaller{i} }     }     \Big)m_\mathsmaller{i} +\frac{e_\mathsmaller{i} e_\mathsmaller{k}(\mathbf{n}_\mathsmaller{j}+\mathbf{n}_\mathsmaller{k})\cdot \mathbf{v}_\mathsmaller{k}}{2r_\mathsmaller{ki}(1-\mathbf{n}_\mathsmaller{k} \cdot \mathbf{v}_\mathsmaller{k}) } \notag \\&&+\frac{\varepsilon \gamma_{\mathsmaller{i}}(1+\mathbf{n}_\mathsmaller{j} \cdot \mathbf{v}_\mathsmaller{i}) \theta_{\mathsmaller{ki}} }{2r_\mathsmaller{ki}} +\frac{\varepsilon \gamma_{\mathsmaller{i}}(1+\mathbf{n}_\mathsmaller{j} \cdot \mathbf{v}_\mathsmaller{i}) \theta_{\mathsmaller{ji}} }{2r_\mathsmaller{ji}}\,,  \label{CWE2i}
\end{eqnarray}
}
 where the function {\small $\theta_{\mathsmaller{\alpha \beta}} : \mathbf{v}_{\mathsmaller{\alpha}} \in \mathbb{R}^3 \rightarrow \mathbb{R}$} is defined from the lightcone falling either in the backward iterated segment or in the forward iterated segment,
{\small
\begin{eqnarray}
\theta_{\mathsmaller{\alpha \beta}} \equiv \frac{\sqrt{1-\mathbf{v}^{\mathsmaller{2}}_{\mathsmaller{\alpha}} }}{(1\pm \mathbf{n}_{\mathbf{\alpha}}\cdot \mathbf{v}_{\mathsmaller{\alpha}} )}=\frac{-B_\mathsmaller{\beta}+\sqrt{C_\mathsmaller{\beta}+B_{\mathsmaller{\beta}}^{2}}}{(1 \pm \mathbf{n}_\mathsmaller{\alpha}\cdot \mathbf{v}_\mathsmaller{\alpha})} \geq 0,
\label{defteta}
\end{eqnarray}}
where {\small $B_\mathsmaller{\beta}$} and {\small $C_\mathsmaller{\beta}$} are given respectively by  (\ref{defB}) and (\ref{defC}), for either {\small $ \beta=i$} and {\small $\alpha \in (k,j) $}, or {\small $\beta=j$} and  {\small $\alpha=(i,\ell)$}, or {\small $\beta=k$} and {\small $\alpha=(s,i)$}. Equation (\ref{defteta}) involves the normal {\small $\mathbf{n}_\mathsmaller{\alpha}$} coming from either the backward iterated segment or from the forward iterated segment, and {\small $\theta_{\mathsmaller{\alpha \beta}}$} is a function of the segment. The corresponding energetic condition using (\ref{WE1}) and (\ref{WE2}) for {\small $\mathbf{P}_\mathsmaller{k}$} and {\small $\mathscr{E}_{\! \mathsmaller{\, k}}$} is
{\small
\begin{eqnarray}
\mathbb{C}_\mathsmaller{k}\equiv  \mathscr{E}_{\! \mathsmaller{\, k}}-\mathbf{n}_\mathsmaller{s} \cdot \mathbf{P}_{\! \mathsmaller{\, k}} -\bar{u}_\mathsmaller{k}&=&\Big( \frac{1-\mathbf{n}_\mathsmaller{s} \cdot \mathbf{v}_\mathsmaller{k} }{  \sqrt{1-\mathbf{v}^{\mathsmaller{2}}_\mathsmaller{k} }     }     \Big)m_{\! \mathsmaller{\, k}} -\frac{e_\mathsmaller{k} e_\mathsmaller{i}(\mathbf{n}_\mathsmaller{s}+\mathbf{n}_\mathsmaller{i})\cdot \mathbf{v}_\mathsmaller{i}}{2r_\mathsmaller{ki}(1+\mathbf{n}_\mathsmaller{i} \cdot \mathbf{v}_\mathsmaller{i}) } \notag \\&&+\frac{\varepsilon \gamma_{\mathsmaller{k}}(1-\mathbf{n}_\mathsmaller{s} \cdot \mathbf{v}_\mathsmaller{k}) \theta_{\mathsmaller{ik}} }{2r_\mathsmaller{ki }} +\frac{\varepsilon \gamma_{\mathsmaller{k}}(1-\mathbf{n}_\mathsmaller{s} \cdot \mathbf{v}_\mathsmaller{k}) \theta_{\mathsmaller{sk}} }{2r_\mathsmaller{ks}}\,.  \label{CWE2k}
\end{eqnarray}
}

Observations:{\color{darkgray}(i)} Eq. (\ref{CWE2i}) involves the lightcone distance from {\color{red}$b_\mathsmaller{ps}$} to {\color{red}$b_\mathsmaller{e1}$}, which is calculated by integrating over the red segment of Fig. \ref{figODE} until the breaking point. In the same way, (\ref{CWE2k}) involves the lightcone distance from breaking point {\color{red}$b_\mathsmaller{e1}$} to breaking point {\color{red}$b_\mathsmaller{p1}$}, another integration over the red segment, {\color{darkgray}(ii)} the right-hand sides of (\ref{CWE2i}) and (\ref{CWE2k}) are continuous functions of the velocities defining large-velocity segments on each side of {\color{red}$b_\mathsmaller{es}$} and {\color{red}$b_\mathsmaller{ps}$}, henceforth the \textbf{boundary layer} conditions and the last consistency tests be satisfied by the a priori velocity discontinuities.

 \section{Elastic collisions of the one-dimensional problem}
 \label{elastic_boundarylayer}

This section is designed to prove that collisions-at-a-distance have solutions in a few simple cases. The transformations of \S \ref{appendix}-\ref{LG} define a unique breaking-point-frame where the velocity of each particle reflects upon collision. However, one can not simultaneously control the other particle's velocity discontinuity. For that reason, there are several inequivalent classes and we shall not exhaust the long list of collisions-at-a-distance. Examples of broken extrema with corners are found in \cite{Gelfand} pgs.61-63, and in Ref. \cite{Daniel}. Here we focus on collisions-at-a-distance where both velocities flip sign, henceforth elastic collisions.  For the avoidance of a head-on collision at the speed of light, as discussed in \S \ref{ODEsec}-\ref{perturbed_algorithm}, an important outcome of the collision is \textbf{if} the velocity of relative approximation changes sign or not. We classify two types of collisions; {\color{darkgray}(i)} \textbf{mutual recoil}, when both initially opposite velocities change sign and thus the velocity of relative approximation changes sign, and {\color{darkgray}(ii)} \textbf{sticky collision}, when both initial velocities were pointed to the same direction and \textit{after} the collision \textbf{both} velocities flip to the other direction, in which case the relative approximation might either change sign or not. Because the equations of motion are time-reversible, and in order to make the continuity conditions more symmetric, we henceforth restrict to initial segments that can be iterated forward \textbf{and} backward at least once, which is the case if we are dealing with a second iterate of the method of steps as illustrated in Fig. \ref{figODE}. In this section we shall use the alternative \textit{absolute} velocity angles defining the velocity along {\small $\hat{\mathbf{x}}$} instead of the velocity (\ref{def_hyperphi}) along {small $\hat{\mathbf{n}}_\mathsmaller{\alpha}$}, i.e.,
{\small
\begin{eqnarray}
  \mathbf{v}_\mathsmaller{\alpha} \equiv v_\mathsmaller{\alpha} \hat{\mathbf{x}} \equiv \tanh{\varphi_\mathsmaller{\alpha}} \, \hat{\mathbf{x}}. \label{def_hyperphi2}
\end{eqnarray}} 
Assuming the instantaneous situation is in the setup defined below Eq. (\ref{gamaphi}) we have {\small $\mathbf{n}_\mathsmaller{e}=\hat{\mathbf{x}}$} and {\small $\mathbf{n}_\mathsmaller{p}=-\hat{\mathbf{x}}$}, and the absolute and relative angles (\ref{def_hyperphi2}) and (\ref{def_hyperphi}) are related by {\small $(\varphi_\mathsmaller{e},\varphi_\mathsmaller{p})=(\phi_\mathsmaller{e},-\phi_\mathsmaller{p})$}. The velocity of relative approximation in lightcone is
{\small
\begin{eqnarray}
\frac{dr_\mathsmaller{ki}}{dt_\mathsmaller{i}}=\frac{d(t_\mathsmaller{i}-t_\mathsmaller{k})}{dt_\mathsmaller{i}}=1-\frac{dt_\mathsmaller{k}}{dt_\mathsmaller{i}}=\frac{v_\mathsmaller{i}-v_\mathsmaller{k}}{1-v_\mathsmaller{k}},
\end{eqnarray}
}
where we have used (\ref{time_rate}). The ellipsoids of Lemmas \ref{paraboloid_lemma} and \ref{asy_paraboloid_lemma} degenerate into a half-line in the one-dimensional case. The case $\varepsilon=0$ is a boundary-value problem, but still an important limit case for the algebraic equations that must have positive roots. 
Unlike the theory of \S \ref{continuationsec} that uses two times on the forward iterated segment, the time-symmetric theory developed next uses one time on the backward iterated segment and one time on the forward iterated segment. The backward iterated segment is henceforth called the pre-image segment, illustrated in Fig. \ref{figODE} by the time domains in lightcone {\small $[T_\mathsmaller{-1}, T_\mathsmaller{1}]$} and {\small $[T_\mathsmaller{-2}, T_\mathsmaller{0}]$}. To avoid traversing the breaking point in the backward direction, we start from the pre-image and proceed to the first and second iterates. Here we introduce a notation to stress the dependence \textit{on the segment} of the elimination provided by Lemmas \ref{paraboloid_lemma} and \ref{asy_paraboloid_lemma}. Again, we abuse of the subscript {\scriptsize $\circledS$} (from segment) to indicate that the value of the denoted quantity depends on the \textbf{segment} {\mbox{\scriptsize $\circledS$}} $ \in \mathcal{FSH}$ defined in \S \ref{ODEsec}-\ref{perturbed_algorithm} by Eq. (\ref{FSH}). Among the quantities obtained by integrating until the breaking point, the limiting values of the velocity variables {\small $\varphi_\mathsmaller{\alpha}$} before each breaking point are functions of the segments illustrated in red in Fig. \ref{figODE}, {\small $\varphi_\mathsmaller{k}(t_\mathsmaller{k} \in [T_\mathsmaller{0}, {\color{red}b_{es}}])$} and {\small $\varphi_\mathsmaller{i}(t_\mathsmaller{i} \in [T_\mathsmaller{1}, {\color{red}b_{ps}}]$)}, namely, $\Phi_\mathsmaller{k} : \mathcal{FSH} \rightarrow \hat{C}^{1}[T_\mathsmaller{0}, {\color{red}b_{es}}] $, $\Phi_\mathsmaller{i} : \mathcal{FSH} \rightarrow \hat{C}^{1}[T_\mathsmaller{1}, {\color{red}b_{ps}}] $ and  $\Phi_\mathsmaller{j} : \mathcal{FSH} \rightarrow \hat{C}^{1}[T_\mathsmaller{2}, {\color{red}b_{e1}}] $ as defined by
{\small
\begin{eqnarray}
\varphi_\mathsmaller{k}|_{t_\mathsmaller{k} = {\color{red}b_{es}^{\mathsmaller{-}}}} &=& \varphi_{\mbox{\tiny$\circledS$} k  } = \Phi_\mathsmaller{k}(\varphi_\mathsmaller{k}(t_\mathsmaller{k} \in [T_\mathsmaller{0}, {\color{red}b_{es}}]),\varphi_\mathsmaller{i}(t_\mathsmaller{i} \in [T_\mathsmaller{1}, {\color{red}b_{ps}}] ) ),\label{segmentk} \\
\varphi_\mathsmaller{i}|_{t_\mathsmaller{i} = {\color{red}b_{ps}^{\mathsmaller{-}}}} &=& \varphi_{\mbox{\tiny $\circledS$} i } =  \Phi_\mathsmaller{i}(\varphi_\mathsmaller{k}(t_\mathsmaller{k} \in [T_\mathsmaller{0}, {\color{red}b_{es}}]),\varphi_\mathsmaller{i}(t_\mathsmaller{i} \in [T_\mathsmaller{1}, {\color{red}b_{ps}}] ) ), \label{segmenti} \\
\varphi_\mathsmaller{j}|_{t_\mathsmaller{j} = {\color{blue}b_{e1}^{\mathsmaller{-}}}} &=& \varphi_{\mbox{\tiny$\circledS$}j} =  \Phi_\mathsmaller{j}(\varphi_\mathsmaller{k}(t_\mathsmaller{k} \in [T_\mathsmaller{0}, {\color{red}b_{es}}]),\varphi_\mathsmaller{i}(t_\mathsmaller{i} \in [T_\mathsmaller{1}, {\color{red}b_{ps}}] ) ). \label{segmentj}
\end{eqnarray}
}
Also functions of the \textbf{segment} \mbox{\scriptsize $\circledS$}  $\in \mathcal{FSH}$ are the distances  {\small$r_\mathsmaller{\alpha i}$} between breaking points. We define the forward ratio of the segment by 
{\small
\begin{eqnarray}
\delta_{\mbox{\tiny $\circledS$}j}(\varphi_\mathsmaller{k}(t_\mathsmaller{k} \in [T_\mathsmaller{0}, {\color{red}b_{es}}]),\varphi_\mathsmaller{i}(t_\mathsmaller{i} \in [T_\mathsmaller{1}, {\color{red}b_{ps}}] ) ) \equiv \frac{r_\mathsmaller{ik}}{r_\mathsmaller{ij}},  \; \label{forwardNumbs}
\end{eqnarray}}
while the backward ratio of the segment is defined by
 {\small
 \begin{eqnarray}
 \delta_{\mathsmaller{\circledS} s}(\varphi_\mathsmaller{k}(t_\mathsmaller{k} \in [T_\mathsmaller{0}, {\color{red}b_{es}}]),\varphi_\mathsmaller{i}(t_\mathsmaller{i} \in [T_\mathsmaller{1}, {\color{red}b_{ps}}] ) ) \equiv \frac{r_\mathsmaller{ki}}{r_\mathsmaller{ks}} ,\label{backwardNumbs}
\end{eqnarray}}
which is calculated using the segment element \mbox{\scriptsize $\circledS$} $ \in \mathcal{FSH}$ (red in Fig. \ref{figODE}) by either forward or backward integration until the breaking point at mid-segment illustrated in Fig. \ref{figODE}. Lemma \ref{paraboloid_lemma} is obtained calculating the segment-dependent value of {\small $\mathbf{P}_\mathsmaller{i}$} defined by (\ref{Psplit}) using the segment functions on the left-hand side of the breaking point. Equating the left-hand side of (\ref{Psplit}) to the right-hand side of (\ref{Psplit}) when {\small $\varepsilon=0$} yields
{\small
\begin{eqnarray}
M_\mathsmaller{i} \sinh \varphi_{{\mathsmaller{\circledS}} i}+\frac{e_\mathsmaller{i}e_\mathsmaller{k}}{4}  (e^{2\varphi_{\mathsmaller{\circledS} k} }-e^{2\varphi_\mathsmaller{k} })=M_\mathsmaller{i} \sinh \varphi_\mathsmaller{i}+\frac{e_\mathsmaller{i} e_\mathsmaller{j} \delta_{\mathsmaller{\circledS} j } }{4} (e^{-2\varphi_{\mathsmaller{\circledS} j} }-e^{-2\varphi_\mathsmaller{j} }),\label{senzaPzero}
\end{eqnarray}
}
where {\small $\varphi_\mathsmaller{k} \equiv \varphi_\mathsmaller{k}|_{t_\mathsmaller{k}=b^{\mathsmaller{+}}_{\mathsmaller{es}} }$}, {\small $\varphi_\mathsmaller{i} \equiv \varphi_\mathsmaller{i}|_{t_\mathsmaller{i}=b^{\mathsmaller{+}}_{\mathsmaller{ep}}}$} and {\small $\varphi_\mathsmaller{j} \equiv \varphi_\mathsmaller{j}|_{t_\mathsmaller{j}=b^{\mathsmaller{+}}_{\mathsmaller{p1}}}$}, while $\varphi_{\mathsmaller{\circledS}k} $\,, $\varphi_{\mathsmaller{\circledS}i}$\, and $\varphi_{\mathsmaller{\circledS}j}$ are the left-hand-side values defined respectively by equations (\ref{segmentk}), (\ref{segmenti}) and (\ref{segmentj}). Equation (\ref{senzaPzero}) has a unique subluminal solution for each given {\small $(\varphi_\mathsmaller{k},\varphi_\mathsmaller{i})$}, as proved in Lemma \ref{paraboloid_lemma}. The \textbf{continuous-velocity} solution is the trivial root of (\ref{senzaPzero}) obtained  by setting {\small $(\varphi_\mathsmaller{k},\varphi_\mathsmaller{i})=(\varphi_{\mathsmaller{\circledS} k}, \varphi_{\mathsmaller{\circledS} i}) $}. We henceforth denote {\small $\theta_\mathsmaller{ji } = e^{-\varphi_{\mathsmaller{\circledS}j}} \equiv \theta_{\mathsmaller{\circledS} j}$} as the function of the segment defined by Eq. (\ref{defteta}). The unique velocity after the breaking point is calculated by solving Eq. (\ref{senzaPzero}) for the value of $e^{-2\mathsmaller{\varphi_\mathsmaller{j}}}$ after the breaking point, yielding a function of {\small $(\varphi_\mathsmaller{k},\varphi_\mathsmaller{i}, \circledS)$}, where \mbox{\scriptsize $\circledS$} $ \in \mathcal{FSH}$ and
{\small
\begin{eqnarray}
e^{-2\varphi_\mathsmaller{j} }= \theta^{\mathsmaller{2}}_{\mathsmaller{\circledS} j} +\frac{e_\mathsmaller{i} e_\mathsmaller{k}(e^{2\varphi_\mathsmaller{k} }-e^{2\varphi_{\mathsmaller{\circledS} k} } ) +4M_\mathsmaller{i}( \sinh{\varphi_\mathsmaller{i}}-\sinh{\varphi_{\mathsmaller{\circledS} i}})  } {e_\mathsmaller{i} e_\mathsmaller{j} \delta_{\mathsmaller{\circledS} j }  } >0. \label{positiveSLemma}
\end{eqnarray} 
}
The positivity of (\ref{positiveSLemma}) corresponds to the half-line of Lemma \ref{paraboloid_lemma}. The energetic corner conditions at {\small $\varepsilon= 0$} are found substituting the hyperbolic velocity-angles (\ref{def_hyperphi}) into (\ref{CWE2i}) and (\ref{CWE2k}), yielding the continuous functions
{\small
\begin{eqnarray}
\lambda^{\mathsmaller{o}}_{\mathsmaller{ik}} (\varphi_\mathsmaller{i}, \varphi_\mathsmaller{k}) &\equiv&r_\mathsmaller{ki} \mathbb{C}_\mathsmaller{i} +\frac{e_\mathsmaller{i} e_\mathsmaller{k}}{2}= M_\mathsmaller{i} e^{\varphi_{\mathsmaller{i}}}+\frac{e_\mathsmaller{i} e_\mathsmaller{k} e^{2\varphi_\mathsmaller{k}}}{2}, \label{poly1zero}\\
\lambda^{\mathsmaller{o}}_{\mathsmaller{ki}} (\varphi_\mathsmaller{i}, \varphi_\mathsmaller{k}) &\equiv& r_\mathsmaller{ki}\mathbb{C}_\mathsmaller{k} +\frac{e_\mathsmaller{i} e_\mathsmaller{k}}{2}= M_\mathsmaller{k} e^{\varphi_{\mathsmaller{k}}}+\frac{e_\mathsmaller{i} e_\mathsmaller{k} e^{2\varphi_\mathsmaller{i}}}{2}, \label{poly2zero}
\end{eqnarray}
}
where we have defined scaled masses by 
{\small
\begin{eqnarray}
M_\mathsmaller{\alpha} \equiv r_\mathsmaller{ki}m_\mathsmaller{\alpha}; \; \;\; \alpha \in (i,k). \label{scaledmass}
\end{eqnarray}
}
The continuity of (\ref{poly1zero}) can be used to eliminate {\small $(e^{2\varphi_\mathsmaller{k} }-e^{2\varphi_{\mathsmaller{\circledS} k} } ) $} from the right-hand side of (\ref{positiveSLemma}), thus yielding a condition involving only {\small $\varphi_\mathsmaller{j}$} and {\small $\varphi_\mathsmaller{i}$}, i.e.,
 {\small
 \begin{eqnarray}
e^{-2\varphi_\mathsmaller{j}}+\frac{2M_\mathsmaller{i} e^{-\varphi_\mathsmaller{i} }}{e_\mathsmaller{i} e_\mathsmaller{j} \delta_{\mathsmaller{\circledS} j} }=e^{-2\varphi_{\mathsmaller{\circledS} j}}+\frac{2M_\mathsmaller{i} e^{-\varphi_{\mathsmaller{\circledS} i} }}{e_\mathsmaller{i} e_\mathsmaller{j} \delta_{\mathsmaller{\circledS} j} }  \equiv J_{\mathsmaller{\circledS} j}, \label{double-rooted_zero}
 \end{eqnarray}
 }
 where {\small $J_{\mathsmaller{\circledS} j}$} is the function of the segment \mbox{\scriptsize $\circledS$} $ \in \mathcal{FSH}$ calculated at the left-hand side of the breaking point by the last equality of (\ref{double-rooted_zero}). Notice that (\ref{poly1zero}) and (\ref{poly2zero}) are each independent of one next-neighbor velocity, either {\small $\varphi_\mathsmaller{j}$} or {\small $\varphi_\mathsmaller{s}$}, a degeneracy of the {\small $\varepsilon=0$} case that reduces the study of velocity discontinuities to solving the right-hand side of (\ref{poly1zero}) and (\ref{poly2zero}) with given left-hand-side values {\small $\lambda^{\mathsmaller{o}}{\mathsmaller{\circledS ik}} $} and {\small $\lambda^{\mathsmaller{o}}{\mathsmaller{\circledS ki}} $}.  Elimination of {\small $e^{\varphi_\mathsmaller{i}} $} from (\ref{poly1zero}) and substitution into (\ref{poly2zero}) yields a polynomial of the fourth degree with coefficients depending on {\small $(M_\mathsmaller{i}, M_\mathsmaller{k},\lambda^{\mathsmaller{o}}_{\mathsmaller{\circledS} ik},\lambda^{\mathsmaller{o}}_{\mathsmaller{\circledS} ki}, e_\mathsmaller{i} e_\mathsmaller{k}, \varepsilon)$}. In order to have a velocity discontinuity, the quartic polynomial must have at least \textbf{two} positive roots {\small $e^{\varphi_\mathsmaller{k}}\in \mathbb{R}_\mathsmaller{+}$} and {\small $ e^{\tilde{\varphi}_\mathsmaller{k}} \in \mathbb{R}_\mathsmaller{+}$}. In the following we start from the case {\small $\varepsilon=0$} as a limit case for the polynomial that must have two positive roots when {\small $\varepsilon \neq 0$}.

\begin{lem} The elastic \textbf{collision-at-a-distance} {\small $(\varphi_{\mathsmaller{\circledS} i},\varphi_{\mathsmaller{\circledS} k}) \rightarrow (\mathsmaller{-}\varphi_{\mathsmaller{\circledS} i},\mathsmaller{-}\varphi_{\mathsmaller{\circledS} k}) $} at {\small $\varepsilon=0$} is a \textbf{mutual recoil} for the electron-electron case,  {\small $e_\mathsmaller{i} e_\mathsmaller{k}=1$}, and a \textbf{sticky collision} for the electron-proton case, {\small $e_\mathsmaller{i}e_\mathsmaller{k}=\mathsmaller{-}1$}. 
\label{zero_symmetric_lemma}
\end{lem}
\begin{proof} Inside this proof we drop the sub-index \mbox{\scriptsize $\circledS$}. Because the left-hand side of (\ref{poly1zero}) is continuous, at the breaking point of trajectory $i$, the elastic continuity of (\ref{poly1zero}), {\small $\lambda^{\mathsmaller{o}}_{\mathsmaller{ik}}(\varphi_\mathsmaller{i}, \varphi_\mathsmaller{ik})-\lambda^{\mathsmaller{o}}_{\mathsmaller{ik}}(\mathsmaller{-}\varphi_\mathsmaller{i},\mathsmaller{-}\varphi_\mathsmaller{k})=0$}, yields
{\small
\begin{eqnarray}
M_\mathsmaller{i} \sinh(\varphi_\mathsmaller{i})&=&-\frac{e_\mathsmaller{i}e_\mathsmaller{k}}{2}\sinh(2\varphi_\mathsmaller{k})= -e_\mathsmaller{i}e_\mathsmaller{k} \gamma_\mathsmaller{k} \sinh{\varphi_\mathsmaller{k}}, \label{positive1}
\end{eqnarray}}
where we have used (\ref{gamaphi}). Analogously, for the breaking point of trajectory $k $, the elastic continuity of (\ref{poly2zero}), {\small $\lambda^{\mathsmaller{o}}_{\mathsmaller{ki}}(\varphi_\mathsmaller{i}, \varphi_\mathsmaller{k})-\lambda^{\mathsmaller{o}}_{\mathsmaller{ki}}(\mathsmaller{-}\varphi_\mathsmaller{i},\mathsmaller{-}\varphi_\mathsmaller{k})=0$}, yields
{\small
\begin{eqnarray}
M_\mathsmaller{k} \sinh(\varphi_\mathsmaller{k})&=&-\frac{e_\mathsmaller{i}e_\mathsmaller{k}}{2}\sinh(2\varphi_\mathsmaller{i})= -e_\mathsmaller{i}e_\mathsmaller{k}\gamma_\mathsmaller{i} \sinh{\varphi_\mathsmaller{i}}, \label{positive2}
\end{eqnarray}}
where again we used {\small $\gamma_\mathsmaller{i}=\cosh(\varphi_\mathsmaller{i})>0$} and {\small $\varepsilon=0$}. When {\small $e_\mathsmaller{i}e_\mathsmaller{k}=-1$}, Eqs. (\ref{positive1}) and (\ref{positive2}) determine that {\small $\varphi_\mathsmaller{k}$} and {\small $\varphi_\mathsmaller{i}$} have the same sign, and we have a sticky collision. Otherwise, when {\small $e_\mathsmaller{i}e_\mathsmaller{k}=1$}, Eq. (\ref{positive1}) defines a mutual recoil where the {\small $\varphi$}'s have opposite signs. Equations (\ref{positive1}) and (\ref{positive2}) are homogeneous linear equations relating {\small $\sinh(\varphi_\mathsmaller{k})$} and {\small $\sinh(\varphi_\mathsmaller{i})$}, and the vanishing of the {\small $2\times 2$} determinant can be expressed as
{\small 
\begin{eqnarray}
\frac{\gamma_\mathsmaller{i}}{M_\mathsmaller{i}}=\frac{M_\mathsmaller{k}}{\gamma_\mathsmaller{k}}. \label{detgigk}
\end{eqnarray}
}
Squaring both (\ref{positive1}) and (\ref{positive1}), using that {\small $\sinh^\mathsmaller{2}(\varphi_\mathsmaller{\alpha}) =\cosh^\mathsmaller{2}(\varphi_\mathsmaller{\alpha}) -1=\gamma_{\mathsmaller{\alpha}}^{\mathsmaller{2}} -1$},  and multiplying both sides by (\ref{detgigk}) yields 
{\small
\begin{eqnarray}
M_\mathsmaller{i} \gamma_\mathsmaller{i}(\gamma_{\mathsmaller{i}}^\mathsmaller{2}-1)=M_\mathsmaller{k} \gamma_\mathsmaller{k}(\gamma_{\mathsmaller{k}}^\mathsmaller{2}-1) \equiv 2M_\mathsmaller{i} M_\mathsmaller{k} J, \label{cubic}
\end{eqnarray}
}
where in the last term we have introduced the positive quantity $J$ to parametrize both cubics. Any solution of (\ref{cubic}) yields a solution to  (\ref{positive1}) and (\ref{positive2}), which can easily be found with Cardano's formula. Equation (\ref{cubic}) has a real root when $J>0$, which gives \textbf{two} values for each $\varphi_\mathsmaller{\alpha}$ by inverting the hyperbolic cosine. For sufficiently large $J$,  the two solution pairs are approximated by
{\small
\begin{eqnarray}
\varphi_{\mathsmaller{i}} &\approx& \pm  \frac{1}{3} \log_{e} (16J M_{\mathsmaller{k}})  , \label{ge1} \\
\varphi_{\mathsmaller{k}} &\approx& \mp \frac{e_\mathsmaller{i} e_\mathsmaller{k}}{3} \log_{e} (16JM_\mathsmaller{i} ) . \label{ge2}
\end{eqnarray}
}
Eliminating {\small $J$} from (\ref{ge1}) and (\ref{ge2}) yields the approximate \textit{electromagnetic boundary layer condition}
{\small
\begin{eqnarray}
\gamma_\mathsmaller{k} =\mathpzc{h} \, \gamma_\mathsmaller{i}, \label{largeL_sol}
\end{eqnarray}
}
where
\begin{eqnarray}
\mathpzc{h} \equiv (\frac{m_\mathsmaller{i}}{m_\mathsmaller{k}})^\mathsmaller{1/3}.\label{defhbar}
\end{eqnarray}
Condition (\ref{largeL_sol}) holds when (\ref{ge1}) and (\ref{ge2}) are defined with {\small $J \gg \frac{\mathpzc{h}^\mathsmaller{3/2}}{3\sqrt{3}}$}. As long as inequality (\ref{positiveSLemma}) is satisfied, 
the solution pair (\ref{ge1}) and (\ref{ge2}) \textbf{proves} the existence of elastic collisions-at-a-distance from {\small $(\varphi_{\mathsmaller{\circledS} k},\varphi_{\mathsmaller{\circledS} i})$} to {\small $(-\varphi_{\mathsmaller{\circledS} k},-\varphi_{\mathsmaller{\circledS} i})$} in both cases.
 \end{proof}
 
 The energetic corner conditions when {\small $\varepsilon \neq 0$} are found by substituting the hyperbolic velocity-angles (\ref{def_hyperphi}) into (\ref{CWE2i}) and (\ref{CWE2k}), yielding
{\small
\begin{eqnarray}
\lambda_\mathsmaller{ik} (\varphi_\mathsmaller{i}, \varphi_\mathsmaller{k}) &\equiv&r_\mathsmaller{ki} \mathbb{C}_\mathsmaller{i} +\frac{e_\mathsmaller{i} e_\mathsmaller{k}}{2}= M_\mathsmaller{i} e^{\varphi_{\mathsmaller{i}}}+\frac{e_\mathsmaller{i} e_\mathsmaller{k} e^{2\varphi_\mathsmaller{k}}}{2}+\frac{\varepsilon e^{\varphi_\mathsmaller{i}} e^{\varphi_\mathsmaller{k}}}{2} +\varepsilon \delta_{\mathsmaller{\circledS} j}e^{\varphi_\mathsmaller{i}} \theta_\mathsmaller{ji}, \label{poly1}\\
\lambda_\mathsmaller{ki} (\varphi_\mathsmaller{i}, \varphi_\mathsmaller{k}) &\equiv& r_\mathsmaller{ki}\mathbb{C}_\mathsmaller{k} +\frac{e_\mathsmaller{i} e_\mathsmaller{k}}{2}= M_\mathsmaller{k} e^{\varphi_{\mathsmaller{k}}}+\frac{e_\mathsmaller{i} e_\mathsmaller{k} e^{2\varphi_\mathsmaller{i}}}{2}+\frac{\varepsilon e^{\varphi_\mathsmaller{i}} e^{\varphi_\mathsmaller{k}}}{2}+\varepsilon \delta_{\mathsmaller{\circledS} s} e^{\varphi_\mathsmaller{k}} \theta_\mathsmaller{sk}. \label{poly2}
\end{eqnarray}
}
 
\begin{lem} There is a unique {\small $ e^{\varphi_\mathsmaller{i}} \theta_\mathsmaller{ji}(\varphi_\mathsmaller{k},\varphi_\mathsmaller{i},\mbox{\scriptsize $\circledS$}) \equiv e^{\varphi_\mathsmaller{i}-\varphi_\mathsmaller{j}}$} given by 
 \label{lemma_Uniqe_ps}
 {\small
 \begin{eqnarray}
e^{\varphi_\mathsmaller{i}} \theta_\mathsmaller{ji}=e^{\varphi_\mathsmaller{i}-\varphi_\mathsmaller{j}}=\frac{\varepsilon_\mathsmaller{*}}{2} +\frac{1}{2}\sqrt{\varepsilon_{\mathsmaller{*}}^{\mathsmaller{2}} +8M^{\mathsmaller{*}}_{\mathsmaller{i}}e^{\varphi_\mathsmaller{i}}+\frac{4\varepsilon_\mathsmaller{*}}{\delta_{\mathsmaller{\circledS} j}}e^{\varphi_\mathsmaller{k} +\varphi_\mathsmaller{i}} +4J_{\mathsmaller{\circledS} j} e^{2\varphi_\mathsmaller{i}}  },
\label{EQTS}
 \end{eqnarray}
 }
 where {\small $M^{\mathsmaller{*}}_{\mathsmaller{i}} \equiv -\frac{M_\mathsmaller{i}}{e_\mathsmaller{i} e_\mathsmaller{j}}$ } and {\small $J_\mathsmaller{\circledS j}$} is a function of the element  \mbox{\scriptsize ${\circledS}$} $\in \mathcal{FSH}$.
\end{lem}
 \begin{proof} The perturbed version of (\ref{double-rooted_zero}) is
 calculated from (\ref{Psplit}) with $\varepsilon \neq 0 $ using the segment functions on the left-hand side of the breaking point and equating it to the to the right-hand side of (\ref{Psplit}), yielding 
 {\small
 \begin{eqnarray}
J_{\mathsmaller{\circledS} j} &\equiv& e^{-2\varphi_{\mathsmaller{\circledS} j}}-\varepsilon^\mathsmaller{*}e^{-(\varphi_{\mathsmaller{\circledS} i}  +\varphi_{\mathsmaller{\circledS} j})}+\frac{2M_\mathsmaller{i} e^{-\varphi_{\mathsmaller{\circledS} i} }}{e_\mathsmaller{i} e_\mathsmaller{j} \delta_{\mathsmaller{\circledS} j} } -\frac{\varepsilon_\mathsmaller{*}}{\delta_{\mathsmaller{\circledS} j}}e^{-\varphi_{\mathsmaller{\circledS} i}+\varphi_{\mathsmaller{\circledS} k}}\notag \\ &=&e^{-2\varphi_\mathsmaller{j}}   -\varepsilon^\mathsmaller{*}e^{-(\varphi_\mathsmaller{i}+ \varphi_\mathsmaller{j})}+\frac{2M_\mathsmaller{i} e^{-\varphi_\mathsmaller{i} }}{e_\mathsmaller{i} e_\mathsmaller{j} \delta_{\mathsmaller{\circledS} j} }-\frac{\varepsilon_\mathsmaller{*}}{\delta_{\mathsmaller{\circledS} j}}e^{-\varphi_\mathsmaller{i}+\varphi_\mathsmaller{k}} , \label{double-rooted_epsilon}
 \end{eqnarray}
 }
 where we have used the continuity of (\ref{poly1}) to eliminate {\small $(e^{2\varphi_\mathsmaller{k} }-e^{2\varphi_\mathsmaller{\circledS k} } ) $}, just like in the derivation of (\ref{double-rooted_zero}). Equation (\ref{double-rooted_epsilon}) is a quadratic equation for {\small$e^{-\varphi_\mathsmaller{j}} $} with a unique subluminal solution  {\small $\varphi_\mathsmaller{j}=\varphi_\mathsmaller{j}(\varphi_\mathsmaller{k},\varphi_\mathsmaller{i},J_{\mathsmaller{\circledS} j} )$}, and the value  of {\small $J_{\mathsmaller{\circledS} j}$} must be such that {\small $(\varphi_\mathsmaller{k},\varphi_\mathsmaller{i})$} is inside the half-line domain of Lemma \ref{asy_paraboloid_lemma}. 
 \end{proof}
  Lemma \ref{lemma_Uniqe_ps} holds also with the index shift {\small $(j,i,k) \rightarrow (s,k,i)$}; the corner conditions have advance and delay, and thus couple the segment to its pre-image as well. As illustrated in Fig. \ref{figODE}, when the pre-image is iterated once to obtain the segment, there is a Weirstrass-Erdmann condition to cross the breaking points {\color{Goldyellow} $b_\mathsmaller{e-1}$} and {\color{Goldyellow} $b_\mathsmaller{p-1}$}. Iteration of the pre-image yields {\small $\varphi_\mathsmaller{s}=\varphi_\mathsmaller{s}(\varphi_\mathsmaller{k},\varphi_\mathsmaller{i},J_{\mathsmaller{\circledS} s} )$}, where {\small $J_{\mathsmaller{\circledS} s}$} is calculated from the left-hand side of the breaking points  {\color{Goldyellow} $b_\mathsmaller{e-1}$} and {\color{Goldyellow} $b_\mathsmaller{p-1}$} by the analogous of (\ref{double-rooted_epsilon}) with the shift {\small $(j,i,k) \rightarrow (s,k,i)$}. These are the two energetic corner conditions to be satisfied by {\small $(\varphi_\mathsmaller{k},\varphi_\mathsmaller{i})$}. Summarizing; the energetic corner conditions amount to two isolated conditions on {\small $(\varphi_\mathsmaller{k},\varphi_\mathsmaller{i})$} plus the restriction that  {\small $J_{\mathsmaller{\circledS} j}$} and {\small $J_{\mathsmaller{\circledS} s}$} are such that the last equality of (\ref{double-rooted_epsilon}) and the respective pre-image condition are satisfied on both sides of the breaking point in order to be able to eliminate {\small $\theta_\mathsmaller{ji}=e^{-\varphi_\mathsmaller{j}}$} from (\ref{poly1}) and {\small $\theta_\mathsmaller{sk} =e^{-\varphi_{\mathsmaller{s}}} $} from (\ref{poly2}).
In order to allow a velocity discontinuity, the algebraic conditions obtained from (\ref{poly1}) and (\ref{poly2}) must have at least \textbf{two} positive roots, i.e., {\small $(e^{\varphi_\mathsmaller{k}}, e^{\varphi}_\mathsmaller{i})  \in \mathbb{R}_\mathsmaller{+} \times \mathbb{R}_\mathsmaller{+}$} and {\small $(e^{\tilde{\varphi_\mathsmaller{k}}}, e^{\tilde{\varphi}_\mathsmaller{i}})  \in \mathbb{R}_\mathsmaller{+} \times \mathbb{R}_\mathsmaller{+}$}. 
   The extension of Lemma \ref{zero_symmetric_lemma} when $\varepsilon \neq 0$ is tricky; mutual recoils are simpler because  the outer-cone Lema \ref{outside-cone-Lemma} grants a limit of large recoiling velocities. In the electron-proton case, the outer-cone Lema \ref{outside-cone-Lemma} does not apply and the forward and backward ratios, $\delta_{\mathsmaller{\circledS} j}$ and $\delta_{\mathsmaller{\circledS} s}$, are difficult to control.  In the following we treat only the electron-electron collision for small $\varepsilon \neq 0$ and initial segment being a second iterate with sufficiently small ratios (\ref{forwardNumbs}) and (\ref{backwardNumbs}) such that each last term on the right-hand sides of (\ref{poly1}) and (\ref{poly2}), {\small $\delta_{\mathsmaller{\circledS} j} e^{\varphi_\mathsmaller{i}-\varphi_\mathsmaller{j}}$} and {\small $\delta_{\mathsmaller{\circledS} s} e^{\varphi_\mathsmaller{k}-\varphi_\mathsmaller{s}}$}, is ignorable by Lemma \ref{outside-cone-Lemma}.
\begin{lem} The elastic electron-electron collision-at-a-distance {\small $(\varphi_\mathsmaller{i},\varphi_\mathsmaller{k}) \rightarrow (\mathsmaller{-}\varphi_\mathsmaller{i},\mathsmaller{-}\varphi_\mathsmaller{k}) $} is a \textbf{mutual recoil} if {\small $|\varepsilon| <2\min(\mathpzc{h}^\mathsmaller{3},\frac{1}{\mathpzc{h}^\mathsmaller{3}}) $}. 
\label{ee_symmetric_lemma}
\end{lem}
\begin{proof}  Since {\small $e_\mathsmaller{i}e_\mathsmaller{k} =1$}, we have {\small $\varepsilon_{\mathsmaller{*}}=-\varepsilon$}. We start from the continuity of the left-hand side of (\ref{poly1}), {\small $\lambda_\mathsmaller{ik}(\varphi_\mathsmaller{i}, \varphi_\mathsmaller{ik})-\lambda_\mathsmaller{ik}(\mathsmaller{-}\varphi_\mathsmaller{i},\mathsmaller{-}\varphi_\mathsmaller{k})=0$}, yielding
{\small
\begin{eqnarray}
\big( M_\mathsmaller{i}+\frac{\varepsilon}{2}\gamma_\mathsmaller{k} \big)\sinh(\varphi_\mathsmaller{i})&=& -e_\mathsmaller{i}e_\mathsmaller{k}\big(  \gamma_\mathsmaller{k} -\frac{\varepsilon_\mathsmaller{*}}{2}\gamma_\mathsmaller{i}  \big)\sinh{\varphi_\mathsmaller{k}}, \label{eps1}
\end{eqnarray}}
where again we have used (\ref{gamaphi}). Analogously, for the breaking point of trajectory {\small $k$} we must have {\small $\lambda_\mathsmaller{ki}(\varphi_\mathsmaller{i}, \varphi_\mathsmaller{k})-\lambda_\mathsmaller{ki}(\mathsmaller{-}\varphi_\mathsmaller{i},\mathsmaller{-}\varphi_\mathsmaller{k})=0$}, yielding
{\small
\begin{eqnarray}
\big( M_\mathsmaller{k}+\frac{\varepsilon}{2}\gamma_\mathsmaller{i}  \big)\sinh(\varphi_\mathsmaller{k})&=&-e_\mathsmaller{i}e_\mathsmaller{k}\big( \gamma_\mathsmaller{i} -\frac{\varepsilon_\mathsmaller{*} }{2}\gamma_\mathsmaller{k} \big)\sinh{\varphi_\mathsmaller{i}}, \label{eps2}
\end{eqnarray}}
where {\small $\gamma_\mathsmaller{\alpha}=\cosh(\varphi_\mathsmaller{\alpha})>0$}. Equations (\ref{eps1}) and (\ref{eps2}) are linear equations relating $\sinh(\varphi_\mathsmaller{k})$ and $\sinh(\varphi_\mathsmaller{i})$, and the vanishing of the $2\times 2$ determinant can be expressed either as
{\small 
\begin{eqnarray}
\frac{\big( \gamma_\mathsmaller{i} -\frac{\varepsilon_\mathsmaller{*} }{2}\gamma_\mathsmaller{k} \big)}{\big( M_\mathsmaller{i}+\frac{\varepsilon}{2}\gamma_\mathsmaller{k} \big)}=\frac{\big( M_\mathsmaller{k}+\frac{\varepsilon}{2}\gamma_\mathsmaller{i}  \big)}{\big( \gamma_\mathsmaller{k} -\frac{\varepsilon_\mathsmaller{*} }{2}\gamma_\mathsmaller{i} \big)}, \label{hfeps}
\end{eqnarray}
}
or explicitly
{\small
\begin{eqnarray}
\gamma_\mathsmaller{i} \gamma_\mathsmaller{k}=M_\mathsmaller{i} M_\mathsmaller{k} +\frac{\varepsilon}{2}(M_\mathsmaller{i}\gamma_\mathsmaller{i} +M_\mathsmaller{k}\gamma_\mathsmaller{k}) +\frac{\varepsilon_{\mathsmaller{*}}}{2}(\gamma_{\mathsmaller{i}}^{\mathsmaller{2}}+\gamma_{\mathsmaller{k}}^{\mathsmaller{2}}).
\label{deteps}
\end{eqnarray}
}
Equation (\ref{hfeps}) reduces to (\ref{detgigk}) when $\varepsilon = 0$.  There are two cases to consider:
\begin{enumerate}
\item If $\varepsilon>0$, we have $\varepsilon_\mathsmaller{*}=-|\varepsilon|$ and $\varepsilon=|\varepsilon|$, and Eq. (\ref{eps1}) predicts opposite signs for $\varphi_\mathsmaller{i}$ and $\varphi_\mathsmaller{k}$, which are mutual recoils.
\item If $\varepsilon<0$, we use $\varepsilon_\mathsmaller{*}=|\varepsilon|$ and $\varepsilon=-|\varepsilon|$ to re-arrange Eq. (\ref{eps1}) as
{\small
\begin{eqnarray}
\frac{\sinh(\varphi_\mathsmaller{i})}{\sinh(\varphi_\mathsmaller{k})}=-\frac{2}{|\varepsilon|} \Big( \frac{\gamma_\mathsmaller{k}-\frac{|\varepsilon|}{2}\gamma_\mathsmaller{i} }{\gamma_\mathsmaller{k} -\frac{2M_\mathsmaller{i}}{|\varepsilon|}} \Big). \label{rearrange}
\end{eqnarray}
}
The positivity of (\ref{rearrange}) and the positivity of the respective re-arranged version of (\ref{eps2}) requires both
{\small
\begin{eqnarray}
 \min\big(\frac{2M_\mathsmaller{i}}{|\varepsilon|}, \frac{|\varepsilon| \gamma_\mathsmaller{i}}{2}\big) <\gamma_\mathsmaller{k}< \max \big(\frac{2M_\mathsmaller{i}}{|\varepsilon|}, \frac{|\varepsilon| \gamma_\mathsmaller{i}}{2}\big), \\
 \min\big(\frac{2M_\mathsmaller{k}}{|\varepsilon|}, \frac{|\varepsilon| \gamma_\mathsmaller{k}}{2}\big) <\gamma_\mathsmaller{i}< \max \big(\frac{2M_\mathsmaller{k}}{|\varepsilon|}, \frac{|\varepsilon| \gamma_\mathsmaller{k}}{2}\big).
 \end{eqnarray}
}
The two alternatives are; {\color{darkgray}(i)} {\small  $\gamma_\mathsmaller{k}> \frac{|\varepsilon|\gamma_\mathsmaller{i}}{2}$} and {\small $|\varepsilon|<2\mathpzc{h}^\mathsmaller{3}$} with $\mathpzc{h}$ defined by (\ref{defhbar}) or {\color{darkgray}(ii)} {\small $\gamma_\mathsmaller{i}> \frac{|\varepsilon|\gamma_\mathsmaller{k}}{2}$} and {\small $|\varepsilon|<2\mathpzc{h}^\mathsmaller{-3}$}, with $\mathpzc{h}$ defined again by (\ref{defhbar}). Therefore, we have a mutual recoil until {\small $|\varepsilon| < 2\min{ ( \mathpzc{h}^\mathsmaller{3}}, \frac{1}{\mathpzc{h}^{\mathpzc{3}}}) $}.
\end{enumerate}
\end{proof} 

Squaring either (\ref{eps1}) or (\ref{eps2}), and multiplying both sides by (\ref{hfeps}) yields 
{\small
\begin{eqnarray}
\Big( \gamma_\mathsmaller{\alpha}+ \frac{\varepsilon}{4M_\mathsmaller{\alpha}}( \varepsilon_{\mathsmaller{*}} \gamma_{\mathsmaller{\alpha}}^{\mathsmaller{2}} + \varepsilon M_\mathsmaller{i} \gamma_\mathsmaller{i}+ \varepsilon M_\mathsmaller{k} \gamma_\mathsmaller{k}  ) + \frac{1}{2}(\varepsilon M_\mathsmaller{k}-\varepsilon_{\mathsmaller{*}}\gamma_\mathsmaller{k})\Big)(\gamma_{\mathsmaller{\alpha}}^\mathsmaller{2}-1) \equiv 2J  \mathpzc{h}^{\mathsmaller{\mp 3/2}}, \label{quartic}
\end{eqnarray}
}
where the plus sign is for $\alpha=k$ and the minus sign is for $\alpha=i$. Equation (\ref{quartic}) is a quartic polynomial generalizing the cubic polynomial (\ref{cubic}), which introduces a singular root for $\varepsilon \neq 0$.

\section{Discussions and conclusion }
\label{Sumasection}

%\subsection{Discussions and conclusion}
%\label{discussions}
\begin{enumerate}
\item The energetic corner conditions are the fundamental mechanism for setting the scale based on the magnitudes of the boundary layer. A simplified ODE model without Lorentz-invariance was studied in \cite{simple}. Equation (\ref{equation_aj}) has a nonlinear gyroscopic term (\ref{gyroscopic}) that includes the resonances used in \cite{JDE2} with the provisional Chemical Principle criterion to estimate magnitudes. Resonances were also used in Refs. \cite{cinderela, old_helium} to estimate atomic magnitudes. 
\item The cubic root of the mass ratio appeared in Ref. \cite{double-slit} and Eq. (\ref{defhbar}). Both are applications of the Weierstrass-Erdmann corner conditions; Ref. \cite{double-slit} used {\small $\varepsilon$}-\textbf{VE} to estimate magnitudes for double-slit diffraction, see Eqs. (21) and (22) of \cite{double-slit}, while in Eq. (\ref{defhbar}), the cubic root appeared in the boundary layer condition. Our {\small $\varepsilon$}-\textbf{VE} model has two parameters, {\small $(m_\mathpzc{p}/m_\mathpzc{e})$} and {\small $\varepsilon$}, a far simpler theory than the standard model. Equations (\ref{largeL_sol}) and (\ref{defhbar}) predict a collisional boundary layer at {\small $\Big( \frac{1-\mathbf{v}_{\mathpzc{p}}^{\mathsmaller{2}} }{1-\mathbf{v}_{\mathpzc{e}}^{\mathsmaller{2}} } \Big)=\big( \frac{m_\mathsmaller{p}}{m_\mathsmaller{e}} \big)^{\mathsmaller{2/3}} \simeq 149.947$}. 
\item By inspecting (\ref{defineutron}) we find that the fixed-segment of phase locking exists only for {\small $\varepsilon_\mathsmaller{*}>0$}, which requires {\small $\varepsilon>0$} for the attractive case and $\varepsilon<0$ for the repulsive case. The former is due to the simple way functional (\ref{Fokker}) was defined here. As suggested in Ref. \cite{Martinez}, one can redefine the original functional (\ref{Fokker}) to include an {\small $\varepsilon$}-strong charge {\small $\mathpzc{q}_\mathsmaller{\alpha}$} for each particle chosen such that {\small $\frac{\varepsilon}{\varepsilon_\mathsmaller{*}} \equiv -\frac{\mathpzc{q}_\mathsmaller{i} \mathpzc{q}_\mathsmaller{j}}{e_\mathsmaller{i} e_\mathsmaller{j} }=1$} in the re-defined version of (\ref{Fokker}). The former has {\small $\varepsilon$} with the same sign for both the attractive and the repulsive cases when {\small $\varepsilon_\mathsmaller{*}>0$}. Our setup can generate several models to be studied numerically, e.g., a model for the exclusion principle, a model for Cooper pairs, and a model for collisions seen in bubble chambers and particle accelerators. The subtle differences between cases should be studied numerically and used in physics to model nature. Some may differ from the standard model (or not).
\item Reference \cite{Schild} discusses a failed attempt to model the neutron with pure electrodynamics.  An application of {\small $\varepsilon$}-\textbf{VE} is to model the neutron with the fixed segments of theorem \ref{neutronic}. The model should calculate the deuterium mass using a small {\small $\varepsilon$} to avoid perturbing electrodynamics too much. The potential accomplishment of {\small $\varepsilon$}-\textbf{VE} would be an economic theory with a \textbf{single} parameter modeling the neutron \textbf{and} atomic physics \cite{cinderela,JDE2}, a serious divergence-free contender for the standard model of particle physics. 
\item The orbits with constant velocities of theorem \ref{neutronic} have vanishing far-fields within {\small $\varepsilon$}-\textbf{VE} because the accelerations are zero, which is a strong condition to vanish the far-fields. The weak condition to vanish the far-fields within pure electrodynamics requires only discontinuous velocities\cite{minimizer}.
\item  It would be helpful to repeat our studies using the full action of \cite{Martinez} keeping the \textbf{two} parameters plus the electromagnetic interaction.  Non-zero angular momentum collisions-at-a-distance should be investigated as well.
\end{enumerate}

   \section{Appendices}
   \label{appendix}

 \subsection{The semiflow when {\small $\varepsilon \neq 0$} and qualitative differences }
 \label{referees}
 
 NDDEs need a \textbf{segment of trajectory} to start from. In contrast, the initial condition for an ODE is the meager set of initial positions and initial velocities (first-category in {\small $\mathbb{R}$}). Initial segments for NDDEs are \textbf{segments} defined on closed intervals (second-category in {\small $\mathbb{R}$}). Moreover, initial segments for NDDEs have the extra freedom of a countable number of velocity discontinuities, henceforth the photonic series, whereby the charges collide at-a-distance by exchanging velocity kicks in lightcone. Velocity discontinuities trigger further collisions in the next segment and can prevent the head-on instability.  A second surprising difference is that NDDEs have solutions with \textit{serrated} accelerations that are only continuous and nowhere differentiable. In contrast, the ODEs of classical mechanics have mainly {\small $C^\infty$} solutions. Likewise for Driver's problem with repulsive interaction \cite{Driver}, as shown in the next theorem. The two-body problem with attractive interaction was studied in \cite{EFY2}, and the bounded {\small $C^\infty$} trajectories all led to head-on collisions at the speed of light. More interesting to physical modeling in general, the possibility of serrated solutions is a blow to the Lorentz-Dirac equation with a renormalized mass\cite{old_helium,Dirac}. Early twentieth-century works expanded deviating arguments in power series to yield a semiflow at the expense of carrying a renormalized mass. Besides breaking the time-reversibility, the self-interaction third-derivative term of the Lorentz-Dirac equation of motion\cite{Dirac} does not even make sense for a serrated orbit of {\small $C^2(\mathbb{R})$}\cite{old_helium,Dirac}. To show that any non-zero $\varepsilon$ makes a qualitative difference, we include a simple theorem on the repulsive case with $\varepsilon=0$ \cite{Driver}.

\begin{theorem} If a one-dimensional repulsive electromagnetic orbit belongs to {\small $C^2(\mathbb{R})$}, then it belongs to {\small $C^\infty(\mathbb{R})$}.
 \label{nosemiflow_theorem}

\begin{proof} Electrodynamics is defined by (\ref{1D_unfold}) with {\small $\varepsilon=0$}. Using Eq. (\ref{1D_unfold}) with {\small $\varepsilon=0$} and {\small $e_\mathsmaller{i}e_\mathsmaller{j}=1$} we obtain
{\small
\begin{eqnarray}
\negthickspace \negthickspace  \,\dot\phi_\mathsmaller{i}   &=& \frac{-1}{m_\mathsmaller{i} \cosh{\phi_\mathsmaller{i}}} \Big( \frac{ e^{-2\phi_\mathsmaller{j}}}{  2r^{\mathsmaller{2}}_{\mathsmaller{ji}}} +\frac{ e^{2\phi_\mathsmaller{k}}}{ 2 r^{\mathsmaller{2}}_{\mathsmaller{ki}}}\Big). \label{euler_1Dzero}
\end{eqnarray}}
\textbf{If} {\small $\phi_\mathsmaller{j}$} and {\small $\phi_\mathsmaller{k}$} belong to {\small $C^2(\mathbb{R})$}, we can take another derivative of the right-hand side of (\ref{euler_1Dzero}), and thus particle {\small $i$}'s trajectory belongs to {\small $C^3(\mathbb{R})$}. The same is concluded from the electronic equation of motion if {\small $\phi_\mathsmaller{s}$} and {\small $\phi_\mathsmaller{i}$} are in {\small $C^2(\mathbb{R})$}, i.e., that particle {\small $k$}'s trajectory belongs to {\small $C^3(\mathbb{R})$} as well. Successively, ad infinitum, we show that the orbit belongs to {\small $C^\infty(\mathbb{R})$}.
\end{proof}
\end{theorem}

On the other hand, when {\small $\varepsilon \neq 0$}, we can \textbf{not} take a derivative of (\ref{matlab1D}) for a generic orbit of {\small $C^2(\mathbb{R})$}. The derivative of the right-hand side of (\ref{matlab1D}) when {\small $\varepsilon \neq 0$} includes the derivative of {\small $\dot \phi_\mathsmaller{k}$} and {\small $\dot \phi_\mathsmaller{i}$}, whose arguments fall on the red segment illustrated in Fig. \ref{figODE}. If the red segments illustrated in Fig. \ref{figODE} belong to {\small $C^2(\mathbb{R})$}, the accelerations {\small $\dot \phi_\mathsmaller{k}$} and {\small $\dot \phi_\mathsmaller{i}$} can be continuous nowhere differentiable serrated functions. The same impossibility manifests in the general case, where the component {\small $ \hat{\mathcal{R}}_{\mathsmaller{j}}\cdot \mathbf{a}_{j}$} of the most advanced acceleration is given in terms of the past accelerations by (\ref{scalarprods}). For {\small $\varepsilon \neq 0$}, the {\small $ \hat{\mathcal{R}}_{\mathsmaller{j}}\cdot \mathbf{a}_{j}$} component will be a serrated function when the past segment is serrated. Unlike in Driver's problem\cite{Driver}, serrated orbits can be solutions for one-dimensional motion with attractive interaction when {\small $\varepsilon \neq 0$}.

 \subsection{The outer-cone distances of a mutual-recoil collision} 
 \label{outer-cone}
 Our next result is useful in a numerical perturbation theory to avoid a head-on collision at the speed of light. It shows that a mutual-recoil collision is optimal in the sense that the outer lightcone distances {\small $r_\mathsmaller{ij}$} and {\small $r_\mathsmaller{ks}$} are much larger than the internal lightcone distance {\small $r_\mathsmaller{ik}$} when the recoiling velocities are large. %The former ensures that the segment ratios (\ref{forwardNumbs}) andare small for large-velocity mutual recoils.
\begin{lem} For a mutual-recoil collision, the forward and the backward segment ratios, (\ref{forwardNumbs}) and  (\ref{backwardNumbs}), are approximated by 
{\small 
\begin{eqnarray} 
  \frac{r_\mathsmaller{\nu \alpha}}{r_\mathsmaller{\beta  \alpha}} =\frac{(1-||\bar{\mathbf{v}}_\mathsmaller{\beta})||)}{(1+||\bar{\mathbf{v}}_\mathsmaller{\beta})||)}.
\end{eqnarray}
}
\label{outside-cone-Lemma}
\end{lem}
\begin{proof} We start from $(\nu,\alpha,\beta)=(k,i,j)$ and assume particle {\small $k$} is in the past lightcone of particle {\small $i$} at time {\small $t_\mathsmaller{i}=0$}. By choosing the origin on the breaking point of particle {\small $i$} at time {\small $t_\mathsmaller{i}=0$} we can extrapolate particle {\small $k$}'s trajectory until the future lightcone of event {\small $(t_\mathsmaller{i},\mathbf{x}_\mathsmaller{i})=(0,0)$} by
{\small
\begin{eqnarray}
\mathbf{x}_\mathsmaller{j}(t_\mathsmaller{j})=r_\mathsmaller{ki}\hat{\mathbf{x}} + (t_\mathsmaller{j}+r_\mathsmaller{k i})\bar{\mathbf{v}}_\mathsmaller{j} ,  \notag
\end{eqnarray}}
where {\small $\bar{\mathbf{v}}_\mathsmaller{j}$} is the average velocity on the segment and the extrapolation is valid from {\small $t_\mathsmaller{j} \geq -r_\mathsmaller{ki}$} until the future lightcone time {\small $t_{\mathsmaller{j}}$}, which
according to (\ref{defi}) is
{\small
\begin{eqnarray}
t _{\mathsmaller{j}}=|| \mathbf{x}_\mathsmaller{j}(t_{\mathsmaller{j}})-0||=r_\mathsmaller{ki} + (t_{\mathsmaller{j}}+r_\mathsmaller{k i})(\hat{\mathbf{x}} \cdot \bar{\mathbf{v}}_\mathsmaller{j} ) , \notag 
\end{eqnarray}}
yielding the future lightcone time and inverse lightcone distance to be
{\small
\begin{eqnarray}
\negthickspace \negthickspace \negthickspace \negthickspace \negthickspace \negthickspace t_{\mathsmaller{j}}&=&\frac{(1+ \hat{\mathbf{x}} \cdot \bar{\mathbf{v}}_\mathsmaller{j})r_\mathsmaller{ki}}{(1-\hat{\mathbf{x}} \cdot \bar{\mathbf{v}}_\mathsmaller{j}) } ,\notag \\
\frac{1}{r_\mathsmaller{ji}}&=& \Big( \frac{1-\hat{\mathbf{x}} \cdot \bar{\mathbf{v}}_\mathsmaller{j} }{1+\hat{\mathbf{x}} \cdot \bar{\mathbf{v}}_\mathsmaller{j} }\Big) \frac{1}{r_\mathsmaller{ki}} \ll  \frac{1}{r_\mathsmaller{ki}},\label{ignore}
\end{eqnarray}
}
where the last sign holds when {\small $0<(1-\hat{\mathbf{x}} \cdot \bar{\mathbf{v}}_\mathsmaller{j}) \ll 1$}.
Formula (\ref{ignore}) holds for the backward segment ratio (\ref{backwardNumbs}) by using the shift {\small $(j,i,k) \rightarrow (s,k,i)$} and setting $\hat{\mathbf{x}} \rightarrow -\hat{\mathbf{x}}$.
\end{proof}

 \subsection{The action of the one-dimensional Lorentz group} 
   \label{LG}
 Lorentz transformations take hyperbolas into hyperbolas, and the coordinate transformation from the synchronized clocks {\small $(t_\mathsmaller{i}, x_\mathsmaller{i})$} of an inertial frame into the synchronized clocks {\small $(\bar{t}_\mathsmaller{i}, \bar{x}_\mathsmaller{i})$} of another inertial frame with boost velocity $-B$ is
 {\small
 \begin{eqnarray}
 \bar{t}_\mathsmaller{i}=\frac{t_\mathsmaller{i}-Bx_\mathsmaller{i}}{\sqrt{1-B^\mathsmaller{2}}}, \qquad \, \; \;\;\;\bar{x}_\mathsmaller{i}=\frac{x_\mathsmaller{i}-Bt_\mathsmaller{i}}{\sqrt{1-B^\mathsmaller{2}}}.  \label{LT}
 \end{eqnarray}
 }
 Notice that (\ref{LT}) preserves the light-cone condition, i.e., if {\small $(t_\mathsmaller{i}-t_\mathsmaller{k})^\mathsmaller{2}=(x_\mathsmaller{i}-x_\mathsmaller{k})^\mathsmaller{2}$} in the original frame and also {\small $(\bar{t}_\mathsmaller{i}-\bar{t}_\mathsmaller{k})^\mathsmaller{2}=(\bar{x}_\mathsmaller{i}-\bar{x}_\mathsmaller{k})^\mathsmaller{2}$}. The light-cone condition is preserved but the distance in lightcone changes, i.e., 
 {\small
 \begin{eqnarray}
 \bar{r}_{\mathsmaller{ki}}^{\mathsmaller{2}}=(\bar{t}_\mathsmaller{i}-\bar{t}_\mathsmaller{k})^\mathsmaller{2} =\Big(\frac{(t_\mathsmaller{i}-t_\mathsmaller{k})-B(x_\mathsmaller{i}-x_\mathsmaller{k})}{\sqrt{1-B^\mathsmaller{2}}}      \Big)^\mathsmaller{2}=e^{-2\varphi_B} r^{\mathsmaller{2}}_{\mathsmaller{ki}} ,  
 \end{eqnarray}
 }
 where in the last equality we have introduced the boost angle by {\small $B\equiv \tanh{\varphi_B}$} and used the convention of Fig. \ref{zorro} that {\small $i$} is in the future lightcone of {\small $k$}, namely {\small $(t_\mathsmaller{i}-t_\mathsmaller{k})=+(x_\mathsmaller{i}-x_\mathsmaller{k})$}. We stress that there is no such thing as a center of mass frame in the theory of relativity, and the boost transformation is to be applied to \textit{each} breaking point separately. To describe the collision from another frame, we express the left- and right- velocities \textit{at} the breaking point in terms of the pre-image with a boost parameter {\small $B$}, as follows. The Lorentz group transforms the velocities according to
{\small
\begin{eqnarray}
\negthickspace \negthickspace \negthickspace \negthickspace \negthickspace \negthickspace \negthickspace \negthickspace \negthickspace\negthickspace \negthickspace \negthickspace\negthickspace \negthickspace \negthickspace \bar{v}_\mathsmaller{i} =\frac{{v}_\mathsmaller{i}-B}{1-B{v}_\mathsmaller{i}}, \label{group}
\end{eqnarray}} 
where {\small $B$} is the (boost) parameter and {\small $\bar{v}_\mathsmaller{i}$} is the image of {\small ${v}_\mathsmaller{i}$} under the group action. Using (\ref{def_hyperphi2}), we express the boost parameter {\small $B$} and the velocity {\small $\bar{v}_\mathsmaller{i}$} as
{\small
\begin{eqnarray}
B&\equiv& \tanh{\varphi_\mathsmaller{B}}, \label{defboost}\\
\bar{v}_\mathsmaller{i}&\equiv& \tanh{\bar{\varphi}_\mathsmaller{i}} \label{def-preimage}.
\end{eqnarray}}
 Using the addition formulas for hyperbolic sines and cosines together with (\ref{def_hyperphi}), (\ref{defboost}) and (\ref{def-preimage}), the group action transforms Eq. (\ref{group}) into
{\small
\begin{eqnarray}
\tanh{\bar{\varphi}_\mathsmaller{i}}=\tanh{({\varphi}_\mathsmaller{i}-\varphi_\mathsmaller{B})} , \label{tangentaddition}
\end{eqnarray}}
thus showing that a change of inertial frame simply shifts the velocity angle by the boost angle, i.e.,
{\small
\begin{eqnarray}
\bar{\varphi}_\mathsmaller{i}={\varphi}_\mathsmaller{i}-\varphi_\mathsmaller{B}. \label{shiftphi}
\end{eqnarray}}

 \subsection{Lagrangian extension for an external electromagnetic field} 
 \label{external}
 
 In order to include an external electromagnetic field {\small $\big( E_{ext}(t,\mathbf{x}),B_{ext}(t,\mathbf{x}) \big)$} into the equation of motion  (\ref{Singular_equation_aj}), one can add an external-force to the definition (\ref{defDDE}) of {\small $\Lambda_{\mathsmaller{\, i}}^{\mathsmaller{\circledS}}$}, i.e.,
 {\small
 \begin{eqnarray}
\Lambda_{\mathsmaller{\, i}}^{\mathsmaller{\circledS}}& \rightarrow & \Lambda_{\mathsmaller{\, i}}^{\mathsmaller{\circledS}} +e_{\! \mathsmaller{\, i}} \Big(  \mathbf{E}_{ext}(t,\mathbf{x}_i) +\mathbf{v}_{\! \mathsmaller{\, i}} \times \mathbf{B}_{ext}(t,\mathbf{x}_i) -(\mathbf{v}_{\! \mathsmaller{\, i}} \cdot \mathbf{E}_{ext}(t,\mathbf{x}_i))\mathbf{v}_{\! \mathsmaller{\, i}} \Big). \label{Proca}  
 \end{eqnarray}}
 The Lagrangian description of the external force is obtained by adding a linear function of each charge's velocity to the {\small $\varepsilon$}-strong functional (\ref{Fokker}). The oscillatory term is called the Proca Lagrangian in Ref. \cite{Jackson} when the external forcing is an electromagnetic wave. %Notice that an external wave can exist only in the three dimensional case.  As an artifact of simplified modeling, it is possible to include an external oscillatory electric field already in the one-dimensional case of Eq. (\ref{matlab1D}), but the resulting motion might not display the same features of the three-dimensional case.


\begin{thebibliography}{99}



\bibitem{cinderela} J.  De Luca, {\em Variational Electrodynamics of Atoms}, \textit{Progress In Electromagnetics Research B} {\bf 53} (2013), 147-186 (40pp).

\bibitem{JDE1}J. De Luca, \emph{Equations of Motion for Variational Electrodynamics}, \textit{Journal of Differential Equations} \textbf{260} (2016), 5816-5833 (18pp).

\bibitem{JDE2}J. De Luca, \emph{Chemical Principle and PDE of Variational Electrodynamics}, \textit{Journal of Differential Equations} \textbf{268} (2019), 272-300 (29pp).
\bibitem{Mallet-Paret1}J. Mallet-Paret, \emph{Generic properties of retarded functional differential equations}, \textit{Bull. Amer. Math. Soc.} \textbf{81} (1975), 750-752, J. Mallet-Paret, \emph{Generic periodic solutions of functional differential equations}, \textit{Journal of Differential Equations} \textbf{25} (1977), 163-183.

\bibitem{Mallet-Paret_bound-layer} J. Mallet-Paret and R. Nussbaum, \emph{Boundary layer phenomena for differential-delay equations with state-dependent time lags}, \textit{J. Reine Angew. Math.} \textbf{477} (1996), 129-197 and \emph{Boundary layer phenomena for differential-delay equations with state-dependent time lags-III}, \textit{Journal of Differential Equations} \textbf{189} (2003), 640-692.

\bibitem{JackHale} J. Hale, \emph{Theory of Functional Differential Equations},
Springer-Verlag (1977), J. Hale and S. M. Verduyn Lunel \emph{Introduction to
Functional Differential Equations}, Springer-Verlag, New York (1993).


\bibitem{Nicola2}N. Guglielmi and E. Hairer, \emph{Numerical approaches for state-dependent neutral-delay equations with discontinuities}, \textit{Mathematics and Computers in Simulation} \textbf{95} (2013), 2-12 and G. Fusco and N. Guglielmi, \emph{A regularization for discontinuous differential equations with application to state-dependent delay differential equations of neutral-type}, \textit{Journal of Differential Equations} \textbf{250} (2011), 3230-3279.

\bibitem{EFY2}E. B. Hollander and J. De Luca, \emph{Regularization of the collision in the electromagnetic two-body problem}, \textit{Physical Review E} \textbf{14} (2004), 1093-1104 (12pp).

\bibitem{Hans}C. M. Andersen and Hans C. von Baeyer, \emph{Almost Circular Orbits in Classical Action-at-a-Distance Electrodynamics}, \textit{Physical Review D} \textbf{5} (1972), 802.


\bibitem{Martinez} D. J. Louis-Martinez, \emph{Relativistic non-instantaneous action-at-a-distance interactions}, \textit{Physics Letters B} \textbf{632} (2006) 733-739.

\bibitem{JMP2009} J. De Luca, \emph{Variational principle for the Wheeler-Feynman electrodynamics}, \textit{Journal of Mathematical Physics} \textbf{50} (2009), 062701 (24pp).

\bibitem{Driver}R. D. Driver, \emph{Can the future influence the present?}, {\it Physical Reviw D} \textbf{19} (1979) 1098-1107. 

\bibitem{CAM}J. De Luca, A. R. Humphries and S. B. Rodrigues, \emph{Finite-element boundary value integration of Wheeler-Feynman electrodynamics}, \textit{Journal of Computational and Applied Mathematics} \textbf{236} (2012), 3319-3337.

\bibitem{Daniel}D. C. De Souza and J. De Luca, \emph{Solutions of the Wheeler-Feynman equations with discontinuous velocities}, \textit{Chaos: An Interdisciplinary Journal of Nonlinear Science} \textbf{25} (2015), 013102 (10pp).



\bibitem{no-interaction} G. Marmo, G. N. Mukunda, and E. C. G. Sudarshan, \emph{Lagrangian proof of the no-interaction Theorem}, \textit{Phys. Rev. D} \textbf{30}, 2110-2116 (1984).

\bibitem{Jackson} J. D. Jackson, \emph{Classical Electrodynamics}, John Wiley and Sons, New York (1975).

\bibitem{Daniel_Mike}D. C. De Souza and M. C. Mackey, \emph{Response of an oscillatory differential delay equation to a periodic stimulus,} {\it Journal of Mathematical Biology}  \textbf{78} (2019) 1637-1679.

\bibitem{Gelfand}I. M. Gelfand and S. V. Fomin, \emph{Calculus of Variations}, Dover, New York (2000),  pgs.61-63.

\bibitem{Jianhong}H. Shu, W. Xu, X.-S. Wang and J. Wu, \emph{Complex dynamics in a delay differential equation with two delays in tick growth with diapause}, {\it Journal of Differential Equations }\textbf{269} (2020) 10937-10963. 


\bibitem{EFY1}E. B. Hollander and J. De Luca, \emph{Two-degree-of-freedom Hamiltonian for the time-symmetric two-body problem of the relativistic action-at-a-distance electrodynamics}, \textit{Physical Review E} \textbf{67}, 026219 (2003) (15pp). 

%\bibitem{Politi}J. De Luca, N. Guglielmi, A. Humphries and A. Politi, \textit{J. Phys. A} \textbf{43} (2010) 205103 (20pp).

\bibitem{Schild}A. Schild, \emph{Electromagnetic two-body problem}, \textit{Phys. Rev.} \textbf{131}, (1963) 2762.

\bibitem{Aarseth}S. J. Aarseth and K. Zare, \emph{A regularization of the three-body problem}, {\it Celestial mechanics} \textbf{10} (1974) 185-205.


\bibitem{Cheeger}J. Cheeger, \emph{Differentiability of Lipschitz Functions on Metric spaces}, {\it GAFA Geometric and Functional Analysis}, \textbf{9} (1999) 428-517.

\bibitem{Brault}A. Brault and A. Lejay, \emph{The non-linear sewing lemma II: Lipischitz continuous formulation} \textit{Journal of Differential Equations} \textbf{293} (2021) 482-519. 

\bibitem{simple} J. De Luca, \emph{Simple dynamical system with discrete bound states}, {\it Phys. Rev. E} {\bf 62} (2000), 2060-2067.

\bibitem{old_helium} J. De Luca, \emph{Electrodynamics of helium with retardation and self-interaction}, {\it Phys. Rev. Lett.} {\bf 80} (1998), 680-683 and  J. De Luca, \emph{Electrodynamics of a two-electron atom with retardation and self-interaction}, {\it Phys. Rev. E} {\bf 58} (1998), 5727-5741.

\bibitem{double-slit} J. De Luca, {\em Electromagnetic models to complete quantum mechanics}, {\it Journal of  Computational and Theoretical Nanoscience} {\bf 8}, (2011) 1040-1051.





%\bibitem{simple} J. De Luca, \emph{Simple dynamical system with discrete bound states}, {\it Phys. Rev. E} {\bf 62} (2000), 2060-2067.

%\bibitem{pre_hydrogen} J. De Luca, \emph{Stiff three-frequency orbit of the hydrogen atom}, \textit{Physical Review E} \textbf{73} (2006), 026221 (17pp).



%\bibitem{Daniel_Mike}D. C. De Souza and Michael C. Mackey, \emph{Response of an oscillatory differential delay equation to a periodic stimulus,} {\it Journal of Mathematical Biology}  \textbf{78} (2019) 1637-1679.


%\bibitem{Jianhong}H. Shu, W. Xu, X.-S. Wang and J. Wu, \emph{Complex dynamics in a delay differential equation with two delays in tick growth with diapause}, {\it Journal of Differential Equations }\textbf{269} (2020) 10937-10963. 



%\bibitem{old_helium} J. De Luca, \emph{Electrodynamics of helium with retardation and self-interaction}, {\it Phys. Rev. Lett.} {\bf 80} (1998), 680-683 and  J. De Luca, \emph{Electrodynamics of a two-electron atom with retardation and self-interaction}, {\it Phys. Rev. E} {\bf 58} (1998), 5727-5741.



%\bibitem{pre_hydrogen} J. De Luca, \emph{Stiff three-frequency orbit of the hydrogen atom}, \textit{Physical Review E} \textbf{73} (2006), 026221 (17pp).


\bibitem{minimizer} J. De Luca, \emph{Minimizers with discontinuous velocities for the electromagnetic variational method}, \textit{Physical Review E} \textbf{82} (2010), 026212 (9pp).



\bibitem{Dirac}P. A. M. Dirac, \emph{Classical theory of radiating electrons}, \textit{Proc. Royal Society London}, \textbf{167}, 148-169 (1938).




\end{thebibliography}
\end{document}